\RequirePackage{ifpdf}

\documentclass[english]{JHEP3}
\usepackage[T1]{fontenc}
\usepackage[latin9]{inputenc}
\usepackage{amssymb}
\usepackage{mathdots}
\usepackage[numbers]{natbib}
\usepackage[tikz]{bclogo}

\usepackage{pdflscape}

\makeatletter
\pdfoutput=1

\providecommand{\tabularnewline}{\\}

\usepackage{amsmath,amssymb,amsthm}%,amscd}
\newcommand{\CB}{{\cal B}}

\newcommand{\CL}{{\cal L}}

\newcommand{\CN}{{\cal N}}
\newcommand{\CO}{{\cal O}}

\newcommand{\CR}{{\cal R}}
\newcommand{\CS}{{\cal S}}

\def\BZ{{\mathbb Z}}
\def\BR{{\mathbb R}}
\def\BC{{\mathbb C}}

\def\BS{{\mathbb S}}

\newcommand{\be}{\begin{equation}}
\newcommand{\ee}{\end{equation}}
\newcommand{\ba}{\begin{aligned}}
\newcommand{\ea}{\end{aligned}}
\newcommand{\bea}{\begin{eqnarray}}
\newcommand{\eea}{\end{eqnarray}}
\newcommand{\bean}{\begin{eqnarray*}}
\newcommand{\eean}{\end{eqnarray*}}

\def\r{\right\rangle}

\def\1{\mathbf{1}}
\def\0{|\1\r}

\newcommand{\rme}{{\rm e}}
\newcommand{\rmi}{{\rm i}}
\newcommand{\rmd}{{\rm d}}

\newdimen\tableauside\tableauside=1.0ex
\newdimen\tableaurule\tableaurule=0.4pt
\newdimen\tableaustep
\def\phantomhrule#1{\hbox{\vbox to0pt{\hrule height\tableaurule width#1\vss}}}
\def\phantomvrule#1{\vbox{\hbox to0pt{\vrule width\tableaurule height#1\hss}}}
\def\sqr{\vbox{%
  \phantomhrule\tableaustep
  \hbox{\phantomvrule\tableaustep\kern\tableaustep\phantomvrule\tableaustep}%
  \hbox{\vbox{\phantomhrule\tableauside}\kern-\tableaurule}}}
\def\squares#1{\hbox{\count0=#1\noindent\loop\sqr
  \advance\count0 by-1 \ifnum\count0>0\repeat}}
\def\tableau#1{\vcenter{\offinterlineskip
  \tableaustep=\tableauside\advance\tableaustep by-\tableaurule
  \kern\normallineskip\hbox
    {\kern\normallineskip\vbox
      {\gettableau#1 0 }%
     \kern\normallineskip\kern\tableaurule}%
  \kern\normallineskip\kern\tableaurule}}
\def\gettableau#1{\ifnum#1=0\let\next=\null\else
\squares{#1}\let\next=\gettableau\fi\next}
\preprint{
{\small{\textsf{CERN-PH-TH-2014-118}}}}

\title{Resurgent Analysis of Localizable Observables in Supersymmetric Gauge Theories}

\author{In\^es Aniceto$^{\dagger,}\footnote{From October 2014: Institute of Physics, Jagiellonian University, ul. {\L}ojasiewicza 11, 30--348 Krak\'ow, Poland.}$, Jorge G. Russo$^{\ddagger,\natural}$ and Ricardo Schiappa$^{\sharp,\dagger}$
\\
$^{\dagger}$CAMGSD, Departamento de Matem\'atica, Instituto Superior T\'ecnico, Universidade de Lisboa\\
Av. Rovisco Pais 1, 1049--001 Lisboa, Portugal\\
\\
$^{\ddagger}$Instituci\'o Catalana de Recerca i Estudis Avan\c cats (ICREA),\\
Pg. Lluis Companys, 23, 08010 Barcelona, Spain\\
\\
$^{\natural}$ECM Department and Institute for Sciences of the Cosmos, Facultat de F\'\i sica,\\
Universitat de Barcelona, Mart\'i Franqu\`es 1, E08028 Barcelona, Spain\\
\\
$^{\sharp}$Theory Division, Department of Physics, CERN,\\ 
CH--1211 Gen\`eve 23, Switzerland\\
\\
\email{ianiceto@math.tecnico.ulisboa.pt}, \quad
\email{jorge.russo@icrea.cat}, \quad
\email{schiappa@math.tecnico.ulisboa.pt}

}
\abstract{
Localization methods have recently led to a plethora of new exact results in supersymmetric gauge theories, as certain observables may be computed in terms of matrix integrals. These can then be evaluated by making use of standard large $N$ techniques, or else via perturbative expansions in the gauge coupling. Either approximation often leads to observables given in terms of asymptotic series, which need to be properly defined in order to obtain nonperturbative results. At the same time, resurgent analysis has recently been successfully applied to several problems, \textit{e.g.}, in quantum, field and string theories, precisely to overcome this issue and construct nonperturbative answers out of asymptotic perturbative expansions. The present work uses exact results from supersymmetric localization to address the resurgent structure of the free energy and partition function of Chern--Simons and ABJM gauge theories in three dimensions, and of $\CN=2$ supersymmetric Yang--Mills theories in four dimensions. For each case, the complete structure of Borel singularities is exactly determined, and the relation of these singularities with the large--order behavior of (multi--instanton) perturbative expansions is made fully precise.
}

\keywords{Resurgence, Gauge Theory, Supersymmetry, Localization, Perturbation Theory, Nonperturbative Definitions, Large--Order Analysis, Borel Singularities}

\makeatother

\usepackage{babel}

\begin{document}

%%%%%%%%%%%%%%%%%%%%%%%%%%%%%%%%%%%%%%%%%%%%%%%%%%%%%%%%%%%%%%%%%
%%%%%%%%%%%%%%%%%%%%%%%%%%%%%%%%%%%%%%%%%%%%%%%%%%%%%%%%%%%%%%%%%
\maketitle
%%%%%%%%%%%%%%%%%%%%%%%%%%%%%%%%%%%%%%%%%%%%%%%%%%%%%%%%%%%%%%%%%
%%%%%%%%%%%%%%%%%%%%%%%%%%%%%%%%%%%%%%%%%%%%%%%%%%%%%%%%%%%%%%%%%

\vfill

\eject

\allowdisplaybreaks

%%%%%%%%%%%%%%%%%%%%%%%%%%%%%%%%%%%%%%%%%%%%%%%%%%%%%%%%%%%%%%%%%
%%%%%%%%%%%%%%%%%%%%%%%%%%%%%%%%%%%%%%%%%%%%%%%%%%%%%%%%%%%%%%%%%
\section{Introduction and Summary}
%%%%%%%%%%%%%%%%%%%%%%%%%%%%%%%%%%%%%%%%%%%%%%%%%%%%%%%%%%%%%%%%%
%%%%%%%%%%%%%%%%%%%%%%%%%%%%%%%%%%%%%%%%%%%%%%%%%%%%%%%%%%%%%%%%%

Understanding the strong--coupling dynamics of gauge theories requires detailed knowledge of their nonperturbative structure. In many cases, however, the nonperturbative sector may be very complicated and the study of physical observables must mainly rely upon perturbation theory. This may apparently amount to further complications as often perturbative expansions are asymptotic, with their coefficients displaying factorial growth. However, this divergence of perturbation theory is precisely related to the nature of the nonperturbative effects in the gauge theory under scrutiny, materialized as singularities in the complex Borel plane and in most cases associated with instantons (see, \textit{e.g.}, \citep{ZinnJustin:1980uk}) and renormalons (see, \textit{e.g.}, \citep{Beneke:1998ui}). This essentially implies that there is more to perturbation theory than what first meets the eye.

Due to the aforementioned relation between the asymptotic nature of the perturbative expansion and the nature of nonperturbative effects in a given gauge theory, one may wonder to what extent is the nonperturbative information of the theory \textit{already} encoded in the original asymptotic perturbative series. The precise mathematical technique to carry through this extraction of nonperturbative information out of perturbative data goes under the name of \textit{resurgence} (see, \textit{e.g.}, \citep{Candelpergher:1993np, Delabaere:1999ef, Seara:2003ss, Edgar:2008ga} for introductory mathematical reviews). Recently, resurgence has also been steadily applied to diverse physical problems with many interesting results, \textit{e.g.}, covering from quantum mechanics \citep{ZinnJustin:1981dx, ZinnJustin:1982td, Voros:1983, Voros:1993an, Voros:1994, Delabaere:1990eh, ZinnJustin:2003jt, Jentschura:2004jg, Jentschura:2004ib, Jentschura:2004cg, Dunne:2013ada, Basar:2013eka, Dunne:2014bca} to quantum field theory \citep{Stingl:2002cb, Dunne:2012ae, Dunne:2012zk, Cherman:2013yfa, Cherman:2014ofa}, from gauge theories and random matrices \citep{Marino:2007te, Marino:2008ya, Marino:2008vx, Schiappa:2009cc, Klemm:2010tm, Drukker:2011zy, Aniceto:2011nu, Argyres:2012vv, Argyres:2012ka, Schiappa:2013opa} to string theory \citep{Marino:2006hs, Marino:2007te, Marino:2008ya, Pasquetti:2009jg, Garoufalidis:2010ya, Drukker:2011zy, Aniceto:2011nu, Schiappa:2013opa, Santamaria:2013rua, Grassi:2014cla, Couso-Santamaria:2014iia}. Our focus in this paper is to pursue the programme of developing resurgence within gauge theories with the long--term goal of unraveling generic strong--coupling dynamics out of perturbative data. We begin this research within the simplest settings, supersymmetric gauge theories in three and four dimensions, focusing upon observables which localize into matrix models \citep{Pestun:2007rz}. We shall explore the singular structure in the complex Borel plane, and its relation to nonperturbative effects and the large--order behavior of the perturbative series (in this regard also following older work such as \citep{Bender:1969si, Bender:1990pd, Lipatov:1976ny, ZinnJustin:2002jb}).

Let us consider an asymptotic perturbative expansion of the form
\be
\label{firstasymptseries}
F (g) \simeq \sum_{n=0}^{+\infty}\, g^{n} F_n,
\ee
\noindent
where $F$ is some observable, $g$ is the (small) perturbative parameter, and $F_n$ are the coefficients which grow factorially fast at large order $n$, as $F_n \sim n!$. The standard way to make sense out of such an expansion relies upon resummation methods such as Borel resummation (see, \textit{e.g.}, the discussions in \citep{Aniceto:2011nu, Aniceto:2013fka} for quick reviews). This procedure starts by removing the factorial growth out of the coefficients $F_n$ in order to construct the Borel transform, $\CB[F]$, which essentially focuses upon the subleading growth of the original coefficients (as $\CB[g^n](s) = \frac{s^{n-1}}{(n-1)!}$). Then, the Borel resummation $\CS_\theta F$ involves a Laplace--type integration of the Borel transform, which may be applied along all complex directions $\theta$ of the perturbative parameter except, of course, for those which meet singularities in the Borel plane and the integration is thus ill--defined (these are known as Stokes lines). This fact is at the root of Stokes phenomena---exponentially small contributions which might be absent before crossing a Stokes line, and invisible to the perturbative expansion, cannot be disregarded afterwards as they may grow and take dominance---and as such all exponentially suppressed contributions associated with nonperturbative sectors need to be taken into account in order to get a correct physical result along any direction. That this must be so may be understood precisely at these Stokes directions, where only lateral Borel resummations may be defined\footnote{To be a bit more precise, the Borel resummation $\mathcal{S}_{\theta} F$ is a Laplace transform whose integration contour is taken along the direction $\theta$. If this direction is a Stokes line, one can then define lateral Borel resummations, $\mathcal{S}_{\theta^{\pm}} F$, whose integration contour goes just to the left or just to the right of the Stokes line. The lateral Borel resummations avoid the singularities along the Stokes lines and are thus well--defined.}. In this case the resummation is ambiguous as there is a difference between the two lateral resummations given by an exponentially small contribution. In order to obtain an unambiguous, sensible, physical result, we need to include all nonperturbative sectors of the theory in a very particular way, using resurgent analysis and transseries (see the original physical motivation in \citep{'tHooft:1977am, Bogomolny:1980ur, ZinnJustin:1981dx, ZinnJustin:1982td} and, \textit{e.g.}, the recent general framework on the median resummation in \citep{Aniceto:2013fka}). Indeed, instead of just the familiar perturbative series, one ought to consider a \textit{transseries} solution in order to fully describe the system, in which one performs a formal expansion not only in powers of the perturbative parameter, but also in nonanalytical functions of this parameter such as logarithmic and exponentially suppressed terms. In other words, the transseries takes into account all nonperturbative components of the problem at hand. The intimate connection between perturbative and nonperturbative sectors is then very precisely defined by resurgent analysis. Further, relations between \textit{all} different sectors (perturbative and nonperturbative alike) can be studied exactly. We refer the reader to \citep{Upcoming:2014} for an overview of these methods.

Four--dimensional Yang--Mills theory exhibits both ultraviolet and infrared renormalon singularities, dominant as compared to the instanton singularities, and generically hard to deal with---in fact, already when it comes to producing perturbative large--order data (see, \textit{e.g.}, \citep{Beneke:1998ui}). Turning on supersymmetry makes for milder settings, as first discussed in \citep{Dunne:2012ae} concerning the interplay between resurgence and different numbers of supersymmetries. Renormalon singularities must be absent in superconformal field theories---in the complex Borel plane they would be located at $\sim - \frac{1}{\beta_0} \to - \infty$ at conformality (here $\beta_0$ is the first coefficient of the beta function). For non--conformal theories, infrared renormalon singularities (located on the positive real axis in the complex Borel plane) are expected to be absent in $\CN=2$ theories on a four--sphere, while ultraviolet renormalon singularities (located on the negative real axis) should in principle be present for non--BPS observables. One thus expects that, at least on what concerns asymptotic expansions in the gauge coupling, BPS observables will have simpler resurgent structures. 

Expecting milder Borel singular structures due to supersymmetry, one next asks for classes of observables where large--order data can be produced to very high orders. In non--supersymmetric nonabelian gauge theories, perturbative expansions in the gauge coupling are only feasible to the first few orders. On the other hand, in supersymmetric gauge theories exact results to all orders in the coupling can be obtained by localization techniques. These techniques have been applied for the study of the partition function and Wilson loops in certain (supersymmetric) gauge theories in  three \citep{Beasley:2005vf, Kapustin:2009kz} and four dimensions \citep{Pestun:2007rz}, and they lead to simple expressions for supersymmetric observables expressed in terms of finite--dimensional matrix--model integrals. This thus opens the possibility of obtaining exact results for the coefficients of the asymptotic expansions associated with these observables. As we shall see throughout this paper, this immediately allows for exact resurgent and Borel analyses in supersymmetric gauge theories.

The matrix model formulation of localizable observables usually depends upon two parameters: the rank of the gauge group, $N$, and the gauge coupling\footnote{The coupling constant will differ for each theory. In three dimensions the Chern--Simons coupling $g_s$ is related to the level, $k$, by $g_s=2\pi\rmi/k$; while in four dimensions the Yang--Mills coupling is generically denoted by $g_{\text{YM}}$.} of the theory, $g$. Perturbative analyses may be either at large $N$ or at small coupling---and to each one may analyze convergence properties of the respective expansions and consequently develop their resurgent analysis. Naturally, in the familiar case of a double expansion in $1/N$ and 't~Hooft parameter, a complete transseries representation of the aforementioned observables will have to include nonperturbative effects associated with all parameters. This is a highly nontrivial problem, although steady progress has been recently achieved for the particular case of the ABJM partition function (see \citep{Grassi:2014cla} and references therein). In the present work, we shall follow along these steps and develop the ideas originally laid down in \citep{Russo:2012kj}, but now augmented via the use of resurgent analysis. We shall either focus on the large $N$ expansion at fixed 't~Hooft coupling, whenever exact results are known for the respective coefficients; or otherwise on a small coupling expansion at fixed rank of the gauge group. We shall assume the reader is at least familiar with the resurgent basics outlined in \citep{Upcoming:2014}, in the sense that we shall not review any of these essentials but rather directly use the techniques described in the aforementioned references in what follows.

This paper is organized as follows. We start in section \ref{sec:Chern-Simons-theory} with the analysis of the resurgence properties of three--dimensional gauge theoretic observables. In subsection \ref{sec:CSLS} we present a warm--up (simpler) example, addressing the free energy of Chern--Simons gauge theory on lens spaces, $L(p,1)$, from the standpoint of the large $N$ expansion. The associated Borel transform is shown to be a meromorphic function with double--poles falling upon an infinite (countable) number of distinct Stokes lines. Resurgent analysis further allows us to determine the discontinuities across these singular lines, and to obtain complete large--order formulae recovering the exact free energy coefficients. A slightly different approach is followed in subsection \ref{sec:ABJM}, where we perform a perturbative analysis of $\text{U}(2) \times \text{U}(2)$ ABJM gauge theory localized on $\mathbb{S}^3$, at large Chern--Simons level $k$. This time around there are only two Stokes lines, positive and negative imaginary axes. Again, resurgent analysis precisely determines their associated discontinuities and, once again, we can recover the precise perturbative coefficients out of an exact large--order analysis. In section \ref{sec:4d-SYM} we then turn to the resurgent analysis of four--dimensional gauge theoretic observables, within the context of supersymmetric Yang--Mills (SYM) theories. We first give a brief overview of the known asymptotic properties of the partition function and supersymmetric Wilson loops of  $\mathcal{N}=4$ SYM theory localized on $\mathbb{S}^4$, for both large $N$ and small coupling $g_{\text{YM}}$ cases, in subsection \ref{sec:N4}. This is a particularly simple case, and resurgent analysis is quite straightforward. Things become more intricate as we turn to the analysis of the partition function of $\CN=2$ superconformal $\text{S}\text{U}(2)$ SYM theory on $\mathbb{S}^4$ (with four massless multiplets) at small gauge coupling, which we do in subsection \ref{sec:N2}. Herein we again find that the Borel transform is a meromorphic function but this time around with poles lying along a single Stokes direction: the negative real axis. However, unlike in earlier examples, these singularities are poles of higher and higher degree the further one moves away from the origin. Albeit being more intricate, resurgent analysis still allows for a determination of the Stokes discontinuities, which are verified by matching against the perturbative large--order behavior. Finally, in subsection \ref{sec:N2star}, we outline the same analysis for the case of the $\CN=2^{*}$ SYM theory on $\mathbb{S}^4$, with gauge group $\text{S}\text{U}(2)$. This theory introduces an extra parameter, $M$, representing the mass of the hypermultiplet, and the singularities in the Borel plane will depend on it. In particular, at fixed finite $M$, we find a countable infinity of Stokes lines, each with a single pole. The limiting cases of $M\rightarrow 0$ and $M\rightarrow +\infty$ are also of great interest as they interpolate between superconformal $\mathcal{N}=4$ and pure $\mathcal{N}=2$ SYM theories. We end our analysis in section \ref{sec:physical-int}, where we discuss the (semiclassical) physical interpretation of the many Borel singularities found in our previous  resurgent analyses. We discuss which physical effects give rise to the (asymptotic) factorial growth of the perturbative expansions (thus being responsible for the resurgent properties of the considered observables). In particular, the Borel singularities coincide with resonances that arise when zero modes appear in the Coulomb branch of the gauge theory moduli space. The paper closes in section  \ref{sec:comments}, with some conclusions and a brief outline of some open problems for future research. An appendix collects some technical results needed for the analysis carried through in the main body of the text.

%%%%%%%%%%%%%%%%%%%%%%%%%%%%%%%%%%%%%%%%%%%%%%%%%%%%%%%%%%%%%%%%%
%%%%%%%%%%%%%%%%%%%%%%%%%%%%%%%%%%%%%%%%%%%%%%%%%%%%%%%%%%%%%%%%%
\section{Resurgence in Chern--Simons Gauge Theories}
\label{sec:Chern-Simons-theory}
%%%%%%%%%%%%%%%%%%%%%%%%%%%%%%%%%%%%%%%%%%%%%%%%%%%%%%%%%%%%%%%%%
%%%%%%%%%%%%%%%%%%%%%%%%%%%%%%%%%%%%%%%%%%%%%%%%%%%%%%%%%%%%%%%%%

This section focuses on three--dimensional gauge theories, namely, Chern--Simons gauge theory on lens spaces (and in their limiting case, $\mathbb{S}^{3}$), and ABJM gauge theory on the sphere, $\mathbb{S}^{3}$. The matrix model description of the partition function of these theories is typically of the form \citep{Marino:2002fk, Kapustin:2009kz}
\begin{equation}
\label{eq:Z-with-Z-1-loop}
Z \propto \int_{\mathfrak{g}} \left[\rmd \lambda \right]\, \rme^{- S_{\text{cl}} (\lambda)}\, Z_{\text{1-loop}} (\lambda).
\end{equation}
\noindent
The integration is over the Lie algebra $\mathfrak{g}$ associated with the gauge group $G$, and $S_{\text{cl}} (\lambda)$ comes from the classical action on $\mathbb{S}^{3}$, with $\lambda \in \mathfrak{g}$. The factor $Z_{\text{1-loop}} \left(\lambda\right)$ comes from one--loop determinant in the localization procedure, and is typically given in terms of hyperbolic functions, $\sinh$ and $\cosh$. Our main interest is to analyze the resurgent properties of the above partition function for the aforementioned theories. They depend on the Chern--Simons level, $k$, and the rank of the gauge group, $N$, and the asymptotic properties that we shall analyze will be obtained by fixing one of these parameters and taking the other to be large. Resurgence in these cases will turn out to be somewhat simple, and as such these examples will also provide for an introduction to the upcoming less trivial cases of four--dimensional gauge theories.

%%%%%%%%%%%%%%%%%%%%%%%%%%%%%%%%%%%%%%%%%%%%%%%%%%%%%%%%%%%%%%%%%
\subsection{Chern--Simons Gauge Theory on Lens Spaces}\label{sec:CSLS}
%%%%%%%%%%%%%%%%%%%%%%%%%%%%%%%%%%%%%%%%%%%%%%%%%%%%%%%%%%%%%%%%%

The partition function for Chern--Simons gauge theory, with gauge group $G$ and on a generic three--manifold $M$, is defined as (see, \textit{e.g.}, \citep{Witten:1988hf})
\begin{equation}
Z_{\text{CS}} (M) = \int\left[\rmd A\right]\, {\rme}^{{\rmi} S_{\text{CS}} (A)},
\end{equation}
\noindent
where the Chern--Simons action is given by
\begin{equation}
S_{\text{CS}} (A) = \frac{k}{4\pi} \int_{M} \text{Tr} \left( A \wedge A + \frac{2}{3} A \wedge A \wedge A \right).
\end{equation}
\noindent
In the above expressions $k \in \BZ$ is the Chern--Simons coupling constant and $A$ is a Lie algebra $\mathfrak{g}$--valued gauge connection. It has been known for quite a while that this partition function may be explicitly written as a matrix integral \citep{Marino:2002fk, Aganagic:2002wv}, via the use of localization techniques (see, \textit{e.g.}, \citep{Marino:2011nm} for a review). For the cases of interest in this section we shall consider $G=\text{U}(N)$.

While the simplest possible example deals with Chern--Simons on the three--sphere, we shall consider the slightly more general example of lens spaces $M = L(p,1) \simeq \mathbb{S}^{3}/\mathbb{Z}_{p}$ (which reverts back to the three--sphere when $p=1$). In this case, it was shown in \citep{Marino:2002fk, Aganagic:2002wv, Halmagyi:2003ze, Halmagyi:2003mm} that the Chern--Simons partition function may be written as a multi--cuts matrix integral. One finds $p$ sets of eigenvalues (corresponding to their distribution over $p$ cuts), labeled by an index $I \in \left\{ 0, \ldots, p-1 \right\}$. The measure factor is a product of a self--interacting factor, $D_{1} \left( \lambda \right)$, with a term involving interactions between different sets of eigenvalues, $D_{2} \left( \lambda \right)$. These are given by
\begin{eqnarray}
D_{1} \left( \lambda \right) &=& \prod_{I=0}^{p-1}\, \prod_{1 \le i<j \le N_I} \left( 2 \sinh \left( \frac{\lambda_{i}^{I}-\lambda_{j}^{I}}{2} \right) \right)^{2}, \\
D_{2} \left( \lambda \right) &=& \prod_{0 \le I<J \le p-1}\, \prod_{i=1}^{N_I} \prod_{j=1}^{N_J} \left( 2 \sinh \left( \frac{\lambda_{i}^{I}-\lambda_{j}^{J}+d^{IJ}}{2} \right) \right)^{2},
\end{eqnarray}
\noindent
where $d^{IJ}=\frac{2\pi\rmi}{p} \left(I-J\right)$. The potential term is given by
\be
V(\lambda) = p\, \sum_{I=0}^{p-1} \sum_{i=1}^{N_I} \left(\lambda_{i}^{I}\right)^{2},
\ee
\noindent
and the partition function is finally written as 
\begin{equation}
Z_{\text{CS}}\left( L(p,1) \right) = \frac{\rmi^{-\frac{1}{2}\left(\sum_{I=0}^{p-1} N_{I}^{2}\right)}}{\prod_{I=0}^{p-1} N_{I}!} \int\, \prod_{I=0}^{p-1} \prod_{i=1}^{N_I} \frac{\rmd \lambda_{i}^{I}}{2\pi}\, D_{1}\left( \lambda \right) D_{2}\left( \lambda \right)\, \rme^{-\frac{1}{2g_{s}} V\left( \lambda\right)}, \qquad g_s \equiv \frac{2\pi \rmi}{k}.
\end{equation}
\noindent
Do notice that the partition function explicitly depends upon the $\left\{ N_{I}\right\} $, which are the number of eigenvalues in each cut $I$. The (partial) 't~Hooft parameters are $t_{I}=g_{s} N_{I}$, depending on the number of eigenvalues in each cut, and $t=g_{s}N$ the total 't~Hooft coupling. From the standpoint of the original Chern--Simons gauge theory, this split of eigenvalues across multiple cuts should be understood as an expansion of the gauge theoretic partition function around a non--trivial flat connection, with the following symmetry--breaking pattern of the original gauge group:
\be
\text{U} \left(N\right) \rightarrow \text{U} \left(N_{0}\right) \times \cdots \times \text{U} \left(N_{p-1}\right).
\ee

Let us briefly note that it is possible to find an hermitian formulation of the above lens space matrix integral, generalizing to arbitrary $p$ a result which is well--known for the conifold. Changing variables as
\be
z_{i}^{I} := \exp \left( \lambda_{i}^{I} + \frac{t}{p} + \frac{2\pi\rmi}{p} I\right),
\ee
\noindent
the integration contours for the eigenvalues get shifted from $\left( 0,+\infty \right)$ onto $( 0, \rme^{\frac{2\pi\rmi}{p} I} \infty )$, and one obtains
\begin{equation}
Z_{\text{CS}}\left( L(p,1) \right) = \CN\, \int\, \prod_{I=0}^{p-1} \prod_{i=1}^{N_I} \frac{\rmd z_{i}^{I}}{2\pi}\, \Delta_{1}^2 \left( z \right) \Delta_{2}^2 \left( z \right)\, \rme^{-\frac{p}{2g_{s}}\, \sum_{I=0}^{p-1} \sum_{i=1}^{N_I} \left( \log z_{i}^{I} - \frac{2\pi\rmi}{p}\, I \right)^{2}}.
\end{equation}
\noindent
In this expression we already find the usual form of the Vandermonde determinants; one has:
\begin{eqnarray}
\Delta_{1}^2 \left( z \right) &=& \prod_{I=0}^{p-1}\, \prod_{1 \le i<j \le N_I} \left( z_{i}^{I}-z_{j}^{I} \right)^{2}, \qquad \Delta_{2}^2 \left( z \right) = \prod_{0 \le I<J \le p-1}\, \prod_{i=1}^{N_I} \prod_{j=1}^{N_J} \left( z_{i}^{I}-z_{j}^{J} \right)^{2}, \\
\CN &:=& \frac{\rmi^{-\frac{1}{2} \left( \sum_{I=0}^{p-1} N_{I}^{2} \right)}\, \rme^{-\frac{1}{2 p} g_{s}N^{3}}\, \rme^{- \frac{2\pi\rmi}{p}\, N\, \sum_{I=0}^{p-1} I\, N_{I}}}{\prod_{I=0}^{p-1} N_{I}!}.
\end{eqnarray}
\noindent
It is now clear that the two Vandermonde determinants combine into a single determinant, and that the $\frac{2\pi\rmi}{p}$ factor in the potential can be conveniently absorbed into the logarithm to yield the partition function
\begin{equation}
Z_{\text{CS}}\left( L(p,1) \right) = \CN\, \int\, \prod_{i=1}^{N} \frac{\rmd z_{i}}{2\pi}\, \Delta^2 \left( z \right)\, \rme^{-\frac{1}{2 p g_{s}}\, \sum_{i=1}^{N} \left( \log z_{i}^p \right)^{2}}.
\end{equation}
\noindent
In this expression, there is no longer any explicit reference to the $p$ cuts. The Chern--Simons partition function on $L(p,1)$, with gauge group $G=$U$(N)$, thus has an hermitian matrix model representation of the type
\begin{equation}
Z = \frac{1}{\text{vol}\left(\text{U}(N)\right)} \int \rmd M\, \rme^{-\frac{1}{g_{s}}\, \text{Tr}\, \frac{1}{2p} \left( \log M^p \right)^{2}}.
\end{equation}
\noindent
Notice that the critical points of the potential are precisely the $p$th roots of identity, as expected.

The large--order and Borel analysis of the partition function of Chern--Simons theory on a three--sphere was addressed in \citep{Pasquetti:2009jg, Aniceto:2011nu}, albeit closer in spirit to the topological string Gopakumar--Vafa integral representation \citep{Gopakumar:1998ii} (the coefficients in the genus expansion of its large $N$ free energy $F = \log Z$ are known exactly, see, \textit{e.g.}, \citep{Marino:2004eq, Pasquetti:2009jg}). Here, we shall follow a purely resurgence viewpoint within the gauge theory, focusing on the general case of Chern--Simons gauge theory on lens spaces $L(p,1)$. Also in this case an explicit formula for the coefficients in the (topological) genus expansion of its large $N$ free energy is known exactly, for the one--cut/trivial flat--connection solution \citep{Garoufalidis:2008id} (but see also \citep{Gopakumar:1998ii, Marino:2002fk, Marino:2004uf}). With 't~Hooft coupling $t=g_sN$, this is\footnote{We are ignoring contributions at genus $g=0,1$; not particularly relevant from a Borel analysis standpoint.}
\be\label{eq:CSLS-freeen}
F_g (t) =  \frac{B_{2g}\, B_{2g-2}}{2g \left( 2g-2 \right) \left( 2g-2 \right)!} +\frac{B_{2g}}{2g \left( 2g-2 \right)!}\, p^{2-2g}\, \text{Li}_{3-2g} \left( \rme^{-\frac{t}{p}} \right) , \qquad g \ge 2,
\ee
\noindent
with ${\text{Li}}_{n} (z)$ the index $n$ polylogarithm. The first term is associated with the so--called constant map contribution \citep{Marino:1998pg, Faber:1998}, fully discussed in \citep{Pasquetti:2009jg}, and we shall drop it in the following. What remains is the second term, the ``true'' free energy of the Chern--Simons gauge theory on $L(p,1)$ (and which we shall denote by $\widehat{F}_g (t)$ in the following). This contribution grows factorially fast,
\be
\left.
\begin{aligned}
B_{2g} &\sim \left( 2g \right)! \\
\lim_{|t| \to 0} {\text{Li}}_{3-2g} \left( \rme^{-\frac{t}{p}} \right) &\sim \Gamma \left(2g-2\right) \left( \frac{t}{p} \right)^{2-2g}
\end{aligned}
\,\, \right\rbrace\quad
\Rightarrow \quad \widehat{F}_g \sim \left(2g-2\right)!\, \left( 1+\frac{1}{2g-2} \right) t^{2-2g},
\ee
\noindent
thus rendering the perturbative genus expansion asymptotic. 

\begin{figure}
\centering{}
\includegraphics[scale=0.7]{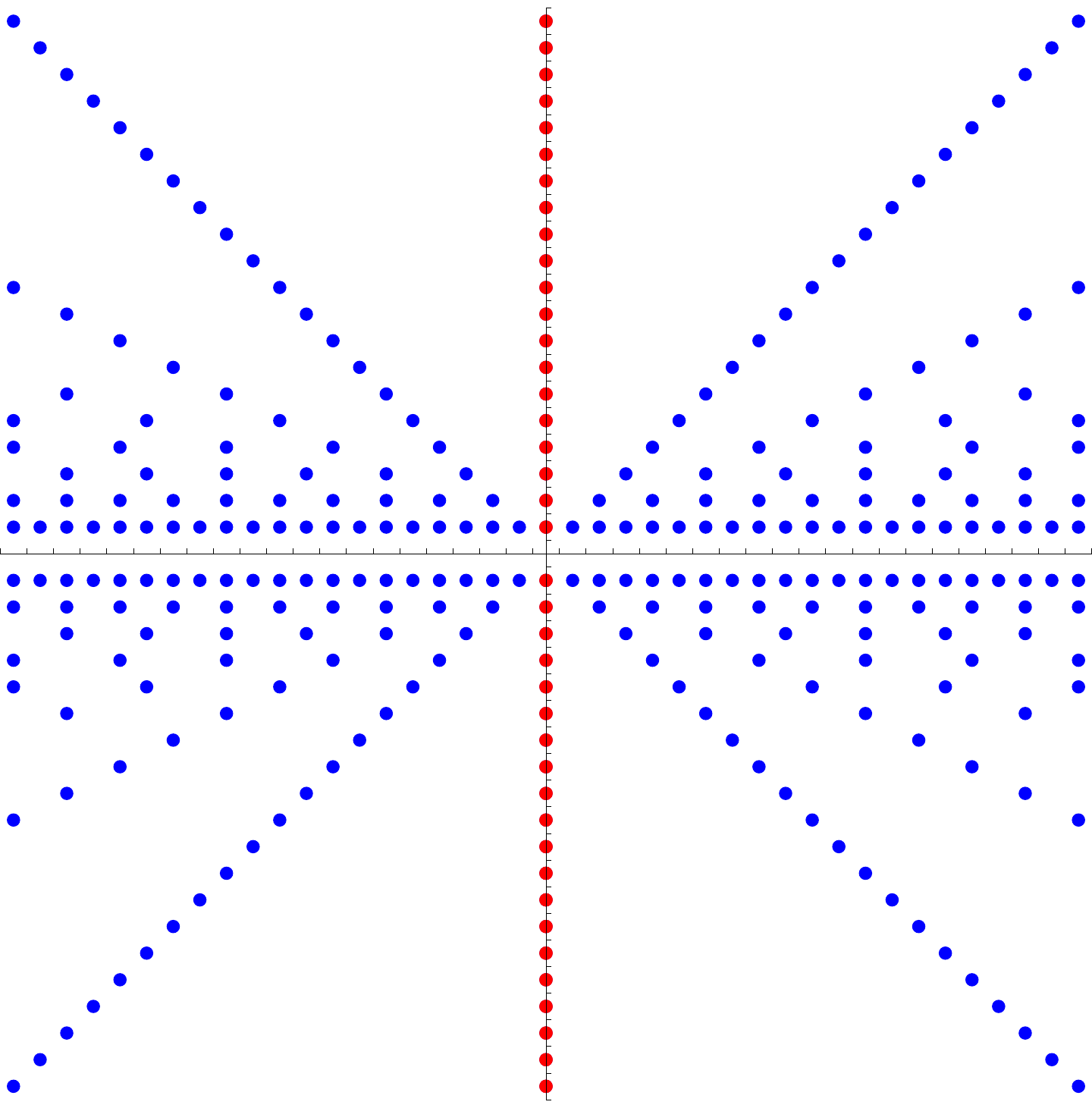}
\caption{Singularities in the complex Borel plane for Chern--Simons gauge theory on $L(2,1)$ with $t=4\pi$. The singularities in red are associated with the Gaussian or $c=1$ string contribution; see \citep{Pasquetti:2009jg} for details.
\label{fig:CSS3Borel}}
\end{figure}

The computation of the Borel transform of this asymptotic series follows standard procedures (herein first discussed right after \eqref{firstasymptseries}, but see as well the discussions in \citep{Aniceto:2011nu, Aniceto:2013fka, Upcoming:2014}), and in our case is thus given by
\begin{eqnarray}
\label{BoreldefinitionCSS3}
\mathcal{B}[\widehat{F}](s) &\equiv& \sum_{g=2}^{+\infty} \frac{\widehat{F}_{g}(t)}{(2g-3)!}\, s^{2g-3} = \frac{1}{s}\, \sum_{\ell \in\mathbb{Z}}\, \sum_{g=2}^{+\infty} \frac{B_{2g}}{2g\left(2g-2\right)!} \left(\frac{s}{t+2\pi\rmi p \ell}\right)^{2g-2} \\
&=&
\frac{1}{s}\, \sum_{\ell \in\mathbb{Z}} \left\{ -\frac{1}{12} + \left(\frac{t+2\pi\rmi p\ell }{s}\right)^{2} - \frac{1}{4} \left(\sinh \left( \frac{s}{2 \left( t+2\pi\rmi p\ell \right)} \right) \right)^{-2} \right\}. 
\end{eqnarray}
\noindent
Here, we have used the representation of the polylogarithm as a sum over residues (also extensively used in \citep{Pasquetti:2009jg})
\be
\label{eq-Polylogs-and-residues}
{\text{Li}}_{3-2g} \left( \rme^{-t} \right) = \Gamma \left( 2g-2 \right) \sum_{\ell \in \BZ} \left( t+2\pi\rmi \ell \right)^{2-2g}, \qquad g \ge 2.
\ee
\noindent
Understanding the singularity structure of the Borel transform will yield the resurgence properties of the Chern--Simons free energy on lens spaces. One immediately reads off that the singularities of $\CB [\widehat{F}] (s)$ are located at\footnote{Do notice that there is no singularity at the origin.}
\be
\label{eq:sing-CSLS}
\omega_{n\ell } = 2\pi\rmi n \left( t+2\pi\rmi p \ell \right) \equiv |n|\, A_\ell (t), \quad n \in \BZ \setminus \{ 0 \},
\ee
\noindent
for all $\ell \in \BZ$. Figure \ref{fig:CSS3Borel} shows a schematic representation of these singularities in the complex Borel plane. One sees that they arrange themselves along radial directions as multiples of the ``basic'' instanton\footnote{Throughout we shall use the word ``instanton'' a bit loosely, simply implying nonperturbative exponential contributions. The physical origin of these nonperturbative effects will be later addressed in section \ref{sec:physical-int}.} actions $\pm A_\ell (t)$, $\pm A_\ell^\dagger (t)$, now with $\ell =1,2,3,\ldots$,
\be
\left| \omega_{n \ell } \right| = |n| \left( 2 \pi \sqrt{t^2 + 4\pi^2 p^2 \ell^2} \right) \equiv |n|\, \left|A_\ell (t)\right|, 
\ee
\noindent
and with direction\ $\arg A_\ell (t) $ defined in each quadrant in the complex plane as follows
\vspace{10pt}
\begin{center}
\begin{tabular}{c|c|c}
Quadrant & $n,\, \ell $ & $\arg A_{\ell }(t)$\tabularnewline
\hline 
1st & $n\ge1,\, \ell >0$ & $\theta_{\ell }$\tabularnewline
2nd & $n\ge1,\, \ell <0$ & $\pi-\theta_{\ell}$\tabularnewline
3rd & $n\le-1,\, \ell >0$ & $\pi+\theta_{\ell }$\tabularnewline
4th & $n\le-1,\, \ell <0$ & $-\theta_{\ell }$\tabularnewline
\end{tabular}
\end{center}
\vspace{10pt}
\noindent
where $\theta_{\ell }=\arctan \left( \frac{t}{2 \pi p \left| \ell \right|} \right)$. Note that $\ell=0$ corresponds to the directions $\arg A_{0}(t) = \pm \pi / 2$, and that for $\ell \ne 0$ the possible directions are bounded by $0 < \arg A_\ell (t) \le \arctan \left( \frac{t}{2 \pi p} \right)$ for the first quadrant (and correspondingly in the others).

%%%%%%%%%%%%%%%%%%%%%%%%%%%%%%%%%%%%%%%%%%%%%%%%%%%%%%%%%%%%%%%%%
%%%%%%%%%%%%%%%%%%%%%%%%%%%%%%%%%%%%%%%%%%%%%%%%%%%%%%%%%%%%%%%%%

Near each of these many singularities, the Borel transform behaves as
\be
\label{BorelsingularCSS3}
\left. \CB [\widehat{F}] (s) \right|_{\omega_{n\ell }} = - \frac{\omega_{n\ell }}{\left( 2\pi\rmi n \right)^2 \left( s - \omega_{n\ell } \right)^{2}} + \frac{1}{\left( 2\pi\rmi n \right)^2 \left( s - \omega_{n\ell }  \right)}\, + \text{holomorphic},
\ee
\noindent
The singularities are thus simple and double poles, and the Borel transform is meromorphic, \textit{i.e.}, the Borel surface is essentially $\BC$ rather than a higher genus Riemann surface (as is the case in, for instance, matrix models \citep{Aniceto:2011nu} or string theory \citep{Couso-Santamaria:2014iia}). A similar behavior was already found for Chern--Simons on the sphere \citep{Pasquetti:2009jg, Aniceto:2011nu}. Let us fix $\ell $ and thus a Stokes line (a singular ray in the Borel complex plane), of angle $\theta_\ell = \arg A_\ell (t)$. As mentioned earlier in this paper, there is a discontinuity between left and right Borel resummations along such Stokes line. This discontinuity may be computed by considering the difference between Laplace transforms of the Borel transform $\CB [\widehat{F}] (s)$ just above and just below each singular direction $\theta_\ell $. Given that the structure of singularities of the Borel transform is made up solely of poles (double and single), it is not difficult to see that the discontinuity across each singular direction $\theta_\ell $ will be simply given by a sum over residues at each pole along the Stokes line. One finds:
\be
\label{CSS3discontinuity}
\mathrm{Disc}_{\theta_{\ell }}\widehat{F}(g_{s}) = -\frac{1}{2\pi\rmi\, g_{s}}\sum_{n=1}^{+\infty}\frac{1}{n^{2}}\left(\omega_{n\ell}+g_{s}\right)\,\rme^{-\frac{\omega_{n\ell}}{g_{s}}}.
\ee
\noindent
This is a familiar result \citep{Pasquetti:2009jg, Aniceto:2011nu} with a distinct multi--instanton flavor. Do note that when addressing the resurgent structure of a $1/N$ expansion, the variable $N$ is understood as a continuous variable. Of course that in the present Chern--Simons example it is simple to see that if $N$ is restricted to be an integer, then the nonperturbative factor is $\rme^{-\frac{\omega_{n\ell}}{g_{s}}} =  \rme^{- 2\pi\rmi n \left( N + k p \ell \right)} =1$.

In parallel to the above discussion, for all the examples we shall be later addressing in this paper it will turn out that the structure of Borel singularities will always consist of poles (of different orders), also implying that the corresponding Borel transforms are meromorphic. However, unlike what just happened in our Chern--Simons example, the Stokes discontinuities associated with these cases will not be as simple to evaluate. Instead, one must rely on some basic concepts concerning alien calculus (in this context, see, \textit{e.g.}, \citep{Aniceto:2011nu, Upcoming:2014} for further details). As such, let us pause our resurgent analysis of Chern--Simons gauge theory for a moment and obtain this very same result, \eqref{CSS3discontinuity}, in a slightly different manner. This consists of finding an alternative ``representative'' for the Borel transform, \eqref{BoreldefinitionCSS3}, where its singular structure will be slightly modified (but in a controlled manner) from \eqref{BorelsingularCSS3} into a so--called ``simple'' resurgent form, where the subsequent resurgent and discontinuity analyses follow systematically (see, \textit{e.g.}, \citep{Aniceto:2011nu}). Details will become clear as we proceed in the following.

In \eqref{BoreldefinitionCSS3} we determined the Borel transform of the asymptotic expansion $\widehat{F}(g_s)$ following its definition and effectively removing a factorial growth of $(2g-3)!$. But what if we were to remove a slightly different growth, say $(2g-2)!$ or similar? It is simple to see that this would lead us instead to
\be
\label{BoreldefinitionCSS3-v2}
\CB [g_{s} \widehat{F}] (s) = \sum_{\ell \in \BZ} \left( -\frac{1}{12}\log\left(\frac{s}{t+2\pi\rmi p \ell }\right) - \frac{1}{2} \left( \frac{t+2\pi\rmi p \ell }{s} \right)^2 - \frac{1}{4} \int_0^{\frac{s}{t+2\pi\rmi p \ell }} \frac{\rmd x}{x \sinh^2 \left( \frac{x}{2} \right)} \right).
\ee
\noindent
This Borel transform is related to the original free energy (without the constant map contribution) by a multiplication by $g_{s}$. Further, both Borel transforms \eqref{BoreldefinitionCSS3} and \eqref{BoreldefinitionCSS3-v2} are simply related\footnote{These relations of Borel transforms are thoroughly studied in \citep{Upcoming:2014}.}
\be
\CB [\widehat{F}] (s)=\frac{\rmd}{\rmd s}\CB [g_{s} \widehat{F}] (s).
\ee
\noindent
Next, notice that while the Borel transform of $g_{s} \widehat{F}$ still has singular points at \eqref{eq:sing-CSLS}, just as before, its behavior close to these singularities is now seemingly distinct
\be
\label{BorelsingularCSS3-v2}
\left. \CB [g_s \widehat{F}] (s) \right|_{\omega_{n\ell }} = \frac{\omega_{n\ell }}{\left( 2\pi\rmi n \right)^2 \left( s - \omega_{n\ell } \right)} + \frac{\CB [G] (s)}{\left( 2\pi\rmi n \right)^2}\, \log \left( s - \omega_{n\ell } \right) + \text{holomorphic},
\ee
\noindent
where $\CB [G] (s)$---the Borel transform of $G(g_s) = g_{s}$---is just a convenient way to write $1$. The Borel singularities are now simple poles and logarithmic branch cuts. At first this could lead us to believe that the Borel transform associated with this problem was not meromorphic, which we already know \textit{not} to be the case. But this fact is noticeable in this version of the Borel transform\footnote{\label{footnote-alien-calc}Note that, as presented, this result is the \textit{primitive} of the results in \citep{Pasquetti:2009jg, Aniceto:2011nu}. One could in fact further find either higher primitives, or instead find derivatives, of those results by simply changing the additive factor in the denominator appearing in the definition of the Borel transform \eqref{BoreldefinitionCSS3} \citep{Upcoming:2014}. As explained, our choice in here is just to reproduce formulae for simple resurgent functions.} as there is no resurgent mixing of multi--instanton nonperturbative sectors, given that the Borel transform associated with the logarithmic branch--cut is always the same (constant and entire).

The reason to consider this rewriting is solely to simplify some calculations ahead. As discussed earlier, there is a discontinuity between left and right Borel resummations along Stokes lines; in the present case along the directions $\theta_\ell \equiv \arg A_\ell (t)$. These discontinuities are generically measured by the so--called Stokes automorphism which itself is computed via alien calculus. While in general this automorphism (and subsequent discontinuities) may be hard to evaluate, things turn out to be a bit simpler when dealing with \textit{simple} resurgent functions, which precisely have the singularity structure \eqref{BorelsingularCSS3-v2} (again, see, \textit{e.g.}, \citep{Aniceto:2011nu, Aniceto:2013fka, Upcoming:2014} for details on alien calculus within the present contexts). Very briefly and schematically, simple resurgent functions are characterized by Borel transforms which, near singularities $\omega_{\ell}$ along some Stokes direction $\theta$, behave as
\be
\left. \CB [F] (s) \right|_{\omega_{\ell }} = \frac{\alpha}{2 \pi \rmi \left(s-\omega_{\ell }\right)} + \CB [G_{\ell }] (s-\omega_{\ell })\, \frac{\log \left(s-\omega_{\ell }\right)}{2\pi\rmi} + \text{holomorphic}.
\ee
\noindent
One can then immediately obtain the discontinuity across the Stokes singular direction via
\be
\text{Disc}_{\theta}\, F (g_s) = F (g_s) -\exp \left(\sum_{\ell } \rme^{-\omega_\ell / g_{s}} \Delta_{\omega_{\ell }} \right)\, F (g_s),
\ee
\noindent
where $ \Delta_{\omega_{\ell}} F (g_s) = G_{\ell} (g_s) + \alpha$ is the so--called alien derivative. In other words, the end result essentially states that the discontinuity we are looking for may be read off directly from the singular structure of the above Borel transform \eqref{BorelsingularCSS3-v2}. Taking into account that in this case $\Delta_{\omega_{\ell}}^2\, F = \Delta_{\omega_{\ell}}\, G = 0$, we obtain
\be
\label{CSS3discontinuity-v2}
\mathrm{Disc}_{\theta_{\ell }}\widehat{F}(g_{s}) = -\frac{1}{2\pi\rmi\, g_{s}}\sum_{n=1}^{+\infty}\frac{1}{n^{2}}\left(\omega_{n\ell}+g_{s}\right)\, \rme^{-\frac{\omega_{n\ell }}{g_{s}}}.
\ee
\noindent
Of course this is precisely the same result as the one we retrieved earlier in \eqref{CSS3discontinuity}. Again, let us stress that the sole reason to consider this detour is to simplify some calculations ahead. One could of course do the full resurgent analysis without this change of Borel ``representative'', only that the calculations would be longer and a bit more intricate. In fact, while in this particular case of Chern--Simons gauge theory the alien calculus detour was clearly redundant, in the examples which will later follow this will not be the case: the alien calculus approach via simple resurgent functions will allow us to calculate the Stokes discontinuities in a simple and fast manner.

A particularly nice feature of these Borel meromorphic examples is that one may go past large--order analysis and actually recover the full, exact free energy out of the Stokes discontinuities. This was first noticed in \citep{Pasquetti:2009jg} for Chern--Simons on the sphere. On what concerns Chern--Simons on lens spaces, assuming that the large $N$ free energy has no Borel singularities at infinite coupling (a common yet generically unproved assumption), one may use the Cauchy theorem in order to write a dispersion relation equating the free energy to its Stokes discontinuities,
\be
\label{cauchydispersion}
\widehat{F} (g_s) = \frac{1}{2\pi\rmi}\, \sum_{\ell \in\, \text{all quadrants}}\, \int_0^{\rme^{\rmi \theta_\ell } \cdot \infty} \rmd w\, \frac{\text{Disc}_{\theta_\ell }\, \widehat{F} (w)}{w-g_s}.
\ee
\noindent
Inserting the full set of discontinuities \eqref{CSS3discontinuity-v2} back in \eqref{cauchydispersion} above, one recovers the Chern--Simons free energy \eqref{eq:CSLS-freeen} (naturally, without the constant map contribution). It is important to notice that the symmetry between upper and lower--half Borel planes, \textit{i.e.}, the presence of instanton actions $A_\ell (t)$ and $A^\dagger_\ell (t)$, is responsible for the reality of the original series; while the symmetry between singular directions $\pm \rmi \theta_\ell$, \textit{i.e.}, the presence of instanton actions $\pm A_\ell (t)$, implements back the topological genus expansion. To see this explicitly, let us expand the integrand in powers of $g_{s}$ and use the discontinuities determined above. We obtain
\begin{eqnarray}
\widehat{F}\left(g_{s}\right) &\simeq& \frac{1}{4\pi^{2}}\, \sum_{\ell \in\textrm{all quads}}\, \sum_{g=1}^{+\infty} \left( g_{s}\, \rme^{-\rmi\theta_{\ell }} \right)^{g-1}\, \sum_{n=1}^{+\infty} \frac{1}{n^{2}}\, \int_{0}^{+\infty}\rmd w \left( n \left|A_{\ell }(t)\right| + w \right) \frac{\rme^{-\frac{n \left|A_{\ell }(t)\right|}{w}}}{w^{g+1}} \\
&=& \frac{1}{4\pi^{2}}\, \sum_{\ell \in\textrm{all quads}}\, \sum_{g=1}^{+\infty} \sum_{n=1}^{+\infty} \left( g_{s}\, \rme^{-\rmi\theta_{\ell }} \right)^{g-1}\, \frac{g\, \Gamma (g-1)}{n^{2} \left(n \left|A_{\ell }(t)\right|\right)^{g-1}}.
\end{eqnarray}
\noindent
We can rewrite the sum over all singular directions as 
\be
\sum_{\ell \in\textrm{all quads}} \left( \left|A_{\ell }(t)\right| \rme^{\rmi\theta_{\ell }} \right)^{1-g} = \left( 1 + \left(-1\right)^{g-1} \right) \left\{ \sum_{\ell =1}^{+\infty}\frac{\rme^{\rmi\theta_{\ell}(g-1)} + \rme^{-\rmi\theta_{\ell }(g-1)}}{\left| A_{\ell }(t) \right|^{g-1}} + \frac{\rme^{\rmi(g-1)\frac{\pi}{2}}}{\left(2\pi p t \right)^{g-1}} \right\}.
\ee
\noindent
In this result, summing over the left and right quadrants originated the factor $(1+\left(-1\right)^{g-1})$, resulting in the familiar topological genus expansion (as only odd $g$ survive, implying $g-1$ is even). The sum over the two angles $\pm\theta_{\ell }$ gives rise to a real expansion, as one obtains
\be
\rme^{\rmi\theta_{\ell }\alpha} + \rme^{-\rmi\theta_{\ell }\alpha} = \left(\frac{2\pi}{\left|A_{\ell }(t)\right|}\right)^{\alpha} \Big( \left( 2\pi p\ell  +\rmi t \right)^{\alpha} + \left( 2\pi p\ell  -\rmi t \right)^{\alpha} \Big).
\ee
\noindent
Substituting these results into the previous expansion for $\widehat{F}\left(g_{s}\right)$, one finally obtains
\begin{eqnarray}
\widehat{F} \left(g_{s}\right) &\simeq& 2 \sum_{g=1}^{+\infty} g_{s}^{2g-2}\, \left(-1\right)^{1-g} \left(2g-1\right) \frac{\zeta\left(2g\right)}{\left(2\pi\right)^{2g}}\, \Gamma \left( 2g-2 \right) \sum_{\ell \in \mathbb{Z}} \left( t+2\pi\rmi p \ell \right)^{2-2g}  \nonumber \\
&\simeq&
\sum_{g=2}^{+\infty} g_{s}^{2g-2}\, \frac{B_{2g}}{2g \left(2g-2\right)!}\, p^{2-2g}\, \mathrm{Li}_{3-2g} \left(\rme^{-\frac{t}{p}}\right),
\end{eqnarray}
\noindent
where we have used the representation of the polylogarithm given in \eqref{eq-Polylogs-and-residues}, valid once we remove the $g=1$ term, and also made use of the relation between the zeta function and the Bernoulli numbers,
\be
\left(-1\right)^{g+1} \left(2g-1\right) \frac{\zeta\left(2g\right)}{\left(2\pi\right)^{2g}} = \frac{1}{2}\, \frac{B_{2g}}{2g \left(2g-2\right)!}.
\ee
\noindent
We have thus rederived the original free energies (without the constant map contribution), given in \eqref{eq:CSLS-freeen}, from the nonperturbative discontinuities. The results obtained are exact results, instead of the typical large--order results usually achieved through these methods (see a few examples in, \textit{e.g.}, \citep{Aniceto:2011nu}), and this is due to the meromorphicity of the Borel transform.

Having presented a resurgent analysis of the nonperturbative structure of the large $N$ free energies of Chern--Simons gauge theory localized on lens spaces, we shall next focus on another example: ABJM gauge theory localized on the sphere $\mathbb{S}^{3}$. This time, instead of studying the $1/N$ expansion, we shall fix $N=2$, take the level $k$ to be large and study the standard gauge--theory perturbation series.

%%%%%%%%%%%%%%%%%%%%%%%%%%%%%%%%%%%%%%%%%%%%%%%%%%%%%%%%%%%%%%%%%
\subsection{ABJM Gauge Theory on the Sphere}\label{sec:ABJM}
%%%%%%%%%%%%%%%%%%%%%%%%%%%%%%%%%%%%%%%%%%%%%%%%%%%%%%%%%%%%%%%%%

ABJM gauge theory \citep{Aharony:2008ug} is a superconformal Chern--Simons gauge theory in three dimensions, coupled to matter. It has $\mathcal{N}=6$ supersymmetry and gauge group $G=$ U$\left(N\right)_{k} \times $U$\left(N\right)_{-k}$, with the corresponding actions having Chern--Simons couplings $k$ and $-k$. One may also consider an extension of this theory by generalizing the gauge group to be $G=$ U$\left(N_{1}\right)_{k} \times $U$\left(N_{2}\right)_{-k}$ \citep{Aharony:2008gk}. These theories have been studied via localization techniques (see, \textit{e.g.}, \citep{Marino:2011nm}), and in \citep{Kapustin:2009kz} their partition functions on $M=\mathbb{S}^{3}$ were given matrix integral representations.

The matrix model representing the partition function of ABJM gauge theory on $M=\mathbb{S}^{3}$, with general gauge group $G=$ U$\left(N_{1}\right)_{k} \times $U$\left(N_{2}\right)_{-k}$, may be also obtained from Chern--Simons gauge theory on the lens space $L(2,1)$ (which we discussed earlier) via a supergroup extension \citep{Marino:2009jd}. One can further rewrite such matrix model representation using the Fermi gas approach studied in \citep{Marino:2011eh}. In this way, using the Cauchy identity (see, \textit{e.g.}, \citep{Marino:2011nm}), the ABJM partition function may be finally written as \citep{Kapustin:2010xq}
\be
\label{eq:Z-ABJM-fermi-gas}
Z_{\text{ABJM}} \left( \BS^3 \right) \left[N,k\right] =\frac{1}{ N!}\, \sum_{\sigma\in S_{N}} \left(-1\right)^{\epsilon\left(\sigma\right)} \int\, \prod_{i=1}^{N} \frac{\rmd x_{i}}{2\pi k}\, \frac{1}{\prod_{i} 2 \cosh \left(\frac{x_{i}}{2}\right) 2 \cosh \left(\frac{x_{i}-x_{\sigma(i)}}{2k}\right)}.
\ee
\noindent
The large $N$ behaviour of the free energy $F=\log Z$ has been analyzed in \citep{Drukker:2010nc}, where it was shown that its perturbative series can be determined recursively even though a closed form for the coefficients is not known at present. In the current work, we will instead fix the rank of the gauge group to be $N=2$, and analyze the resurgent properties of the perturbative series obtained when one takes the level $k$ to be large, \textit{i.e.}, considering an asymptotic expansion in $1/k$. Setting $N=2$ the partition function can be taken to the form \citep{Okuyama:2011su, Russo:2012kj}
\be
\label{eq:Part-funct-ABJM-first}
Z_{\text{ABJM}} \left(\BS^{3}\right) \left[2,k\right]  = \frac{1}{8} \int_0^{+\infty} \rmd u\, \frac{u}{\sinh \pi u k}\, \tanh^2 u.
\ee
\noindent
This integral may be computed exactly via residues, giving rise to an expression involving a sum over $k$ different terms \cite{Okuyama:2011su}. However, we are interested in the explicit analytic expression for the complete weak--coupling perturbative series in $1/k$ (which is the same perturbation series that one would obtain by standard Feynman diagram calculations). In this case, using the series expansion for $\tan^2 \pi x$ when $|x|<\frac{1}{2}$,
\be
\tan^2 \pi x = \frac{2}{\pi^2} \sum_{m=2}^{+\infty} \left(2m-1\right) \left(2^{2m}-1\right) \zeta(2m)\, x^{2m-2},
\ee
\noindent
one then finds
\be
\label{eq:Partfunc-ABJM-pert-series}
Z_{\text{ABJM}} \left(\BS^{3}\right) \left[2,k\right] \simeq \sum_{n=2}^{+\infty} \frac{Z_{n}}{k^{2n}},
\ee
\noindent
where the perturbative coefficients are given by
\be
Z_{n} = \left(-1\right)^{n}\, \frac{2}{\left(2\pi\right)^{2n+2}}\, \left(2n-1\right) \left(2n-1\right)! \left(2^{2n}-1\right)^{2}\, \zeta(2n)^{2}.
\label{eq:ABJM-pert-coeff-partfunc}
\ee
\noindent
It is now possible to check the asymptotic behavior of the coefficients $Z_{n}$. For large $n$, we can easily see that
\be
\left|Z_{n}\right| \sim \left(\frac{2}{\pi}\right)^{2n}(2n)!, \qquad n \gg 1,
\ee
\noindent
which means the coefficients grow as $\sim (2n)!$, and which agrees with the results found in \citep{Russo:2012kj}. A closer examination of the coefficients \eqref{eq:ABJM-pert-coeff-partfunc} reveals that they are built out of different components, with distinct subleading exponential behavior. In fact, we can write the $Z_n$ coefficients as a sum of three separate terms of the form
\be
\label{eq:Pert-Zg-diff-actions}
Z_{n}=Z_{n}^{(1)} - 2 Z_{n}^{(2)} + Z_{n}^{(3)}, \qquad \textrm{where} \qquad Z_{n}^{(i)} = \frac{\left(-1\right)^{n}}{2\pi^{2}} \left(2n-1\right) \left(2n-1\right)!\, \zeta(2n)^2\, \frac{1}{A_{i}^{2n}},
\ee
\noindent
where the corresponding ``instanton'' actions are $A_{3} = A_{2}/2 = A_{1}/4 = \frac{\pi}{2}$. The asymptotic behavior of each of these separate terms is then the familiar
\be
Z_{n}^{(i)} \sim \frac{(2n)!}{A_{i}^{2n}}, \qquad n\gg1.
\ee
\noindent
With this split in mind, also the perturbative series for the partition function, \eqref{eq:Partfunc-ABJM-pert-series}, may be naturally divided into three parts, as
\be
\label{eq:Pert-partfunc-diff-actions}
Z_{\text{ABJM}} \left(\BS^{3}\right) \left[2,k\right] = Z^{(1)}(k) - 2 Z^{(2)}(k) + Z^{(3)}(k), \qquad \textrm{where} \qquad Z^{(i)}(k) \simeq \sum_{n=2}^{+\infty} \frac{Z_{n}^{(i)}}{k^{2n}}.
\ee

In order to uncover the resurgent structure of the partition function, we shall next analyze each of the above asymptotic expansions separately. They all have the same type of leading factorial growth, $\sim (2n)!$, and in this case the Borel transform would act on the asymptotic series by changing $k^{-2n} \rightarrow s^{2n-1}/(2n-1)!$. As we discussed earlier, in the example of Chern--Simons gauge theory on lens spaces, an easier and faster way to obtain the Stokes discontinuities associated with the partition--function asymptotic expansion is by addressing instead the series $k^{-1} Z^{(i)}(k)$ (whose Borel transform will be associated with a \textit{simple} resurgent function, for which discontinuity formulae are directly available and applicable \citep{Upcoming:2014}). We shall  return to the precise Borel transform  below. 
For the moment let us thus consider instead 
\bea
\mathcal{B} [k^{-1}Z^{(i)}] (s) &\equiv& \sum_{n=2}^{+\infty} \frac{Z_{n}^{(i)}}{(2n)!}\, s^{2n} = \sum_{n=2}^{+\infty} \frac{\left(-1\right)^{n}}{2\pi^{2}}\, \frac{2n-1}{2n}\, \zeta(2n)^{2} \left(\frac{s}{A_{i}}\right)^{2n}  \\
&=&
\sum_{\ell=1}^{+\infty} \left\{ \frac{1}{24} \left( \frac{s}{\ell A_{i}} \right)^{2} + \frac{1}{4\pi^{2}} \log \left( \sinh \left( \frac{\pi s}{\ell A_{i}} \right) \right) - \frac{s}{4\pi \ell A_{i}} \coth \left( \frac{\pi s}{\ell A_{i}} \right) + \right. \\
&&
\hspace{150pt}
\left.
+ \frac{\ell^{2}}{4\pi^{2}} \left( 1 - \log \left( \frac{s}{A_{i}} \right) \right) \right\},
\label{lastline}
\eea
\noindent
where we made use of the definition of the zeta function and interchanged orders of summation. The terms in the last line \eqref{lastline} will cancel against each other once we consider the Borel transform for the \textit{full} partition function, $\mathcal{B} [k^{-1}Z_{\text{ABJM}} \left(\BS^{3}\right) ] (s)$, corresponding to the sum in \eqref{eq:Pert-partfunc-diff-actions}. One thus obtains:
\be
\label{eq:Borel-ABJM}
\mathcal{B} [k^{-1}Z_{\text{ABJM}} \left(\BS^{3}\right) ] (s) = \frac{s^{2}}{64} + \frac{1}{4\pi^{2}} \sum_{\ell=0}^{+\infty} \left\{ \log \left( \cosh \left( \frac{s}{2\ell+1} \right) \right) - \frac{s}{2\ell+1} \tanh \left( \frac{s}{2\ell+1} \right) \right\}.
\ee

%%%%%%%%%%%%%%%%%%%%%%%%%%%%%%%%%%%%%%%%%%%%%%%%%%%%%%%%%%%%%%%%%
\begin{figure}
\centering{}
\begin{tabular}{cccccc}
\includegraphics[height=7cm]{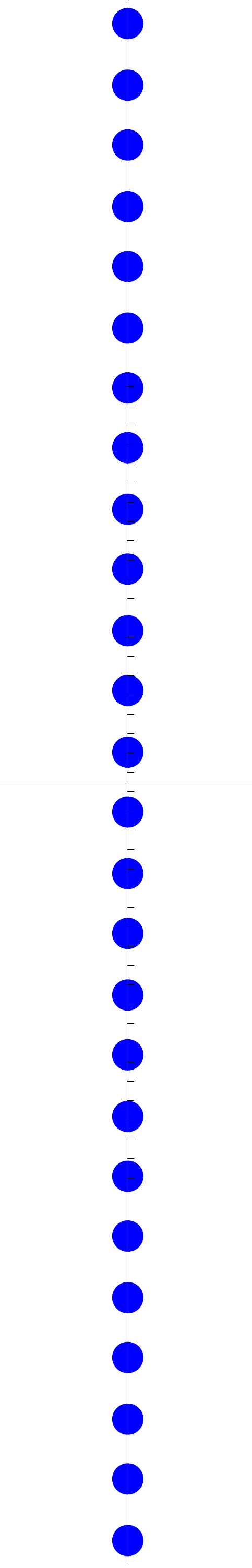} & \qquad\qquad 
\includegraphics[height=7cm]{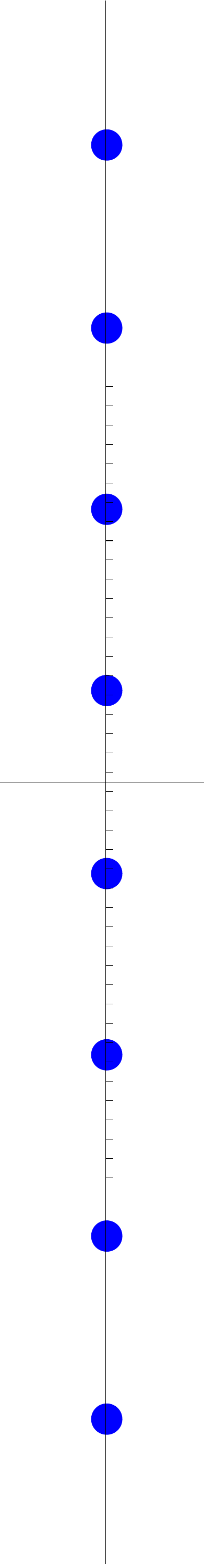} & \qquad\qquad 
\includegraphics[height=7cm]{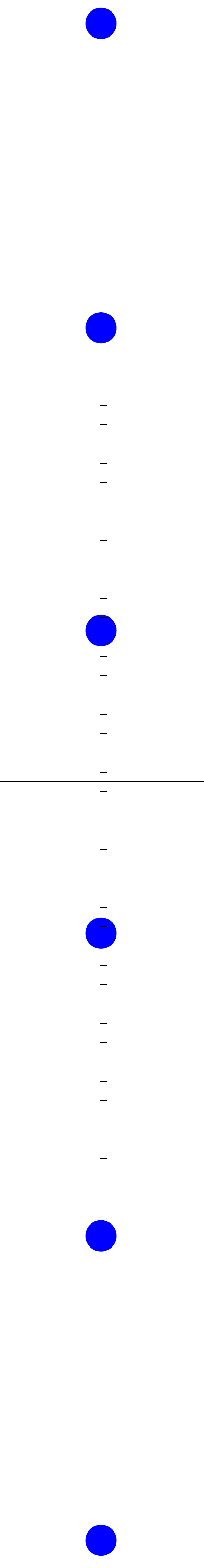} & \qquad\qquad 
\includegraphics[height=7cm]{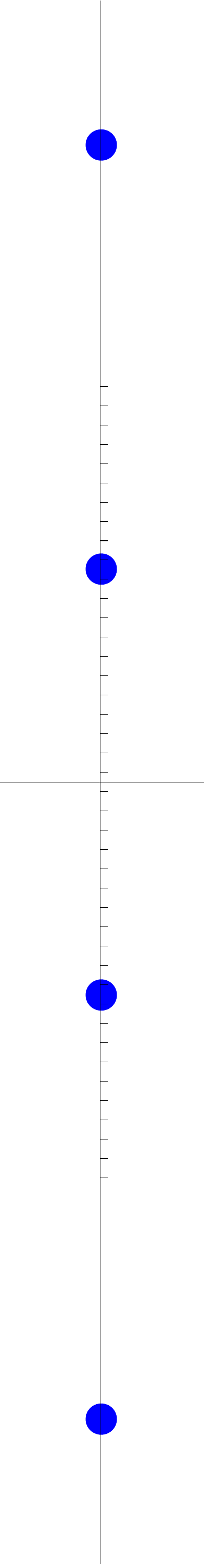} & \qquad\qquad 
\includegraphics[height=7cm]{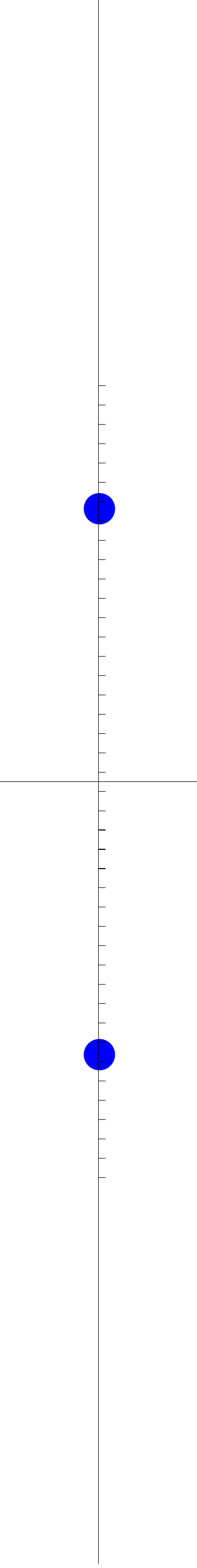} & \qquad\qquad 
\includegraphics[height=7cm]{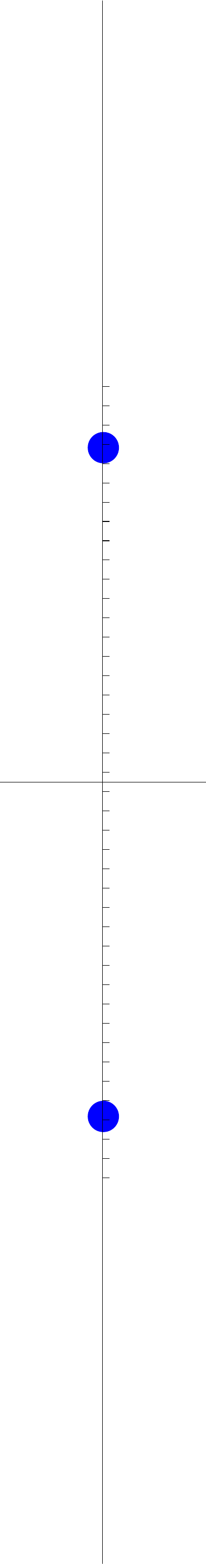}
\tabularnewline
{\small $\ell=0$} & \qquad\qquad 
{\small $\ell=1$} & \qquad\qquad 
{\small $\ell=2$} & \qquad\qquad 
{\small $\ell=3$} & \qquad\qquad 
{\small $\ell=4$} & \qquad\qquad 
{\small $\ell=5$}
\tabularnewline
\end{tabular}
\caption{Singularities in the complex Borel plane for ABJM gauge theory on $\mathbb{S}^3$ (at large level $k$, and with rank of the gauge group $N=2$). Every vertical line in this figure should be seen as the very same imaginary axis, \textit{i.e.}, all singularities are along the same line (they are shown separately just to make manifest the existence of different ``instanton'' actions $A_\ell$, for $\ell=1,2,3,4,5$).
}
\label{fig:ABJM-Borel}
\end{figure}
%%%%%%%%%%%%%%%%%%%%%%%%%%%%%%%%%%%%%%%%%%%%%%%%%%%%%%%%%%%%%%%%%

The Borel singularities of the ABJM partition function are located at
\be
\omega_{\ell m}^{\pm}=\pm\frac{\rmi\pi}{2}\left(2\ell+1\right)(2m+1)=\left(2m+1\right)A_{\ell}^{\pm}, \qquad m,\ell \in \mathbb{Z}_{0}^{+}.
\label{eq:Singularities-ABJM}
\ee
\noindent
This implies that there are only two singular directions, \textit{i.e.}, $\omega_{\ell m}^{\pm} = \left|\omega_{\ell m}\right| \rme^{\pm\rmi\frac{\pi}{2}}$, with
\be
\left| \omega_{\ell m} \right| = \frac{\pi}{2} \left(2\ell+1\right) \left(2m+1\right) \equiv \left(2m+1\right) \left|A_{\ell}\right|, \qquad m,\ell \in \mathbb{Z}_{0}^{+}.
\ee
\noindent
Similarly to what happened with Chern--Simons gauge theory in the previous subsection, in this case one finds singularities at odd multiples of a basic ``instanton'' action, $A_{\ell}^{\pm}$, and these singularities align themselves along the directions $\theta_{\pm}=\arg A_{\ell}^{\pm}=\pm\rmi\frac{\pi}{2}$ corresponding to the imaginary axis. Figure \ref{fig:ABJM-Borel} shows a schematic representation of these singularities. Near each of these singularities, the Borel transform \eqref{eq:Borel-ABJM} behaves as 
\be
\left.\mathcal{B} [k^{-1}Z_{\text{ABJM}}\left(\BS^{3}\right)] (s) \right|_{\omega_{\ell m}} =  \frac{\omega_{\ell m}}{\left(2\pi \rmi\right)^{2} \left(s-\omega_{\ell m}\right)} + \frac{1}{4\pi^{2}}\, \mathcal{B} [G] (s)\, \log \left(s-\omega_{\ell m}\right) + \textrm{holomorphic},
\label{eq:ABJM-Borel-at-sing}
\ee
\noindent
where once again $\mathcal{B} [G] (s)=1$ is the Borel transform of $G(k)=k^{-1}$. In parallel to our earlier discussion, one should not rush into conclusions based on this singularity structure. The Borel transform of the ABJM partition function itself, $Z_{\text{ABJM}}\left(\BS^{3}\right)$, is indeed \textit{meromorphic} as one may check immediately. One finds
\be
\mathcal{B} [Z_{\text{ABJM}}\left(\BS^{3}\right)] (s) = \frac{\rmd}{\rmd s} \mathcal{B} [k^{-1}Z_{\text{ABJM}}\left(\BS^{3}\right)] (s) = \frac{s}{32} -  \frac{s}{4\pi^{2}}\,  \sum_{\ell=0}^{+\infty} \left\{ \frac{1}{\left(2\ell+1\right)^{2}}\, \mathrm{sech}^{2} \left( \frac{s}{2\ell+1} \right) \right\},
\ee
\noindent
which has the exact same singular points \eqref{eq:Singularities-ABJM}, but which now correspond to double and simple poles, rendering the Borel transform of the ABJM partition function as a meromorphic function. Nonetheless, as explained, we proceed with the ``representative'' \eqref{eq:ABJM-Borel-at-sing} for computational reasons.

From the singular behavior \eqref{eq:ABJM-Borel-at-sing} follow the Stokes discontinuities of the partition function along the Stokes directions $\theta_{\pm}=\pm\frac{\pi}{2}$ (this is done via alien calculus; see, \textit{e.g.}, \citep{Aniceto:2011nu, Upcoming:2014} for details, and footnote \ref{footnote-alien-calc} on page \pageref{footnote-alien-calc}). These are simply
\be
\label{discZABJMS3}
\text{Disc}_{\theta_{\pm}}Z_{\text{ABJM}}\left(\BS^{3}\right) (k) = \frac{k}{2\pi\rmi}\,  \sum_{\ell=0}^{+\infty} \sum_{m=0}^{+\infty} \left( \frac{1}{k} - \omega_{\ell m}^{\pm} \right) \rme^{-\omega_{\ell m}^{\pm}k}.
\ee
\noindent
Having these discontinuities at hand one may proceed to the \textit{exact} large--order analysis we have done before, recovering the full ABJM partition function out of the above Stokes discontinuities. Under the usual assumption that there are no discontinuities at zero level, the familiar Cauchy dispersion relation \eqref{cauchydispersion} holds. Note that our small coupling is now $\lambda \equiv \frac{1}{k}$ (which replaces the $g_s$ coupling back in \eqref{cauchydispersion}). Next, expanding the integrand for small $\lambda$ and using the discontinuities derived above (in the variable $w=k^{-1}$), we can easily find\footnote{In this calculation we consider $n\ge2$, discarding $n=1$ (this term is not important for the asymptotic analysis).}
\begin{eqnarray}
Z_{\text{ABJM}}\left(\BS^{3}\right) (\lambda=k^{-1}) &\simeq& \frac{1}{2\pi\rmi}\, \sum_{\pm} \sum_{n=1}^{+\infty} \lambda^{n-1} \int_{0}^{\pm\rmi\infty} \rmd w\, w^{-n}\, \text{Disc}_{\theta_{\pm}} Z_{\text{ABJM}}\left(\BS^{3}\right) (w)  \\
&=& \frac{1}{4\pi^{2}}\, \sum_{n=2}^{+\infty} \lambda^{n-1} \sum_{\pm} \rme^{-\rmi\theta_{\pm}(n-1)} \sum_{\ell=0}^{+\infty} \sum_{m=0}^{+\infty} \frac{\left(n-2\right) \Gamma \left(n-1\right)}{\left|\omega_{\ell m}\right|^{n-1}}.
\end{eqnarray}
\noindent
The sum over the two Stokes directions $\sum_{\pm} \rme^{-\rmi\theta_{\pm}(n-1)} = 2 \cos \left(\left(n-1\right)\frac{\pi}{2}\right)$ is zero if $n$ is even, and is $2 \left(-1\right)^{\frac{n-1}{2}}$ if $n$ is odd. This will allow us to recover the squared power of $k$ in the perturbative expansion of the partition function, \eqref{eq:Partfunc-ABJM-pert-series}. Making use of the precise values for the Borel singularities, \eqref{eq:Singularities-ABJM}, one finally obtains\footnote{Recall that from the definition of the zeta function one has $\sum_{m=0}^{+\infty} \left(2m+1\right)^{-2g} = 2^{-2g} \left(2^{2g}-1\right) \zeta(2g)$.} 
\be
Z_{\text{ABJM}} \left(\BS^{3}\right) (k) \simeq \sum_{n=1}^{+\infty} \left(-1\right)^{n} \frac{2}{\left(2\pi\right)^{2n+2}} \left(2n-1\right) \Gamma\left(2n\right) \left(2^{2n}-1\right)^{2} \zeta(2n)^{2}\, k^{-2n},
\ee
\noindent
which agrees with the original asymptotic expansion, \eqref{eq:Partfunc-ABJM-pert-series} and \eqref{eq:ABJM-pert-coeff-partfunc}, once we discard the first term $n=1$ (which is removed once we take into account the $\frac{s^{2}}{64}$ term in \eqref{eq:Borel-ABJM}). Note that this derivation is \textit{exact} (and not just an approximation at large perturbative order) due to our complete knowledge of the singular structure of the (meromorphic) Borel transform.

Having obtained these results, one could address other localizable observables in both Chern--Simons and ABJM gauge theories, such as supersymmetric Wilson loops. But as it turns out, the resurgent properties of these observables are similar to the ones of the partition function.

%%%%%%%%%%%%%%%%%%%%%%%%%%%%%%%%%%%%%%%%%%%%%%%%%%%%%%%%%%%%%%%%%
%%%%%%%%%%%%%%%%%%%%%%%%%%%%%%%%%%%%%%%%%%%%%%%%%%%%%%%%%%%%%%%%%
\section{Resurgence in Supersymmetric Yang--Mills Theories }\label{sec:4d-SYM}
%%%%%%%%%%%%%%%%%%%%%%%%%%%%%%%%%%%%%%%%%%%%%%%%%%%%%%%%%%%%%%%%%
%%%%%%%%%%%%%%%%%%%%%%%%%%%%%%%%%%%%%%%%%%%%%%%%%%%%%%%%%%%%%%%%%

We shall now turn to four dimensions, and address the partition function of $\CN=2$ supersymmetric gauge theories on the sphere $\mathbb{S}^{4}$, using the results of supersymmetric localization \citep{Erickson:2000af, Drukker:2000rr, Nekrasov:2002qd, Pestun:2007rz}. The simplest gauge theory in four dimensions is maximally supersymmetric $\mathcal{N}=4$ Yang--Mills theory. This case, however, turns out to be quite simple from a resurgence point--of--view; in order to find new resurgent structures---as compared to the previous three--dimensional examples---one needs to consider gauge theories with less supersymmetry. Herein, we shall mainly focus upon $\mathcal{N}=2$ superconformal Yang--Mills theory and in the $\mathcal{N}=2^{*}$ $\text{S}\text{U}(2)$ gauge theory (which is obtained via mass deformation of the $\mathcal{N}=4$ case). Both theories will show a rich structure of singularities in their corresponding Borel transforms.

In four dimensions, the (localized) partition function is of the form \citep{Pestun:2007rz}
\begin{equation}
\label{eq:Z-with-Z-1-loop-and-Z-inst}
Z \propto \int_{\mathfrak{g}} \left[\rmd a\right]\, \rme^{- S_{\text{cl}} (a)}\, Z_{\text{1-loop}}(\rmi a)\, \left| Z_{\text{inst}}\left(\rmi a, g^{2}\right) \right|^{2} .
\end{equation}
\noindent
The structure is similar to \eqref{eq:Z-with-Z-1-loop}. The integration is over the Lie algebra $\mathfrak{g}$ of the gauge group $G$, and $S_{\text{cl}} (a)$ is the classical kinetic action given by $S_{\text{cl}} (a) =  \frac{1}{2g}\,\text{Tr} \left(a\cdot a\right)$. The factor $Z_{\text{1-loop}} \left(\rmi a\right)$ comes from the gauge--fixing determinant in the localization procedure, and is given by an infinite--dimensional product typically written in terms of Barnes $G$--functions. An addition compared to \eqref{eq:Z-with-Z-1-loop} is that there are now gauge--theory instanton contributions represented by the factor $Z_{\text{inst}}\left(\rmi a, g_{s}^{2}\right)$, given by the Nekrasov partition function \citep{Nekrasov:2002qd, Nekrasov:2003rj}. If one considers other observables, say Wilson loops in some representation $\mathcal{R}$, one needs to insert extra factors $\text{Tr}_{\CR} \left( \rme^{2\pi \rmi a} \right)$. Both factors $Z_{\text{1-loop}}$ and $Z_{\text{inst}}$ show nontrivial behavior and need to be addressed in detail.

%%%%%%%%%%%%%%%%%%%%%%%%%%%%%%%%%%%%%%%%%%%%%%%%%%%%%%%%%%%%%%%%%
\subsection{$\mathcal{N}=4$ Supersymmetric Yang--Mills Theory}\label{sec:N4}
%%%%%%%%%%%%%%%%%%%%%%%%%%%%%%%%%%%%%%%%%%%%%%%%%%%%%%%%%%%%%%%%%

The partition function of $\mathcal{N}=4$ $\text{S}\text{U}(N)$ SYM theory on a unit sphere $\mathbb{S}^{4}$ localizes in the usual form, \eqref{eq:Z-with-Z-1-loop-and-Z-inst} \citep{Pestun:2007rz}. However, in this case both one--loop and higher--instanton contributions to the partition function have a trivial role to play as $Z_{\text{1-loop}} = 1$, $Z_{\text{inst}} = 1$ \citep{Okuda:2010ke}. Consequently, the localized partition function is \textit{exactly} given by a Gaussian matrix model. Considering gauge group $\text{U}(N)$, this matrix model is simply 
\begin{equation}
Z_{\text{4SYM}} \left(\mathbb{S}^{4}\right) = \frac{1}{\text{vol}\left(\text{U}(N)\right)} \int \rmd M\, \rme^{-\frac{1}{g_{\text{YM}}^2}\, \text{Tr} M^{2}}.
\end{equation}
\noindent
Besides the partition function, the localization techniques developed in \citep{Pestun:2007rz} further provided for a proof of the formulae in \citep{Erickson:2000af, Drukker:2000rr} concerning the vev of a $\frac{1}{2}$BPS circular Wilson loop. It may be likewise written in a Gaussian matrix model formulation. In a representation $\CR$, this  is given by the expectation value 
\be
\label{wila}
\left\langle W_{\CR} \right\rangle _{\text{4SYM}} = \left\langle \text{Tr}_{\CR} \rme^M \right\rangle.
\ee
\noindent 
In the fundamental representation $\CR = \tableau{1}$, \eqref{wila} reduces to a sum of simple Gaussian integrals, which can be easily evaluated to \citep{Drukker:2000rr} (herein we only present the planar result, see \citep{Drukker:2000rr} for the complete all--orders 't~Hooft expansion)
\be
\left\langle W_{\tableau{1}} \right\rangle_{\text{4SYM}} = \frac{2}{\sqrt{t}}\, I_{1} \left(\sqrt{t}\right) + \CO \left( \frac{1}{N^2} \right),
\ee
\noindent
where $I_{1}(x)$ is a modified Bessel function, for which to all orders in the planar expansion its perturbative series has infinite radius of convergence. The reason for this convergence is a massive cancellation of planar Feynman diagrams. Indeed, it has been shown that only rainbow graphs without internal vertices contribute \cite{Erickson:2000af}. On the other hand \citep{Drukker:2000rr}, expanding in large 't~Hooft coupling the resulting series is indeed asymptotic with its resurgence properties resembling some of the examples appearing in \citep{Upcoming:2014} (and see also \citep{Maldacena:1998im} for the context of large $N$ duality). It would be very interesting to carefully analyze all these different limits and expansions from a resurgence viewpoint, where complete and exact results should be simple to obtain. 

The $1/N$ expansion also exhibits an interesting structure: for the free energy it is asymptotic; while for Wilson loops it is a convergent series. As mentioned, the free energy precisely localizes to a Gaussian matrix model, and the resurgent properties of this Gaussian matrix model were studied in great detail in \citep{Pasquetti:2009jg}. They turn out to be essentially very similar to the case of Chern--Simons gauge theory on lens spaces we addressed earlier; in fact the Borel singularities of the Gaussian matrix model reduce to the red singularities illustrated in figure \ref{fig:CSS3Borel}. Using the results in \citep{Pasquetti:2009jg} alongside our earlier Chern--Simons analysis, it should be a simple exercise for the reader to work out the complete set of details.

More intricate perturbation series with asymptotic $n!$ large--order behavior arise in theories with less supersymmetry, and in this regard some asymptotic properties of Wilson loops in $\CN=2$ theories were already addressed in \cite{Russo:2012kj}. We shall next turn to this class of theories, focusing upon their partition functions (the perturbation series for the vev of the $\frac{1}{2}$BPS circular Wilson loop has similar asymptotic behavior as the partition function, and in fact the same Borel singularities, so here we shall omit an explicit discussion).

%%%%%%%%%%%%%%%%%%%%%%%%%%%%%%%%%%%%%%%%%%%%%%%%%%%%%%%%%%%%%%%%%
\subsection{$\mathcal{N}=2$ Superconformal Yang--Mills Theory}\label{sec:N2}
%%%%%%%%%%%%%%%%%%%%%%%%%%%%%%%%%%%%%%%%%%%%%%%%%%%%%%%%%%%%%%%%%

We shall now focus upon $\mathcal{N}=2$ superconformal Yang--Mills theory with gauge group $\text{S}\text{U}(2)$, which we shall analyze at small gauge coupling, $g_{\text{YM}}$. Due to superconformality, there will be no renormalon contributions to the nonperturbative structure, and one can expect all singular Borel behavior to have its origins in semiclassical effects, as we will discuss later in section \ref{sec:physical-int}. As we shall see, the resurgence properties arising from perturbative expansions in this theory, as well as those originating in each instanton sector $Z_{\text{inst}}$, will be quite different than one might naively have expected for, based on usual instanton analysis. Nonetheless, due to the meromorphicity of the Borel transform we shall soon unveil, it will still be the case that one can find closed form expressions describing all these resurgent nonperturbative properties.

$\mathcal{N}=2$ superconformal Yang--Mills theory is given by standard $\mathcal{N}=2$ supersymmetric Yang--Mills theory with gauge group $\text{S}\text{U}(N_{\text{c}})$ and $N_{\text{f}}=2N_{\text{c}}$ massless multiplets. Its partition function localized on $\mathbb{S}^{4}$ was first computed in \citep{Pestun:2007rz} and from these results one can write down its (diagonal) matrix model representation as
\begin{equation}
Z_{\text{2SYM}} \propto g_{\text{YM}}^{-(N_{\text{c}}^{2}-1)} \int \rmd^{N_{\text{c}}-1}\lambda \prod_{1 \le i<j \le N_{\text{c}}} (\lambda_{i}-\lambda_{j})^{2}\, \rme^{-\frac{2}{g_{\text{YM}}^{2}}\sum_{i=1}^{N_{\text{c}}}  \lambda_{i}^{2}}\, Z_{\text{1-loop}} (\lambda) \left| Z_{\text{inst}} (\lambda,g_{\text{YM}}^{2}) \right|^{2},
\end{equation}
\noindent
where $\sum_{i=1}^{N_{\text{c}}} \lambda_{i} = 0$, and where
\be
Z_{\text{1-loop}} (\lambda) = \frac{\prod_{1 \le i<j \le N_{\text{c}}} H^{2} \left(\lambda_{i}-\lambda_{j}\right)}{\prod_{i=1}^{N_{\text{c}}} H^{2N_{\text{c}}} \left(\lambda_{i}\right)}, \qquad H(x) = \prod_{n=1}^{+\infty} \left( 1+\frac{x^{2}}{n^{2}} \right)^{n} \rme^{-\frac{x^{2}}{n}}.
\ee
\noindent
Note that the function $H(x)$ is related to the Barnes $G$--function, $G(x)$, by
\be
H(x) = \rme^{- \left(1+\gamma\right) x^{2}}\, G \left(1+\rmi x\right) G \left(1-\rmi x\right),
\ee
\noindent
with $\gamma$ the Euler constant. We shall focus on the case of $N_{c}=2$, and perform an analysis for small Yang--Mills coupling $g_{\text{YM}}$. This analysis was initiated in \citep{Russo:2012kj}, where it was shown that the perturbative series in $g_{\text{YM}}$ is asymptotic with its coefficients growing factorially fast. In here we shall study this factorial growth in full detail, in particular addressing the (nonperturbative) Borel singular structure which gives rise to this behavior. Following \citep{Russo:2012kj}, we find that for $N_{c}=2$ the partition function reduces to
\be
\label{eq:2SYM-full-part-func}
\mathcal{Z}_{\text{2SYM}}^{\text{S}\text{U}(2)} \left(g_{\text{YM}}\right) = \frac{128\pi^{5/2}}{g_{\text{YM}}^{3}} \int_{-\infty}^{+\infty} \rmd a\, a^{2}\, \rme^{-\frac{16\pi^{2}}{g_{\text{YM}}^{2}}\, a^{2}}\, \prod_{n=1}^{+\infty}\frac{\left( 1+\frac{4a^{2}}{n^{2}} \right)^{2n}}{\left( 1+\frac{a^{2}}{n^{2}} \right)^{8n}}\, \left|Z_{\text{inst}}^{\text{S}\text{U}(2)}\right|^{2},
\ee
\noindent
where the overall numerical constant was fixed in order for the partition function to be normalized to $1$ in the limit $g_{\text{YM}}\rightarrow0$. The infinite product in the integrand arises from the $Z_{\text{1-loop}}$ factor, and may also be written as a product of Barnes $G$--functions,
\be
\label{eq:2SYM-one-loop-factor}
\prod_{n=1}^{+\infty} \frac{\left( 1+\frac{4a^{2}}{n^{2}} \right)^{2n}}{\left( 1+\frac{a^{2}}{n^{2}} \right)^{8n}} = \frac{\Big( G \left(1+2\rmi a\right)\, G \left(1-2\rmi a\right) \Big)^{2}}{\Big( G \left(1+\rmi a\right)\, G \left(1-\rmi a\right) \Big)^{8}}.
\ee
\noindent
As for the instanton factor, $Z_{\text{inst}}\left(\rmi a\right)$, determined in \citep{Nekrasov:2002qd} (but see also \citep{Alday:2009aq}), it is given schematically by
\be
Z_{\text{inst}} (\rmi a) = 1 + \sum_{k=1}^{+\infty} \rme^{2\pi\rmi k\tau}\, Z_{k} (\rmi a).
\ee
\noindent
In this expression, $\tau=\frac{4\pi\rmi}{g_{\text{YM}}^{2}}+\frac{\theta}{2\pi}$ is the complexified gauge coupling, with $\theta$ the topological charge, and $Z_{k}(a)$ is the $k$--instanton partition function ($Z_{0}(a)\equiv1$). Expanding the integrand component $\left| Z_{\text{inst}} (\rmi a) \right|^{2}$ in both topological charge and instanton number it follows
\begin{eqnarray}
\label{topchargeZ}
\left| Z_{\text{inst}} (\rmi a) \right|^{2} &=& \sum_{k'=0}^{+\infty} \sum_{\pm} \rme^{\pm\rmi\theta k' - \frac{8\pi^{2}}{g_{\text{YM}}^{2}}\, k'}\, 
\widehat{Z}_{k',\pm}^{\theta} \left(\rmi a\right), \\
\label{instchargeZ}
\widehat{Z}_{k',\pm}^{\theta} \left(\rmi a\right) &\equiv& \sum_{k=0}^{+\infty} \rme^{-\frac{16\pi^{2}}{g_{\text{YM}}^{2}}\, k} \left\{ Z_{k+\frac{(1\pm1)}{2}k'} \left(\rmi a\right) Z_{k+\frac{(1\mp1)}{2}k'} \left(-\rmi a\right) \right\}.
\end{eqnarray}
\noindent
It is important to note that each distinct topological sector should be addressed separately, as the resurgent structure of different topological sectors does not mix. Consider some arbitrary theory whose effective coupling has a topological charge, and for which the study of some given observables around each topological and instanton sectors returns an asymptotic series. In this case, general resurgence properties such as ambiguity cancelations will occur independently, within each separate topological sector. This lead to the construction of the resurgence triangle in \citep{Dunne:2012ae} (but see also the final discussion in \citep{Aniceto:2013fka} on general ambiguity cancelation). As we shall see next, in the present case extended supersymmetry will simplify the resurgence triangle even further, in such a way that resurgence properties will occur within each topological charge sector \textit{and} each gauge--theoretic instanton sector. Furthermore, it will be the localization procedure itself which will be responsible for the tower of Borel singularities which lead to resurgence within each of these sectors (see the discussion in section \ref{sec:physical-int}).

To begin with, let us focus upon the zero topological charge sector, corresponding to taking $k'=0$ in the above expansion. Setting the instanton action to be $A=8\pi^{2}$, this results in
\be
\widehat{Z}_{0}^{\theta} (\rmi a) = \sum_{k=0}^{+\infty} \rme^{-\frac{2A}{g_{\text{YM}}^{2}}\, k}\, \left| Z_{k} (\rmi a)\right|^{2}.
\ee
\noindent
This sector includes the perturbative\footnote{Note that any other topological sector will not only have a distinct topological charge, $\rme^{\pm\rmi\theta k'}$, but it will also be exponentially suppressed by at least a factor of $\exp \left(-\frac{A}{g_{\text{YM}}^{2}}\, k'\right)$, with respect to this perturbative series.} series with no instanton corrections, corresponding to $k=0$, $Z_{0} (\rmi a)\equiv1$. The resurgent analysis which now follows can be likewise done for any other topological sector, and the results will be analogous to the ones presented below. Substituting the $Z_{\text{inst}}$ factor by the above zero topological charge sector, in the integrand of the partition function \eqref{eq:2SYM-full-part-func}, we get
\begin{equation}
\mathcal{Z}_{\text{2SYM}}^{\theta=0} \left(g_{\text{YM}}\right) = \sum_{k=0}^{+\infty} \rme^{-\frac{2A}{g_{\text{YM}}^{2}}\, k}\, \mathcal{Z}_{k}^{\theta=0} \left(g_{\text{YM}}\right),
\label{eq:N2-SYM-zero-top-partfunc}
\end{equation}
\noindent
with
\begin{equation}
\mathcal{Z}_{k}^{\theta=0} \left(g_{\text{YM}}\right) \equiv \frac{128\pi^{5/2}}{g_{\text{YM}}^{3}} \int_{-\infty}^{+\infty} \rmd a\, a^{2}\, \rme^{-\frac{2A}{g_{\text{YM}}^{2}}\, a^{2}}\, \prod_{n=1}^{+\infty} \frac{\left( 1+\frac{4a^{2}}{n^{2}} \right)^{2n}}{\left( 1+\frac{a^{2}}{n^{2}} \right)^{8n}}\, \left| Z_{k} (\rmi a) \right|^{2}.
\end{equation}
\noindent
Above we introduced the $k$--instanton partition function, $\mathcal{Z}_{k}^{\theta=0}$, in the zero topological charge sector. Next, within this sector, we may separately analyze the many instanton contributions. Let us first address the perturbative series, $k=0$,
\begin{eqnarray}
\label{eq:N2-SYM-pert-partfunc}
\mathcal{Z}_{0}^{\theta=0} \left(g_{\text{YM}}\right) &=& \frac{128\pi^{5/2}}{g_{\text{YM}}^{3}} \int_{-\infty}^{+\infty} \rmd a\, a^{2}\, \rme^{-\frac{2A}{g_{\text{YM}}^{2}}\, a^{2}}\, \prod_{n=1}^{+\infty} \frac{\left( 1+\frac{4a^{2}}{n^{2}} \right)^{2n}}{\left( 1+\frac{a^{2}}{n^{2}} \right)^{8n}} \\
&\equiv& 
\frac{2}{\sqrt{\pi}} \int_{-\infty}^{+\infty} \rmd\widetilde{a}\, \rme^{- V\left(\widetilde{a},g_{\text{YM}}\right)},
\end{eqnarray}
\noindent
where we changed variables as $\widetilde{a}=\frac{\sqrt{2A}}{g_{\text{YM}}}\, a$. Using \eqref{eq:2SYM-one-loop-factor} we find that the effective potential can be easily determined to be 
\be
\label{eq:N2-SYM-pert-pot}
V \left(\widetilde{a},g_{\text{YM}}\right) = \widetilde{a}^{2} - 2 \log \left(\widetilde{a}\right) - \log \left( \frac{\left( G \left( 1 + 2\rmi \widetilde{a}\, \frac{g_{\text{YM}}}{\sqrt{2A}} \right) G \left( 1  - 2\rmi \widetilde{a}\, \frac{g_{\text{YM}}}{\sqrt{2A}} \right) \right)^{2}}{\left(G \left( 1 + \rmi \widetilde{a}\, \frac{g_{\text{YM}}}{\sqrt{2A}} \right) G \left( 1 - \rmi \widetilde{a}\, \frac{g_{\text{YM}}}{\sqrt{2A}} \right) \right)^{8}} \right).
\ee

\begin{figure}
\centering{}
\begin{tabular}{cc}
\includegraphics[scale=0.6]{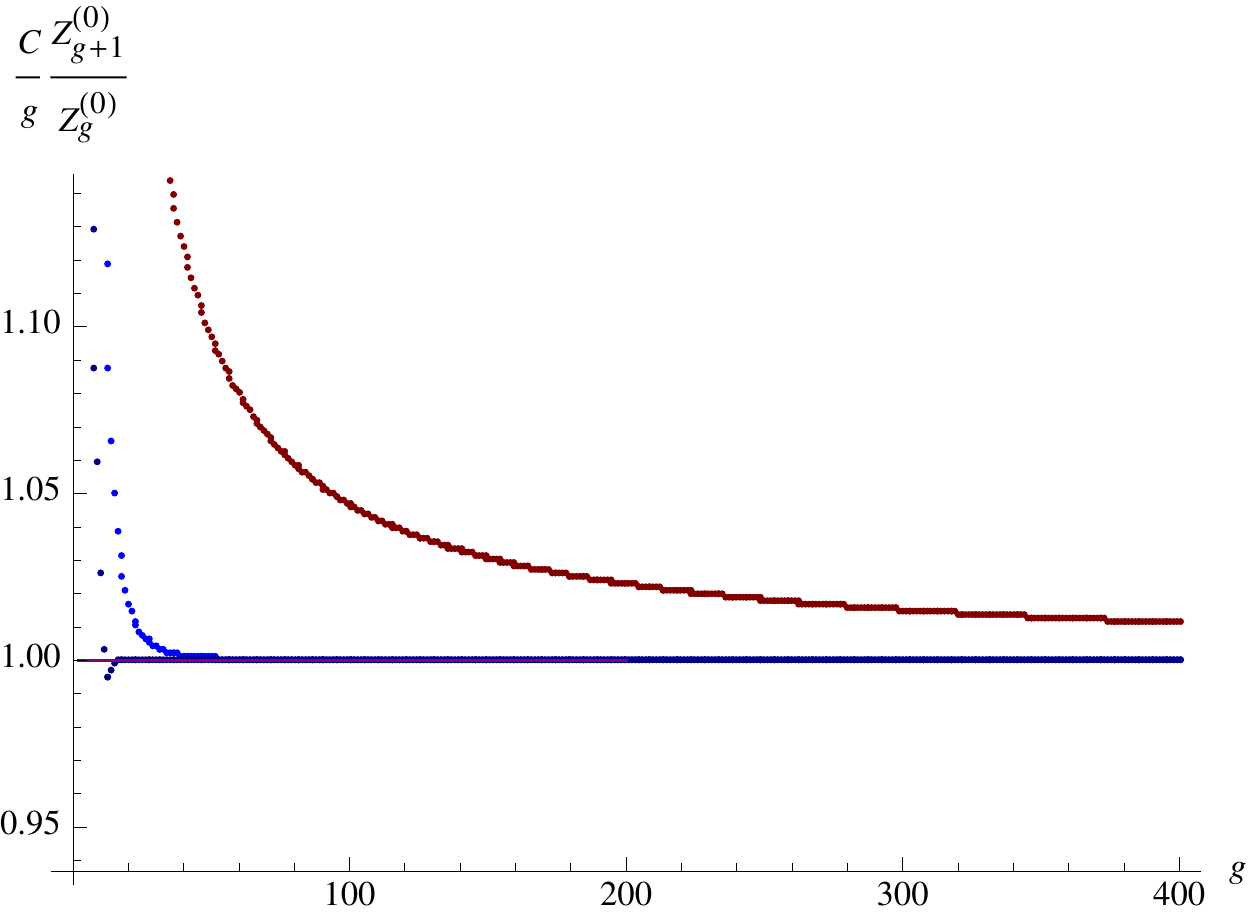} & \includegraphics[scale=0.6]{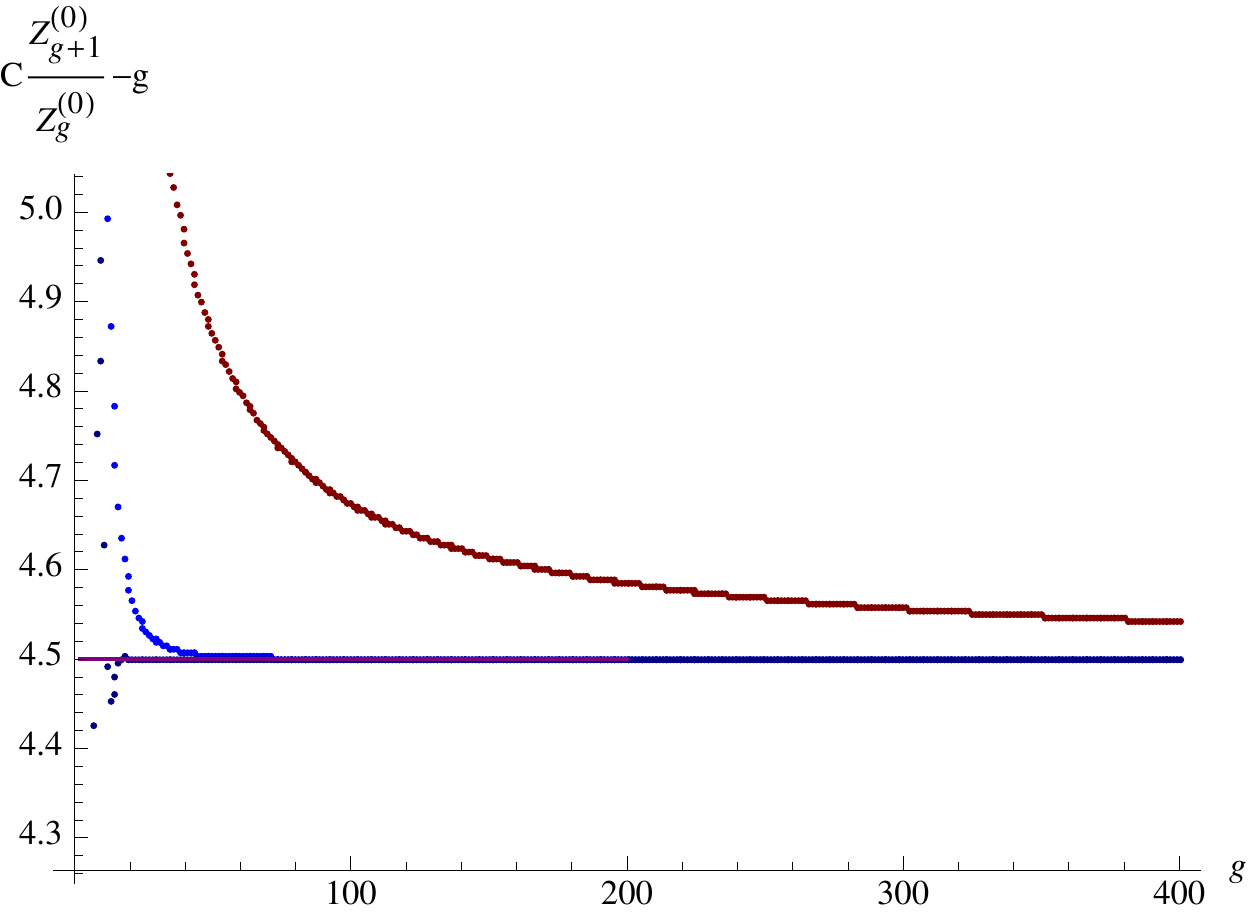}
\end{tabular}
\caption{Numerical analysis of the leading (left plot) and subleading (right plot) behavior of the ratio $Z_{g+1}^{(0)} / Z_{g}^{(0)}$, at large order $g$. In each case we plot the values of the ratio (the red line) and two of its corresponding Richardson transforms (of orders $2$ and $6$, shown in blue). The convergence towards the predicted values (the horizontal lines) is clearly very precise.
\label{fig:N2-pert-0and1loops}}
\end{figure}

The perturbative series can be obtained by taking $g_{\text{YM}}\ll1$, and in fact its  first few terms were already determined analytically in \citep{Russo:2012kj}. However, in order to examine the large--order behavior we need to generate higher order terms. In this sense, let us first note that this perturbative series is of the generic form 
\be
\label{eq:N2-SYM-pertseries}
\mathcal{Z}_{0}^{\theta=0} \left(g_{\text{YM}}\right) \simeq \sum_{g=0}^{+\infty} g_{\text{YM}}^{2g}\, Z_{g}^{(0)}.
\ee
\noindent
Next, performing a numerical calculation of the coefficients $Z_{g}^{(0)}$ in this series by straightforward perturbative evaluation of the integral \eqref{eq:N2-SYM-pert-partfunc}, we find a large--order factorial growth of the type $Z_{g}^{(0)} \sim C^{-g}\, \Gamma (g+\beta) \left( c_0 + c_1 g^{-1} + c_2 g^{-2} + \cdots \right)$, for some $C$, $\beta$, $c_k$. In fact we can simply analyze the ratio of these coefficients, which for large $g$ will obey
\be
\frac{C}{g}\, \frac{Z_{g+1}^{(0)}}{Z_{g}^{(0)}} \sim 1 + \frac{\beta}{g} - \frac{c_1}{c_0\, g^2} + \cdots.
\ee
\noindent
The value of these unknown coefficients may now be determined numerically (in fact further increasing the speed of convergence through the use of Richardson transforms; see, \textit{e.g.}, \citep{Marino:2007te} for a simple introduction in the present large--order context). The left plot of figure \ref{fig:N2-pert-0and1loops} shows the convergence of the above ratio to its leading value, $1$, after making the choice of $C = -2A = -(4\pi)^2$. The right plot shows the convergence of the same ratio, minus its leading value of $1$, in order to find the value $\beta = \frac{9}{2}$ to very high numerical accuracy. These results imply that the leading singularities in the complex Borel plane are expected to be instanton--like, albeit located at $s= -2A$, \textit{i.e.}, in the \textit{negative} real axis, and with value of twice the instanton action of the four--dimensional gauge theory. We also conclude that the factorial growth of the perturbative series goes as $Z_g^{(0)} \sim \Gamma(g + 9/2)$.

%%%%%%%%%%%%%%%%%%%%%%%%%%%%%%%%%%%%%%%%%%%%%%%%%%%%%%%%%%%%%%%%%
%%%%%%%%%%%%%%%%%%%%%%%%%%%%%%%%%%%%%%%%%%%%%%%%%%%%%%%%%%%%%%%%%

\begin{figure}
\centering{}
\includegraphics[scale=0.45]{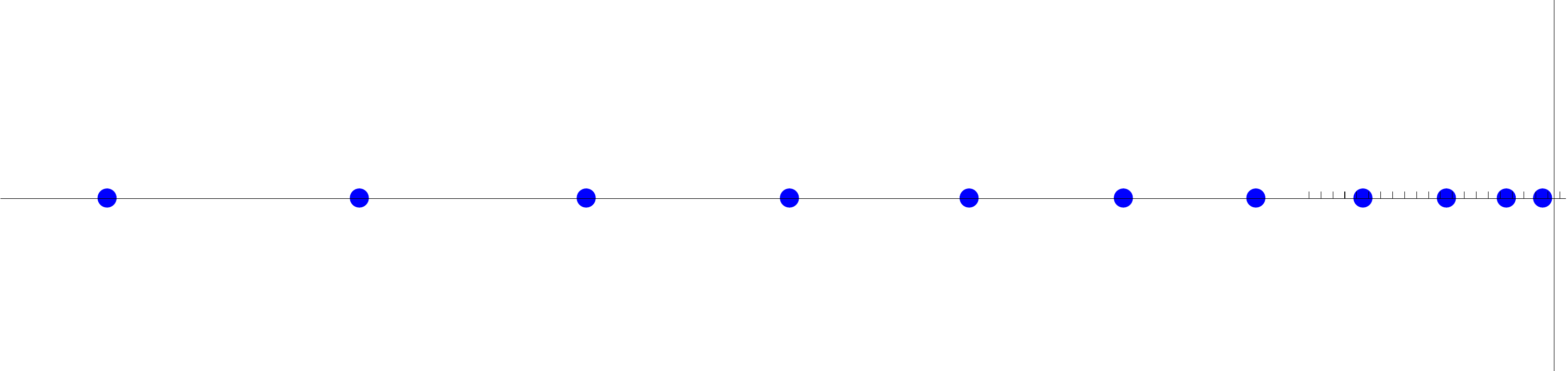}
\caption{Singularities in the complex Borel plane for $\mathcal{N}=2$ superconformal Yang--Mills theory on $\mathbb{S}^4$, when considering small coupling $g_{\text{YM}}$ and rank of the gauge group $N=2$. All singularities lie along the negative real direction, giving rise to the corresponding single Stokes line of this example.}
\label{fig:N2-SYM-Borel}
\end{figure}

One can proceed along this line and extend the numerical analysis to higher orders, thus determining several of the subsequent coefficients. However, a much more interesting approach is the direct analytical analysis of the Borel transform associated with this perturbative series. Notice that this cannot be done directly from the coefficients of the series itself, at least not in any simple way, as the coefficients of this series are not easy to obtain analytically. Nonetheless, there is another approach we shall now discuss which leads to an analytic expression for the Borel transform, thus including its complete singularity structure. Looking back at the integral in \eqref{eq:N2-SYM-pert-partfunc}, we can change variables as $2A\, a^{2}=s$ to obtain 
\be
\left(g_{\text{YM}}^{2}\right)^{3/2} \mathcal{Z}_{0}^{\theta=0} \left(g_{\text{YM}}\right) = \frac{2}{\sqrt{\pi}} \int_{0}^{+\infty} \rmd s\, \rme^{-\frac{s}{g_{\text{YM}}^{2}}}\, \sqrt{s}\, \frac{\Big( G \left( 1+2\rmi\sqrt{\frac{s}{2A}} \right) G \left( 1-2\rmi\sqrt{\frac{s}{2A}} \right) \Big)^{2}}{\Big( G \left( 1+\rmi\sqrt{\frac{s}{2A}} \right) G \left( 1-\rmi\sqrt{\frac{s}{2A}} \right) \Big)^{8}}.
\ee
\noindent
From this expression we can directly read off the Borel transform corresponding to the function $\left(g_{\text{YM}}^{2}\right)^{3/2} \mathcal{Z}_{0}^{\theta=0} \left(g_{\text{YM}}\right)$. Do note that the partition function is a function of the variable $\lambda \equiv g_{\text{YM}}^{2}$, thus this is the correct variable appearing in the exponential of the Borel integrand. As such, what we are interested in analyzing is the singularity structure of
\be
\label{eq:N2-SYM-pert-Borel}
\mathcal{B} [\lambda^{3/2} \mathcal{Z}_{0}^{\theta=0}] (s) \equiv \frac{2 \sqrt{s}}{\sqrt{\pi}}\, \frac{\Big( G \left(1+2\rmi\sqrt{\frac{s}{2A}}\right) G \left(1-2\rmi\sqrt{\frac{s}{2A}}\right) \Big)^{2}}{\Big( G \left(1+\rmi\sqrt{\frac{s}{2A}}\right) G \left(1-\rmi\sqrt{\frac{s}{2A}}\right) \Big)^{8}}.
\ee
\noindent
To do so just recall that we have been frequently interchanging between products of Barnes $G$--functions and infinite products, and we can do so once again to obtain
\be
\label{eq:N2-SYM-look-for-poles-Borel}
\frac{\Big( G \left(1+2\rmi a\right) G \left(1-2\rmi a\right) \Big)^{2}}{\Big( G \left(1+\rmi a\right) G \left(1-\rmi a\right) \Big)^{8}} = \prod_{n=1}^{+\infty} \frac{\left( 1+\frac{4a^{2}}{n^{2}} \right)^{2n}}{\left( 1+\frac{a^{2}}{n^{2}} \right)^{8n}} = \prod_{n=1}^{+\infty} \frac{\left( 1+\frac{4a^{2}}{\left(2n+1\right)^{2}} \right)^{4n+2}}{\left( 1+\frac{a^{2}}{n^{2}} \right)^{4n}}.
\ee
\noindent
Using the infinite--product representation it is now much simpler to identify all the singularities of this function as consisting of poles, located at the points $a^{2} = -n^{2}$, and where each pole is of order $4n$ in the variable $a^{2}$. Applying the same reasoning to the Borel transform, we find that this Borel transform is a meromorphic function (as was the case for the three--dimensional examples studied in the previous section), with poles located at
\be
\label{eq:N2-SYM-sing-pert-Borel}
s_{n} = - 2 A n^{2} \equiv - \left(4\pi\right)^{2} n^{2}, \qquad n \in \mathbb{N},
\ee
\noindent
with each pole at $s=s_{n}$ being of order $4n$. In figure \ref{fig:N2-SYM-Borel} we illustrate the singularity structure of this Borel transform. We can then conclude that all the singularities of the Borel transform lie in the negative real axis, implying there is a single Stokes line along $\theta=\pi$.

%%%%%%%%%%%%%%%%%%%%%%%%%%%%%%%%%%%%%%%%%%%%%%%%%%%%%%%%%%%%%%%%%
%%%%%%%%%%%%%%%%%%%%%%%%%%%%%%%%%%%%%%%%%%%%%%%%%%%%%%%%%%%%%%%%%

The question that naturally follows is how to determine the discontinuity across the aforementioned Stokes line. To do so, let us first analyze the behavior of the Borel transform around each pole,
\be
\label{eq:N2-SYM-borel-taylor-exp-each-pole}
\left. \mathcal{B} [\lambda^{3/2}\mathcal{Z}_{0}^{\theta=0}] (s) \right|_{s_{n}} = \frac{1}{\left(s-s_{n}\right)^{4n}}\, \sum_{\ell=0}^{+\infty} \frac{f_{\ell}^{(n)}}{\ell!} \left(s-s_{n}\right)^{\ell}.
\ee
\noindent
In this expression, the Taylor series around $s_{n}$ with coefficients $f_{\ell}^{(n)}$ is obtained by expanding the product of $\mathcal{B} [\lambda^{3/2}\mathcal{Z}_{0}^{\theta=0}] (s)$ with $\left(s-s_{n}\right)^{4n}$. For each distinct pole $s_{n}$, one obtains a different Taylor expansion. Given our discussions in section \ref{sec:Chern-Simons-theory}, we already know that the Stokes discontinuity is computed via alien calculus, and that there are certain ``Borel representatives'' which may make this calculation easier (even immediate!). In our present example, the above form of the Borel transform around each of its poles, \eqref{eq:N2-SYM-borel-taylor-exp-each-pole}, is not of this simplest form from an alien calculus standpoint. Instead, we would like to rewrite in the form of a simple pole alongside a logarithmic branch--cut, in the same manner as discussed at length in the previous section (but see also \citep{Aniceto:2011nu, Upcoming:2014} for further details). To do this, we first note that taking $4n-1$ primitives of our result leads us to the intended form of the Borel transform. Furthermore, recall from last section that there is a simple relation between integrations (or differentiations) of Borel transforms, and an overall power of $\lambda$ in the original function. This said, we can immediately write
\be
\mathcal{B} [\lambda^{3/2}\mathcal{Z}_{0}^{\theta=0}] (s) = \frac{\rmd^{4n-1}}{\rmd s^{4n-1}}\, \mathcal{B} [\lambda^{3/2}\lambda^{4n-1}\mathcal{Z}_{0}^{\theta=0}] (s),
\ee
\noindent
where we now find the representative we were looking for as
\be
\left. \mathcal{B} [\lambda^{4n+1/2}\mathcal{Z}_{0}^{\theta=0}] (s) \right|_{s_{n}} = \frac{f_{0}^{(n)}}{(-1)^{4n-1} \left(4n-1\right)!}\, \frac{1}{s-s_{n}} + \frac{\mathcal{B} [\psi_{n}] \left(s-s_{n}\right)}{2\pi\rmi}\, \log \left(s-s_{n}\right) + \text{regular}.
\label{eq:N2-SYM-Borel-transf-each-pole-logs}
\ee
\noindent
Here, the Borel transform $\mathcal{B} [\psi_{n}]$ is given by 
\be
\mathcal{B} [\psi_{n}] (s) = \sum_{\ell=1}^{4n-1} \frac{2\pi\rmi\, f_{\ell}^{(n)}}{(-1)^{4n-\ell-1}\, \Gamma (4n-\ell)\, \Gamma (\ell)\, \ell!}\, s^{\ell-1},
\ee
\noindent
and the corresponding function $\psi_{n}$ is then given by
\be
\psi_{n} (\lambda) = \sum_{\ell=1}^{4n-1} \frac{2\pi\rmi\, f_{\ell}^{(n)}}{(-1)^{4n-\ell-1}\, \Gamma (4n-\ell)\, \ell!}\, \lambda^{\ell}.
\label{eq:N2-SYM-Log-term-Borel-transf-each-pole}
\ee
\noindent
Note that this function is not resurgent nor even asymptotic; it is simply a polynomial. It is now this fact which is signaling back the original meromorphicity of the Borel transform. We can next determine the discontinuity of the perturbative partition function across the singular direction $\theta=\pi$, by following a similar procedure to what we have already done twice before. In more technical terms, one would write\footnote{This is only true because the function $\psi_{n} (\lambda)$ is not asymptotic, and thus many expressions simplify as the alien derivative acting on $\psi_{n}$ returns zero.}
\begin{eqnarray}
\text{Disc}_{\pi}\, \mathcal{Z}_{0}^{\theta=0} (\lambda) &=& - \sum_{n=1}^{+\infty}\rme^{-\frac{s_{n}}{\lambda}}\, \Delta_{s_{n}} \mathcal{Z}_{0}^{\theta=0} (\lambda) \\
&=& 
- \sum_{n=1}^{+\infty} \rme^{- \frac{s_{n}}{\lambda}}\, \lambda^{-4n-\frac{1}{2}}\, \Delta_{s_{n}} \left( \lambda^{4n+\frac{1}{2}} \mathcal{Z}_{0}^{\theta=0} (\lambda) \right),
\end{eqnarray}
\noindent
where the alien derivative $\Delta_{s_{n}} \left( \lambda^{4n+1/2} \mathcal{Z}_{0}^{\theta=0} (\lambda) \right)$ can be read off directly from \eqref{eq:N2-SYM-Borel-transf-each-pole-logs} and \eqref{eq:N2-SYM-Log-term-Borel-transf-each-pole} as
\be
\Delta_{s_{n}} \left( \lambda^{4n+\frac{1}{2}} \mathcal{Z}_{0}^{\theta=0} (\lambda) \right) = \sum_{\ell=0}^{4n-1} \frac{2\pi\rmi\, f_{\ell}^{(n)}}{(-1)^{4n-\ell-1}\, \Gamma (4n-\ell)\, \ell!}\, \lambda^{\ell}.
\ee
\noindent
With these results in hand, we can finally determine the discontinuity across the Stokes line as
\be
\label{eq:N2-SYM-Pert-Partfunc-Disc-pi}
\text{Disc}_{\pi}\, \mathcal{Z}_{0}^{\theta=0} (\lambda) = - \sum_{n=1}^{+\infty} \rme^{- \frac{s_{n}}{\lambda}}\, \sum_{\ell=0}^{4n-1} \frac{2\pi\rmi\, f_{\ell}^{(n)}}{(-1)^{4n-\ell-1}\, \Gamma (4n-\ell)\, \ell!}\, \lambda^{\ell-4n-\frac{1}{2}}.
\ee

By now, the reader should already be familiar with the resurgence set--up, where this discontinuity in fact encodes the complete information behind the asymptotic series $\mathcal{Z}_{0}^{\theta=0} (\lambda)$. In particular, we can use it to determine the large--order behavior of the coefficients $Z_{g}^{(0)}$ in \eqref{eq:N2-SYM-pertseries}, making use of the standard Cauchy dispersion relation
\be
\mathcal{Z}_{0}^{\theta=0} (\lambda) = \frac{1}{2\pi\rmi} \int_{0}^{-\infty} \rmd w\,\frac{\text{Disc}_{\pi}\, \mathcal{Z}_{0}^{\theta=0} (w)}{w-\lambda},
\ee
\noindent
and expanding the integrand for small $\lambda$. Substituting into the above expression  the exact result for the discontinuity, \eqref{eq:N2-SYM-Pert-Partfunc-Disc-pi}, evaluating the integral, and comparing with \eqref{eq:N2-SYM-pertseries}, we find
\be
\label{eq:N2-SYM-pertcoeff-prediction}
Z_{g}^{(0)} \simeq \sum_{n=1}^{+\infty} \frac{\Gamma \left( g+4n+\frac{1}{2} \right)}{\left( - 2 A n^{2} \right)^{g+4n+\frac{1}{2}}}\, \sum_{\ell=0}^{4n-1} (-1)^{4n-\ell}\, \frac{\Gamma \left( g+4n+\frac{1}{2}-\ell \right)}{\Gamma \left( g+4n+\frac{1}{2} \right) \Gamma \left(4n-\ell\right)\, \ell!}\, f_{\ell}^{(n)} \left( - 2 A n^{2} \right)^{\ell}.
\ee
\noindent
This expression tells us that in order to understand the large--order behavior of the coefficients $Z_{g}^{(0)}$, at each exponentially suppressed order $\sim n^{-g}$, we only need a limited number of coefficients from the Taylor expansion of the Borel transform around each pole---more precisely, only up to $4n-1$ of these coefficients for each pole $s_{n}$. Making the large--order relation a bit more explicit, one finds:
\begin{eqnarray}
Z_{g}^{(0)} &\simeq& \frac{\Gamma \left( g+\frac{9}{2} \right)}{\left( -2A \right)^{g+\frac{9}{2}}}\, \left\{ \frac{f_{0}^{(1)}}{6} - \frac{A\, f_{1}^{(1)}}{g+\frac{7}{2}} + \frac{2A^{2}\, f_{2}^{(1)}}{\left(g+\frac{5}{2}\right) \left(g+\frac{7}{2}\right)} - \frac{4A^{3}\, f_{3}^{(1)}}{3 \left(g+\frac{3}{2}\right) \left(g+\frac{5}{2}\right) \left(g+\frac{7}{2}\right)} \right\} + \nonumber \\
&& 
+\frac{\Gamma \left( g+\frac{17}{2} \right)}{\left( -8A \right)^{g+\frac{17}{2}}}\, \left\{ \frac{f_{0}^{(2)}}{5040} - \frac{A\, f_{1}^{(2)}}{90 \left(g+\frac{15}{2}\right)} + \frac{4A^{2}\, f_{2}^{(2)}}{15 \left(g+\frac{13}{2}\right) \left(g+\frac{15}{2}\right)} + \cdots - \right. \nonumber \\
&&
\left.
- \frac{131072A^{7}\, f_{7}^{(2)}}{315 \prod_{k=0}^{76} \left(g+\frac{15-2k}{2}\right)} \right\} + \mathcal{O} \left( \left(-18A\right)^{-g-25/2} \right).
\label{eq:N2-SYM-pertseries-coeff-1-2-inst}
\end{eqnarray}
\noindent
Recall that the instanton action is here given by $A=8\pi^{2}$. Table \ref{tab:N2-SYM-Coeff-Taylor-pert-series} in page \pageref{tab:N2-SYM-Coeff-Taylor-pert-series} lists the coefficients $f_{\ell}^{(n)}$ from \eqref{eq:N2-SYM-borel-taylor-exp-each-pole}, which we have computed for the first three poles. As can be seen from that short list, these coefficients are generically transcendental numbers involving zeta numbers.

Note that we now have a very definite prediction for the resurgent large--order behavior of our original perturbative series \eqref{eq:N2-SYM-pertseries}, given by \eqref{eq:N2-SYM-pertcoeff-prediction}, which can be numerically checked in much the same way as we did for the leading factorial growth of the ratio $Z_{g+1}^{(0)}/Z_{g}^{(0)}$. If we readdress this ratio, its leading behavior is given by a series in $\frac{1}{g}$, at large order $g$, which is completely determined by the first line (meaning, with $n=1$) of \eqref{eq:N2-SYM-pertseries-coeff-1-2-inst}. The predicted behavior, up to exponential suppressed terms, will then be
\be
- \frac{2A}{g}\, \frac{Z_{g+1}^{(0)}}{Z_{g}^{(0)}} \simeq 1 + \frac{9}{2g} + \frac{6A\, f_{1}^{(1)}}{g^{2}\, f_{0}^{(1)}} - \frac{3A \left(f_{0}^{(1)} ( 8A\, f_{2}^{(1)} + 7\, f_{1}^{(1)} ) - 12A (f_{1}^{(1)})^{2} \right)}{g^{3}\, (f_{0}^{(1)})^{2}} + \cdots,
\ee
\noindent
where one can read the coefficients $f_{\ell}^{(1)}$ from table \ref{tab:N2-SYM-Coeff-Taylor-pert-series}. Numerically we can check\footnote{Take $\Phi(g) \simeq \sum_{n=0}^{+\infty} c_{n}\, g^{-n}$. Plotting the values of $\Phi(g)$ for very large $g$ will of course verify the leading value of this series, $c_{0}$. To check $c_{1}$ we just need to plot instead $\left( \Phi(g)-c_{0} \right) g$, and to check $c_{2}$ we would plot $\left( \left( \Phi(g)-c_{0} \right) g - c_{1} \right) g$. This procedure may be repeated iteratively for higher and higher $c_{n}$, as long as we have enough precision in our numerical calculations and can evaluate the values of $\Phi(g)$ for high enough order $g$.} each coefficient of this expansion in $\frac{1}{g}$ iteratively, by simply subtracting the previous one and multiplying by a power of $g$. To speed up the convergence, we again use the method of Richardson transforms. Define 
\be
- \frac{2A}{g}\, \frac{Z_{g+1}^{(0)}}{Z_{g}^{(0)}} \simeq \sum_{k=0}^{+\infty} c_{k}^{(1)}\, g^{-k} + \mathcal{O} \left( \left(-8A\right)^{-g} \right).
\ee
\noindent
The leading and subleading coefficients, $c_{0}^{(1)}=1$ and $c_{1}^{(1)}=9/2$, have already been checked, as shown back in figure \ref{fig:N2-pert-0and1loops}. But we can now easily go to higher loops. As an example, let us look at the coefficient\footnote{The predicted value for this coefficient is 
\begin{eqnarray*}
\left. c_{20}^{(1)}\right|_{\text{predicted}} &=& 532998144 \zeta(3)^{6} - 326343370176 \zeta(3)^{5} + 17512240669776 \zeta(3)^{4} - \frac{3364351634340225 \zeta(3)^{3}}{16} + \\
&& 
+ \frac{409926727472490567 \zeta(3)^{2}}{512} - \frac{17984987670868446465 \zeta(3)}{16384} + \frac{62525365950697533681}{131072}.
\end{eqnarray*}} $c_{20}^{(1)}$. The difference between the sixth--order Richardson transform for $g=400$, $\text{RT}_{6} \left(c_{20}^{(1)}\right)$, and the predicted value, is negligible and very strongly confirms our results:
\be
\left. \frac{\text{RT}_{6}\left(c_{20}^{(1)}\right) - c_{20}^{(1)}}{c_{20}^{(1)}} \right|_{g=400} \sim 3.054... \times 10^{-14}.
\ee

\begin{figure}
\centering{}
\begin{tabular}{cc}
\includegraphics[scale=0.6]{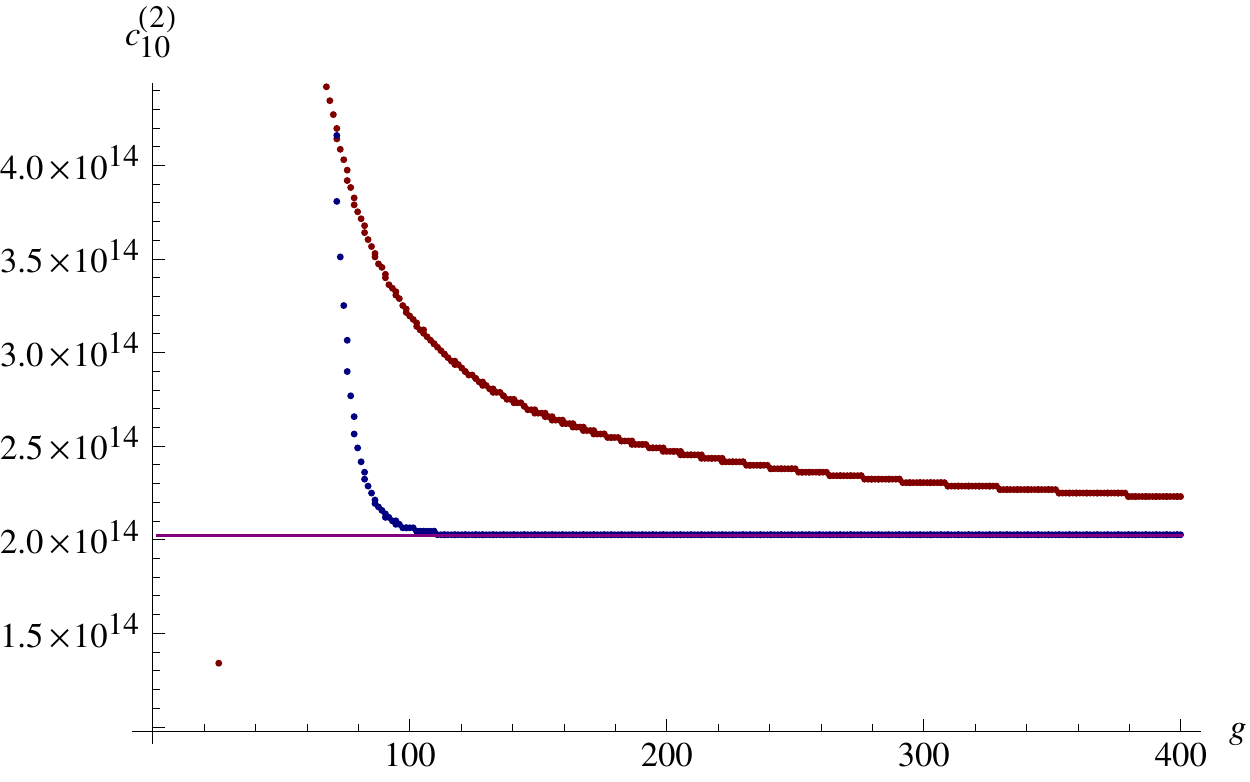} & \includegraphics[scale=0.6]{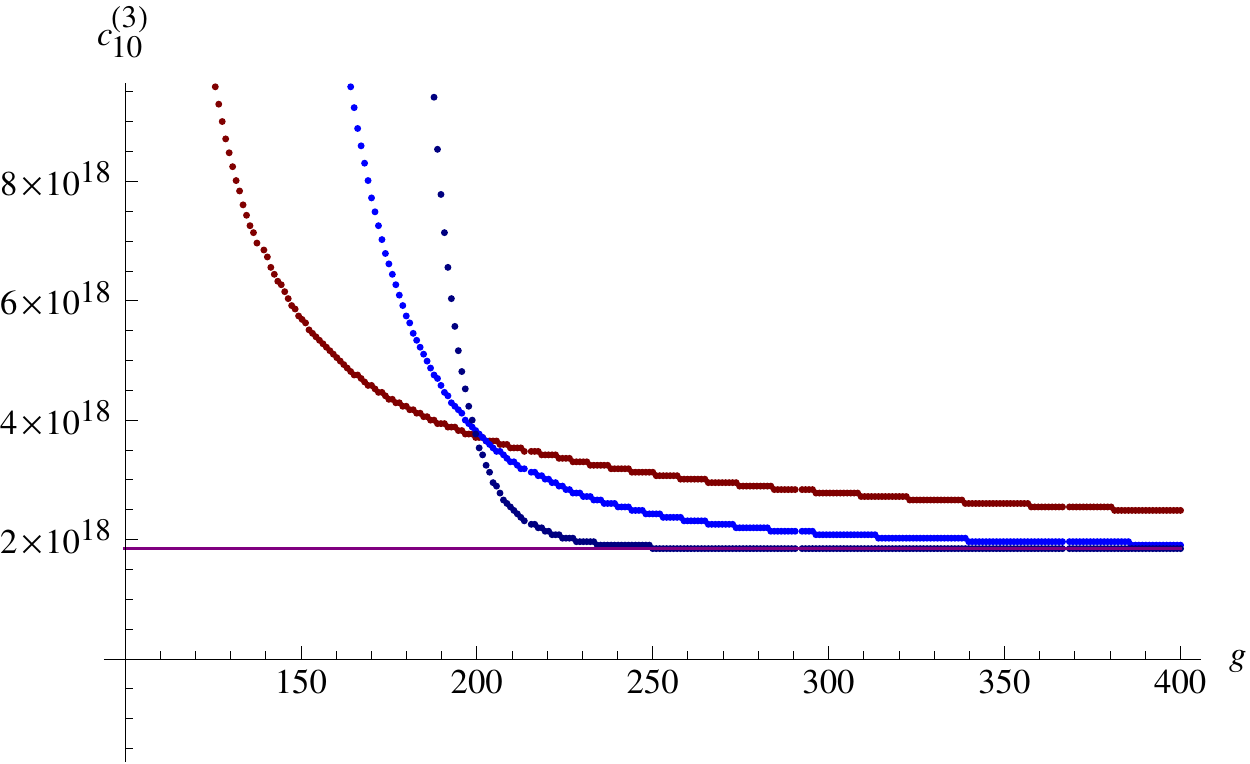}
\end{tabular}
\caption{Numerical analysis of the coefficients multiplying the ``monomial'' $g^{-10}$, for two different exponentially--suppressed orders: the coefficient $c_{10}^{(2)}$ associated with $\left(-8A\right)^{-g}$ (left plot; corresponding to the pole at $s_2 = -8A$) and the coefficient $c_{10}^{(3)}$ associated with $\left(-18A\right)^{-g}$ (right plot; corresponding to the pole at $s_3 = -18A$). For each plot we present the values of the corresponding ratio as explained in the main text (in red) and of two of its corresponding Richardson transforms (of orders 2 and 6, in blue).
\label{fig:N2-pert-2-and-3-inst}
}
\end{figure}

Having very accurately checked the polynomial large--order dependence of the coefficients $Z_{g}^{(0)}$, we can proceed to further check their large--order exponentially--suppressed behavior. This growth is expressed in the second and subsequent lines of \eqref{eq:N2-SYM-pertseries-coeff-1-2-inst}. In order to see these subleading, exponentially--suppressed terms, one first needs to remove the first line of \eqref{eq:N2-SYM-pertseries-coeff-1-2-inst} out of the original coefficients $Z_{g}^{(0)}$. Note that, in general, the (leading) terms to remove consist of an asymptotic series of their own, and resummation methods are required to handle this procedure (see, \textit{e.g.}, \citep{Aniceto:2011nu, Couso-Santamaria:2014iia} for examples within these contexts). It is quite interesting that this is not needed in the present case. In fact, the perturbative series around each fixed nonperturbative sector, $\left( - 2 A n^{2} \right)^{g}$, simply amounts to a \textit{rational} function---the one already found in \eqref{eq:N2-SYM-pertcoeff-prediction}. If we denote the first line in \eqref{eq:N2-SYM-pertseries-coeff-1-2-inst} by
\be
\phi_{1} (g) = \frac{\Gamma \left( g+\frac{9}{2} \right)}{\left( -2A \right)^{g+\frac{9}{2}}}\, \left\{ \frac{f_{0}^{(1)}}{6} - \frac{A\, f_{1}^{(1)}}{g+\frac{7}{2}} + \frac{2A^{2}\, f_{2}^{(1)}}{\left(g+\frac{5}{2}\right) \left(g+\frac{7}{2}\right)} - \frac{4A^{3}\, f_{3}^{(1)}}{3 \left(g+\frac{3}{2}\right) \left(g+\frac{5}{2}\right) \left(g+\frac{7}{2}\right)} \right\},
\ee
\noindent
then, in order to numerically check the (leading) exponentially suppressed behavior of the $Z_{g}^{(0)}$ coefficients, we just need to analyze the ratio 
\begin{eqnarray}
-\frac{8A}{g}\, \frac{Z_{g+1}^{(0)}-\phi_{1} (g+1)}{Z_{g}^{(0)}-\phi_{1} (g)} &\simeq& 1 + \frac{17}{2g} + \frac{56A\, f_{1}^{(2)}}{g^{2}\, f_{0}^{(2)}} - \nonumber \\
&&
- \frac{28 A \left( 3 f_{0}^{(2)} ( 32A\, f_{2}^{(2)} + 5\, f_{1}^{(2)} ) - 112 A ( f_{1}^{(2)} )^{2} \right)}{g^{3} (f_{0}^{(2)})^{2}} + \cdots = \nonumber \\
&\equiv&
\sum_{k=0}^{+\infty} c_{k}^{(2)}\, g^{-k} + \mathcal{O} \left( \left(-18A\right)^{-g} \right).
\end{eqnarray}
\noindent
The left plot of figure \ref{fig:N2-pert-2-and-3-inst} shows a numerical check for the coefficient $c_{10}^{(2)}$. The difference between the sixth--order Richardson transform for $g=400$, $\text{RT}_{6}\left(c_{20}^{(2)}\right)$, and the predicted value\footnote{The predicted value for this coefficient is
\begin{eqnarray*}
\left. c_{10}^{(2)} \right|_{\text{predicted}} &=& - 754325913600 \zeta(3)^{3} + 65551480350720 \zeta(3)^{2} + \frac{791986054060042631783}{1259712} + \\
&&
+ \frac{175}{27} \zeta(3) \left( 295092854784 \zeta(5) - 63615374797129 \right) - 117253785600 \zeta(7) - 25558423343600 \zeta(5).
\end{eqnarray*}
} is again very strongly validating our results,
\be
\left. \frac{\text{RT}_{6} \left( c_{10}^{(2)} \right) - c_{10}^{(2)}}{c_{10}^{(2)}} \right|_{g=400} \sim 1.089... \times 10^{-7}.
\ee

%%%%%%%%%%%%%%%%%%%%%%%%%%%%%%%%%%%%%%%%%%%%%%%%%%%%%%%%%%%%%%%%%
%%%%%%%%%%%%%%%%%%%%%%%%%%%%%%%%%%%%%%%%%%%%%%%%%%%%%%%%%%%%%%%%%

Taking this analysis one step further, to the ``third instanton'' level, let us define the second exponentially--suppressed contribution in \eqref{eq:N2-SYM-pertseries-coeff-1-2-inst} as
\be
\phi_{2} (g) = \frac{\Gamma \left( g+\frac{17}{2} \right)}{\left( -8A \right)^{g+\frac{17}{2}}}\, \left\{ \frac{f_{0}^{(2)}}{5040} - \frac{A\, f_{1}^{(2)}}{90 \left(g+\frac{15}{2}\right)} + \frac{4A^{2}\, f_{2}^{(2)}}{15 \left(g+\frac{13}{2}\right) \left(g+\frac{15}{2}\right)} + \cdots -  \frac{131072A^{7}\, f_{7}^{(2)}}{315 \prod_{k=0}^{76} \left(g+\frac{15-2k}{2}\right)} \right\}.
\ee
\noindent
If we now analyze the ratio
\be
- \frac{18A}{g}\, \frac{Z_{g+1}^{(0)} - \phi_{1} (g+1) - \phi_{2} (g+1)}{Z_{g}^{(0)} - \phi_{1} (g) - \phi_{2} (g)} \simeq \sum_{k=0}^{+\infty} c_{k}^{(3)}\, g^{-k} + \mathcal{O} \left( \left(-2^{5}A\right)^{-g} \right),
\ee
\noindent
we can check the terms which were previously exponentially suppressed by $\left(-3^{2}2A\right)^{-g}$. The right plot of figure \ref{fig:N2-pert-2-and-3-inst} shows a numerical check for the coefficient $c_{10}^{(3)}$, whose difference to the predicted value\footnote{The predicted value for this coefficient is 
\begin{eqnarray*}
\left. c_{10}^{(3)} \right|_{\text{predicted}} &=& - 6202653684955200 \zeta(3)^{3} + \frac{11564852996699371908 \zeta(3)^{2}}{25} - \frac{6869420324055823149 \zeta(5)}{25} + \\
&&
\hspace{-20pt}
+ \frac{1275732601044192825230000161}{250000000} + \frac{24057 \zeta(3) \left( 51103095372000000 \zeta(5) - 6307590904207688089 \right)}{50000} - \\
&&
\hspace{-20pt}
- 5957433657579096 \zeta(7) - 9391359631500 \zeta(9).
\end{eqnarray*}
} in its sixth Richardson transform at $g=400$ is
\be
\left. \frac{\text{RT}_{6} \left( c_{10}^{(3)} \right) - c_{10}^{(3)}}{c_{10}^{(3)}} \right|_{g=400} \sim 5.098... \times 10^{-4}.
\ee

In summary, the (numerical) checks we have performed above provide ample and very strong evidence that indeed the resurgent structure of the perturbative series (\textit{i.e.}, without any Nekrasov instanton corrections) associated with the zero topological--sector partition--function for $\CN=2$ superconformal Yang--Mills theory, explicitly given by the formulae \eqref{eq:N2-SYM-pert-partfunc} and \eqref{eq:N2-SYM-pertseries}, is correctly and fully described by the exact resurgence result \eqref{eq:N2-SYM-pertcoeff-prediction}.

The next natural step is to check if the same resurgent structure holds in the gauge instanton sectors of the theory. Let us consider the sector of zero topological charge
(the generalization to other sectors is straightforward). Recall \eqref{eq:N2-SYM-zero-top-partfunc} where the $k$--instanton sector was defined to be
\be
\mathcal{Z}_{k}^{\theta=0} \left(g_{\text{YM}}\right) = \frac{128\pi^{5/2}}{g_{\text{YM}}^{3}} \int_{-\infty}^{+\infty} \rmd a\, a^{2}\, \rme^{-\frac{2A}{g_{\text{YM}}^{2}}\, a^{2}}\, \prod_{n=1}^{+\infty} \frac{\left( 1+\frac{4a^{2}}{n^{2}} \right)^{2n}}{\left( 1+\frac{a^{2}}{n^{2}} \right)^{8n}}\, \left| Z_{k} (\rmi a) \right|^{2}.
\ee
\noindent
The Nekrasov instanton factors $Z_{k} (a)$ can be determined via some combinatoric calculations, as explained in \citep{Nekrasov:2002qd, Alday:2009aq}, and we have computed the first few expressions up to $k=8$ which may be found in table \ref{tab:Nekrasov-instanton-partfuncs} on page \pageref{tab:Nekrasov-instanton-partfuncs} (but note that each $k$--instanton sector in fact corresponds to having $k$ instantons and $k$ anti--instantons, as each sector is actually exponentially suppressed by $2A$ instead of $A$). These factors are rational functions which, when added into the integrand above, at most cancel the order of the zeroes coming from the one--loop factor \eqref{eq:2SYM-one-loop-factor}, but do not add any extra poles to the ones already present in the denominator of this one--loop factor (see  section \ref{sec:physical-int}).

We can determine the perturbative series around each instanton sector by expanding the integrand above for small $g_{\text{YM}}$ (after a suitable change of variables and in the same way as we did for the original perturbative series)
\be
\label{eq:N2-SYM-pertseries-k-inst}
\mathcal{Z}_{k}^{\theta=0} \left(g_{\text{YM}}\right) \simeq \sum_{g=0}^{+\infty} g_{\text{YM}}^{2g}\, Z_{g}^{(k)}.
\ee
\noindent
This series can be easily shown to be asymptotic, with its coefficients $Z_{g}^{(k)}$ growing factorially fast at large order $g$. Further, the corresponding Borel transform can be determined via the same change of variables as used before, $s = 2A\, a^{2}$ (recall that again $\lambda = g_{\text{YM}}^{2}$ is the effecting coupling), where we now find
\be
\mathcal{B} [\lambda^{3/2}\mathcal{Z}_{k}^{\theta=0} ] (s) = \frac{2 \sqrt{s}}{\sqrt{\pi}}\, \frac{\Big( G \left(1+2\rmi\sqrt{\frac{s}{2A}}\right) G \left(1-2\rmi\sqrt{\frac{s}{2A}}\right) \Big)^{2}}{\Big( G \left(1+\rmi\sqrt{\frac{s}{2A}}\right) G \left(1-\rmi\sqrt{\frac{s}{2A}}\right) \Big)^{8}}\, \left| Z_{k} \left(\rmi\sqrt{\frac{s}{2A}}\right) \right|^{2}.
\ee
\noindent
From \eqref{eq:N2-SYM-look-for-poles-Borel} we already know the order and the location of the poles and zeroes whose origin is in the one--loop factor. As we shall see now, the inclusion of the Nekrasov instanton factor will not change much. Table \ref{tab:Nekrasov-instanton-partfuncs} shows that the poles arising from the instanton factors are at the \textit{same} location as the zeroes arising from the one--loop factor, but that the order of these poles is \textit{lower} than the order of the corresponding zeroes in the one--loop factor. This property applies to more general $\CN=2$ theories and the underlying reason is explained in section \ref{sec:physical-int}. Thus, the location of singularities in the present Borel transforms is the \textit{same} as for the perturbative series \eqref{eq:N2-SYM-sing-pert-Borel}. However, of course, due to the instanton factors, the actual \textit{order} of these singularities will be different. Performing a Laurent expansion of the Borel transforms around each pole $s_{n}$ returns an analogous expansion to \eqref{eq:N2-SYM-borel-taylor-exp-each-pole},
\be
\label{eq:N2-SYM-borel-k-inst-Taylor-each-pole}
\left.\mathcal{B} [\lambda^{3/2}\mathcal{Z}_{k}^{\theta=0}] (s) \right|_{s_{n}} = \frac{1}{\left(s-s_{n}\right)^{4n}}\, \sum_{\ell=0}^{+\infty} \frac{f_{\ell}^{(n)[k]}}{\ell!} \left(s-s_{n}\right)^{\ell},
\ee
\noindent
with the earlier coefficients $f_{\ell}^{(n)}$ of \eqref{eq:N2-SYM-borel-taylor-exp-each-pole} now identified with $f_{\ell}^{(n)[0]}$. These new coefficients $f_{\ell}^{(n)[k]}$ will depend upon the rational functions associated with the instanton factors found in table \ref{tab:Nekrasov-instanton-partfuncs}, but everything else in the analysis which was previously done for the perturbative series essentially translates to the present case. In particular, for each instanton sector the Borel transform is still a meromorphic function. There is a single Stokes line at $\theta=\pi$ (the negative real axis), and the discontinuity across this Stokes line is now given by 
\be
\text{Disc}_{\pi}\, \mathcal{Z}_{k}^{\theta=0} (\lambda) = - \sum_{n=1}^{+\infty} \rme^{- \frac{s_{n}}{\lambda}}\, \sum_{\ell=0}^{4n-1} \frac{2\pi\rmi\, f_{\ell}^{(n)[k]}}{(-1)^{4n-\ell-1}\, \Gamma (4n-\ell)\, \ell!}\, \lambda^{\ell-4n-\frac{1}{2}}.
\ee
\noindent
The large--order behaviour of the coefficients $Z_{g}^{(k)}$ in the series \eqref{eq:N2-SYM-pertseries-k-inst} is finally given by
\be
Z_{g}^{(k)} \simeq \sum_{n=1}^{+\infty} \frac{\Gamma \left( g+4n+\frac{1}{2} \right)}{\left(- 2 A n^{2} \right)^{g+4n+\frac{1}{2}}}\, \sum_{\ell=0}^{4n-1} (-1)^{4n-\ell}\, \frac{\Gamma \left( g+4n+\frac{1}{2}-\ell \right)}{\Gamma \left( g+4n+\frac{1}{2} \right) \Gamma \left( 4n-\ell \right)\, \ell!}\, f_{\ell}^{(n)[k]} \left(- 2 A n^{2} \right)^{\ell}.
\ee

All the numerical checks we have previously carried through may now be performed as well, in order to confirm and support our results for the resurgent behavior of the perturbative expansion of the partition function, around any nonperturbative $k$--instanton sector (herein, at zero topological charge, but also this constraint may be eventually lifted). We have done many of these tests and once again found complete agreement. For completion, table \ref{tab:N2-SYM-Coeff-Taylor-inst-series} on page \pageref{tab:N2-SYM-Coeff-Taylor-inst-series} list some of the relevant coefficients $f_{\ell}^{(n)[k]}$, for the first few instanton sectors $k=1,2$ and poles $n=1,2$.

%%%%%%%%%%%%%%%%%%%%%%%%%%%%%%%%%%%%%%%%%%%%%%%%%%%%%%%%%%%%%%%%%
\subsection{$\mathcal{N}=2^{*}$ Supersymmetric Yang--Mills Theory}\label{sec:N2star}
%%%%%%%%%%%%%%%%%%%%%%%%%%%%%%%%%%%%%%%%%%%%%%%%%%%%%%%%%%%%%%%%%

Having analyzed in detail the superconformal case, we next address a theory without 
conformal symmetry, the $\mathcal{N}=2^{*}$ supersymmetric Yang--Mills theory with gauge group $\text{S}\text{U}(2)$, obtained by a $\mathcal{N}=2$ supersymmetry--preserving mass deformation of $\mathcal{N}=4$. This theory was extensively studied in the context of Seiberg--Witten theory (see, \textit{e.g.}, \cite{Seiberg:1994aj, Dorey:1996ez, D'Hoker:1997ha, Donagi:1995cf, Minahan:1997if, Billo:2014bja}). We shall again consider the weak--coupling perturbative series, and for all values of the mass parameter $M$ in this theory. This parameter is in fact one of its distinguishing features as it interpolates between the limiting case of $M\rightarrow0$, where one recovers the superconformal $\mathcal{N}=4$ SYM theory; and the limiting case of $M\rightarrow+\infty$, where the theory flows to pure $\mathcal{N}=2$ SYM theory, via a renormalization of the coupling. 

The partition function is still of the form \eqref{eq:Z-with-Z-1-loop-and-Z-inst}. Its localization was found in \citep{Pestun:2007rz}, and its dependence upon the mass parameter $M$ later analyzed in detail in \citep{Russo:2012kj}. At large $N$, both the partition function and the vev of the circular Wilson loop were investigated in \cite{Buchel:2013id, Russo:2013qaa, Russo:2013kea, Bobev:2013cja, Chen:2014vka, Marmiroli:2014ssa}, which found a precise match with holography at strong coupling, and the emergence of quantum phase transitions in the decompactification limit at specific values of the 't~Hooft coupling (holographic tests for $\mathcal{N}=2$ pure SYM have also been made \cite{Bigazzi:2013xia}).

Let us now fix the rank of the gauge group to $N=2$. The theory depends upon two parameters, $g_{\text{YM}}$ and $MR$, with $R$ representing the radius of the four--sphere. For most of the discussion we shall use units where $R=1$, restoring the $R$ dependence when convenient. In what follows we keep the parameter $M$ arbitrary but finite, and proceed to analyze the partition function at small gauge coupling. In this case, it is given by
\be
\mathcal{Z}_{2^{*}\text{SYM}}^{\text{S}\text{U}(2)} \left(g_{\text{YM}}\right) =\frac{128\pi^{5/2}}{g_{\text{YM}}^{3}} \int_{-\infty}^{+\infty} \rmd a\, a^{2}\, \rme^{-\frac{16\pi^{2}}{g_{\text{YM}}^{\text{2}}}\, a^{2}}\, Z_{\text{1-loop}} (a,M) \left| Z_{\text{inst}}^{\text{S}\text{U}(2)} \right|^{2},
\label{eq:2star-full-part-func}
\ee
\noindent
where the $Z_{\text{1-loop}}$ factor is given by 
\begin{eqnarray}
\label{eq:2star-one-loop-factor}
Z_{\text{1-loop}} \left(a,M\right) &=& \prod_{n=1}^{+\infty} \frac{\left( 1+\frac{4a^{2}}{n^{2}} \right)^{2n}}{\left( 1+\frac{\left(2a-M\right)^{2}}{n^{2}} \right)^{n} \left( 1+\frac{\left(2a+M\right)^{2}}{n^{2}} \right)^{n} \rme^{-\frac{2M^{2}}{n}}}  \\
&=& 
\frac{\rme^{\left(1+\gamma\right)2M^{2}} \Big( G \left(1+2\rmi a\right) G \left(1-2\rmi a\right) \Big)^{2}}{G \left(1+\rmi\left(2a-M\right)\right) G \left(1+\rmi\left(2a+M\right)\right) G \left(1-\rmi\left(2a-M\right)\right) G \left(1-\rmi\left(2a+M\right)\right)}, \nonumber
\end{eqnarray}
\noindent
with $\gamma$ the Euler constant. In this work we shall only study the zero--instanton sector of this theory, which corresponds to taking $Z_{\text{inst}}^{\text{S}\text{U}(2)} \rightarrow 1$. It would be interesting to proceed with an analysis of higher instanton sectors in future work. Within the present perturbative setting, it is straightforward to perform a perturbative expansion of the partition function \eqref{eq:2star-full-part-func} at small $g_{\text{YM}}$ coupling, which may be done numerically up to very large orders. Formally, we expect this expansion to be of the generic form
\be
\label{eq:2star-pert-series}
\mathcal{Z}_{0} \left(g_{\text{YM}}\right) \simeq \sum_{g=0}^{+\infty} g_{\text{YM}}^{2g}\, Z_{g}^{(0)},
\ee
\noindent
and one can easily check (numerically) that at large order the leading behavior of the above perturbative coefficients is of the type $Z_{g}^{(0)} \sim C^{-g}\, \Gamma(g+\beta)\, \cos \left( \theta_{a} (g+\beta) + \theta_{b} \right)\, \left\{ 1+\mathcal{O}(g^{-1}) \right\}$ for some $C$, $\beta$, $\theta_{k}$. This means that besides the, by now familiar, factorial growth, there is also an oscillatory component. In order to fully understand this behavior from an analytical point--of--view, all we have to do is to proceed in the exact same way as we did before when addressing the superconformal case. Changing variables as $2A\, a^{2}=s$, with $A=8\pi^{2}$, in \eqref{eq:2star-full-part-func} one quickly obtains a representative for the Borel transform of the perturbative series,
\be
\left(g_{\text{YM}}^{2}\right)^{3/2} \mathcal{Z}_{0} \left(g_{\text{YM}}\right) = \int_{0}^{+\infty} \rmd s\, \rme^{-\frac{s}{g_{\text{YM}}^{2}}}\, \mathcal{B} [ (g_{\text{YM}}^{2})^{3/2} \mathcal{Z}_{0}] (s),
\ee
\noindent
with
\be
\mathcal{B} [\lambda^{3/2} \mathcal{Z}_{0}] (s) \equiv \frac{2 \sqrt{s}}{\sqrt{\pi}}\, \rme^{\left(1+\gamma\right) 2M^{2}}\, \frac{\Big( G \left(1+2\rmi\sqrt{\frac{s}{2A}}\right) G \left(1-2\rmi\sqrt{\frac{s}{2A}}\right) \Big)^{2}}{\prod_{\pm} G \left( 1+\rmi\left(2\sqrt{\frac{s}{2A}}\pm M\right) \right) G \left( 1-\rmi\left(2\sqrt{\frac{s}{2A}}\pm M\right) \right)}.
\label{eq:2star-pertBorel}
\ee

The structure of singularities of the above Borel transform can be found by taking into account the infinite--product representation of the $Z_{\text{1-loop}}$ factor in \eqref{eq:2star-one-loop-factor}. In this way, one obtains that the singularities are all poles, located at 
\be
s_{n,\pm} \left(M\right) = \frac{1}{2} A \left( M \pm \rmi n \right)^{2} = \frac{1}{2} A \left( n^{2} + M^{2} \right) \rme^{\rmi\theta_{n,\pm} \left(M\right)}, \qquad n \in \mathbb{N},
\label{eq:2star-sing-pertBorel}
\ee
\noindent
with 
\be
\theta_{n,\pm} \left(M\right) =
\begin{cases}
\pm \arctan \left| \frac{2nM}{M^{2}-n^{2}}\right|, & n \le M \\
\pi \mp \arctan \left| \frac{2nM}{M^{2}-n^{2}} \right|, & n>M
\end{cases},
\label{eq:2star-sing-direc-pertBorel}
\ee
\noindent
and where each pole is of order $n$. The Borel transform is once again a meromorphic function, but the singularity structure in the complex Borel plane is yet different from our earlier examples: we now find a countable infinity of Stokes lines along the directions $\theta_{n,\pm}$, much like in the Chern--Simons example of section \ref{sec:CSLS}, but where each direction only meets a single singularity (a pole of order $n$) at a distance $\left|s_{n,\pm} \left(M\right)\right| = \frac{1}{2} A \left( M^{2}+n^{2} \right)$ from the origin. There are always two poles at the same distance to the origin\footnote{Naturally, the larger the value of $n$ the further away from the origin the corresponding poles are. Also note that if $M$ is an integer, two of the Stokes lines will be precisely at $\theta_{\pm} = \pm\frac{\pi}{2}$, \textit{i.e.},  in the imaginary axis.}, labeled by $n$ and at complex conjugate directions $\theta_{n,\pm}$.

%%%%%%%%%%%%%%%%%%%%%%%%%%%%%%%%%%%%%%%%%%%%%%%%%%%%%%%%%%%%%%%%%
\begin{figure}
\centering{}
\includegraphics[scale=0.7]{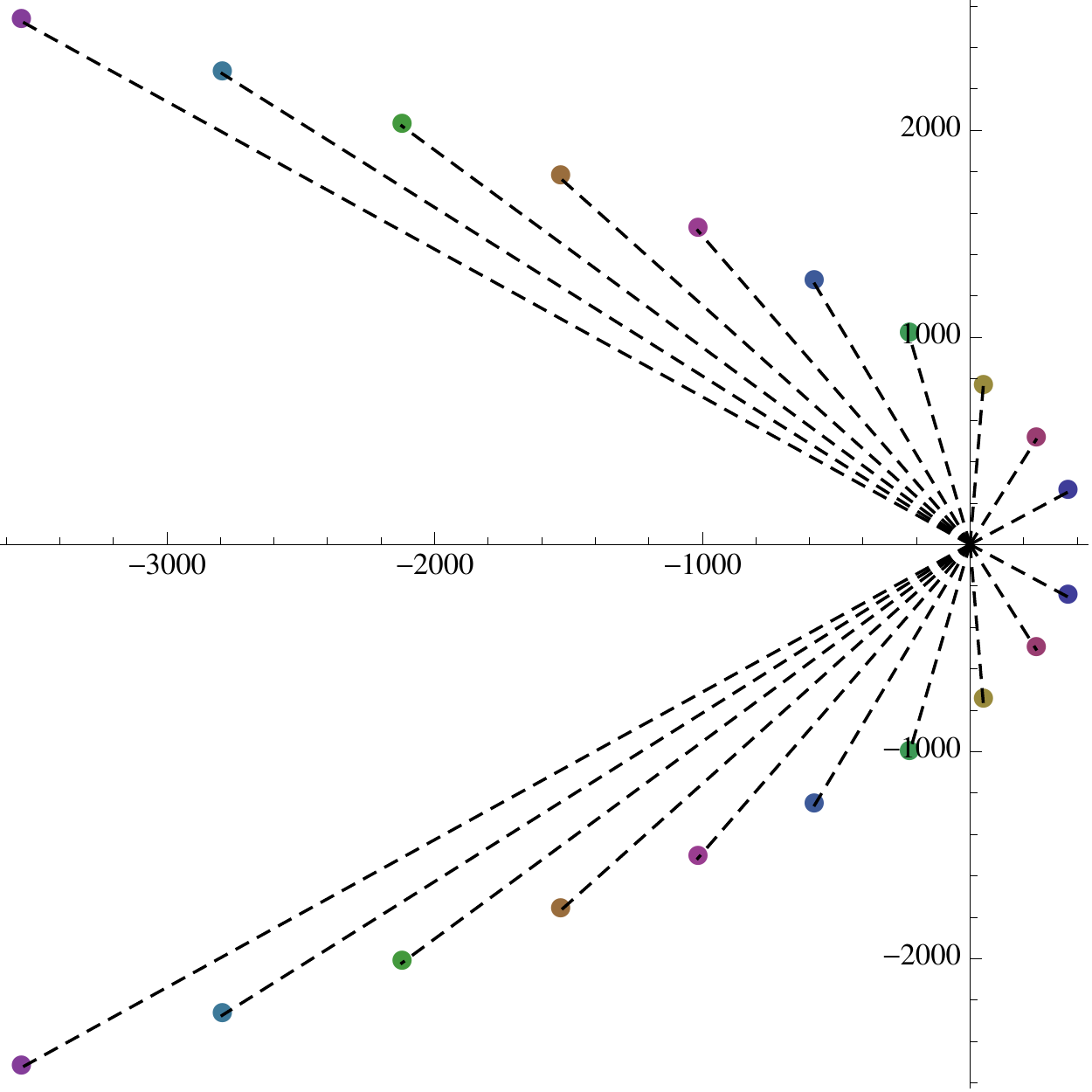}
\caption{Singularities in the complex Borel plane for the $\mathcal{N}=2^{*}$ supersymmetric Yang--Mills theory on $\mathbb{S}^4$. There are infinite countable Stokes lines, each with only one pole. In the plot, the mass parameter was set to $M=3.2$, thus only six of the Stokes lines lie in the first and fourth quadrants (for $n=1,2,3$).}
\label{fig:N2star-SYM-Borel}
\end{figure}
%%%%%%%%%%%%%%%%%%%%%%%%%%%%%%%%%%%%%%%%%%%%%%%%%%%%%%%%%%%%%%%%%

Figure \ref{fig:N2star-SYM-Borel} illustrates the singularity structure of this Borel transform. In particular, it makes the structural dependence of these singularities, with respect to the mass parameter, more evident. In the first and fourth quadrants (positive real part) we have a finite number of singularities (and thus, of Stokes lines). In fact, only poles which satisfy $n<M$ will be in these quadrants---all other poles with $n>M$ will be located in the third and fourth quadrants. 

The limit $M\rightarrow0$ makes all poles converge to the negative real axis, to their new locations $s_{n,\pm} \left(M=0\right) = - \frac{1}{2} A n^{2}$. They become double poles, which exactly cancel with the double zeros found in the numerator of the $Z_{\text{1-loop}}$ factor in \eqref{eq:2star-one-loop-factor}, ensuring that $Z_{\text{1-loop}} \rightarrow 1$ as expected for superconformal $\mathcal{N}=4$ SYM theory.

In the opposite limit, $M\rightarrow+\infty$,  all poles move to infinity\footnote{There is also an overall factor $\rme^{4 M^2 R^2 \log M R}$, which reproduces the expected UV divergence of the partition function originating from zero modes of the one--loop determinant.}. Which is the resulting theory depends on how the coupling $g_{\text{YM}}$ is scaled with $MR$. If the $M=\infty$ limit is taken with $g_{\text{YM}}\to 0$ at the same time, keeping $M R\, \rme^{-2\pi^2/g_{\text{YM}}^2}$ fixed, then the theory flows to  pure $\mathcal{N}=2$ SYM theory \cite{Seiberg:1994aj}. As will be discussed in section \ref{sec:physical-int}, the corresponding perturbative series is convergent, in consistency with the fact that poles move to infinity. If, on the other hand, the $MR\to \infty $ limit is taken with fixed $g_{\text{YM}}$, then one just recovers $\mathcal{N}=2^{*}$ theory on flat spacetime. In this case, the weak--coupling expansion has a finite radius of convergence, again in consistency with the fact that the corresponding Borel transform has no singularities.

We now turn to determining the discontinuities across the existing Stokes lines for fixed finite $M$. In order to determine the Stokes discontinuities, we first need to understand the behavior of the Borel transform around each pole. Akin to before, we find
\be
\left.\mathcal{B} [\lambda^{3/2}\mathcal{Z}_{0}] (s) \right|_{s_{n,\pm}} = \frac{1}{\left(s-s_{n,\pm}\right)^{n}}\, \sum_{\ell=0}^{+\infty}\frac{f_{\ell,\pm}^{(n)}}{\ell!} \left(s-s_{n,\pm}\right)^{\ell}.
\label{eq:2star-borel-taylor-exp-each-pole}
\ee
\noindent
As in the superconformal case, the Taylor series around $s_{n,\pm}$, with coefficients $f_{\ell,\pm}^{(n)}$, is obtained by expanding the product of $\mathcal{B} [\lambda^{3/2}\mathcal{Z}_{0}] (s)$ with $\left(s-s_{n,\pm}\right)^{n}$, and thus each pole will have a different series. Translating this singular structure to the usual ``simple representative'', the discontinuities of the asymptotic (perturbative) partition function along each singular direction, $\theta_{n,\pm}$, may be read off immediately. We shall just follow the same procedure as before. For each Stokes line there is now a single singularity, in which case,
\begin{eqnarray}
\text{Disc}_{\theta_{n,\pm}}\, \mathcal{Z}_{0} (\lambda) &=& -\rme^{-\frac{s_{n,\pm}}{\lambda}}\, \Delta_{s_{n,\pm}} \mathcal{Z}_{0} (\lambda) \\
&=& 
- \rme^{-\frac{s_{n,\pm}}{\lambda}}\, \lambda^{-n-\frac{1}{2}}\, \Delta_{s_{n,\pm}} \left( \lambda^{n+\frac{1}{2}} \mathcal{Z}_{0} (\lambda) \right),
\label{eq:2star-disc-original}
\end{eqnarray}
\noindent
where the alien derivative $\Delta_{s_{n,\pm}} \left( \lambda^{n+1/2} \mathcal{Z}_{0}^{\theta=0} (\lambda) \right)$ can be read off directly from the behavior of the Borel transform (in the adequate representative) around the very same singularity,
\be
\left. \mathcal{B} [\lambda^{n+1/2} \mathcal{Z}_{0}] (s) \right|_{s_{n,\pm}} = \frac{f_{0,\pm}^{(n)}}{(-1)^{n-1} \left(n-1\right)!}\, \frac{1}{s-s_{n,\pm}} + \frac{\mathcal{B} [\psi_{n,\pm}] \left(s-s_{n,\pm}\right)}{2\pi\rmi}\, \log \left(s-s_{n,\pm}\right) + \text{regular}.
\label{eq:2star-Borel-transf-each-pole-logs}
\ee
\noindent
Here $\mathcal{B} [\psi_{n,\pm}]$ is the Borel transform of $\psi_{n,\pm}$, given by
\be
\psi_{n,\pm} (\lambda) = \sum_{\ell=1}^{n-1} \frac{2\pi\rmi\, f_{\ell,\pm}^{(n)}}{(-1)^{n-\ell-1}\, \Gamma (n-\ell)\, \ell!}\, \lambda^{\ell}.
\label{eq:2star-Log-term-Borel-transf-each-pole}
\ee
\noindent
This function is obviously not resurgent; again it is just a polynomial. This basically implies that the original Borel transform is meromorphic. It then follows
\be
\Delta_{s_{n,\pm}} \left( \lambda^{n+\frac{1}{2}} \mathcal{Z}_{0} (\lambda) \right) = \psi_{n,\pm} (\lambda),
\ee
\noindent
and we finally determined the discontinuities across the Stokes lines as
\be
\text{Disc}_{\theta_{n,\pm}}\, \mathcal{Z}_{0} (\lambda) = - \rme^{-\frac{s_{n,\pm}}{\lambda}}\, \sum_{\ell=0}^{n-1} \frac{2\pi\rmi\, f_{\ell,\pm}^{(n)}}{(-1)^{n-\ell-1}\, \Gamma (n-\ell)\, \ell!}\, \lambda^{\ell-n-\frac{1}{2}}.
\label{eq:2star-Pert-Partfunc-Disc-Stokes}
\ee

%%%%%%%%%%%%%%%%%%%%%%%%%%%%%%%%%%%%%%%%%%%%%%%%%%%%%%%%%%%%%%%%%
\begin{figure}[t]
\centering{}
\begin{tabular}{cc}
\includegraphics[scale=0.6]{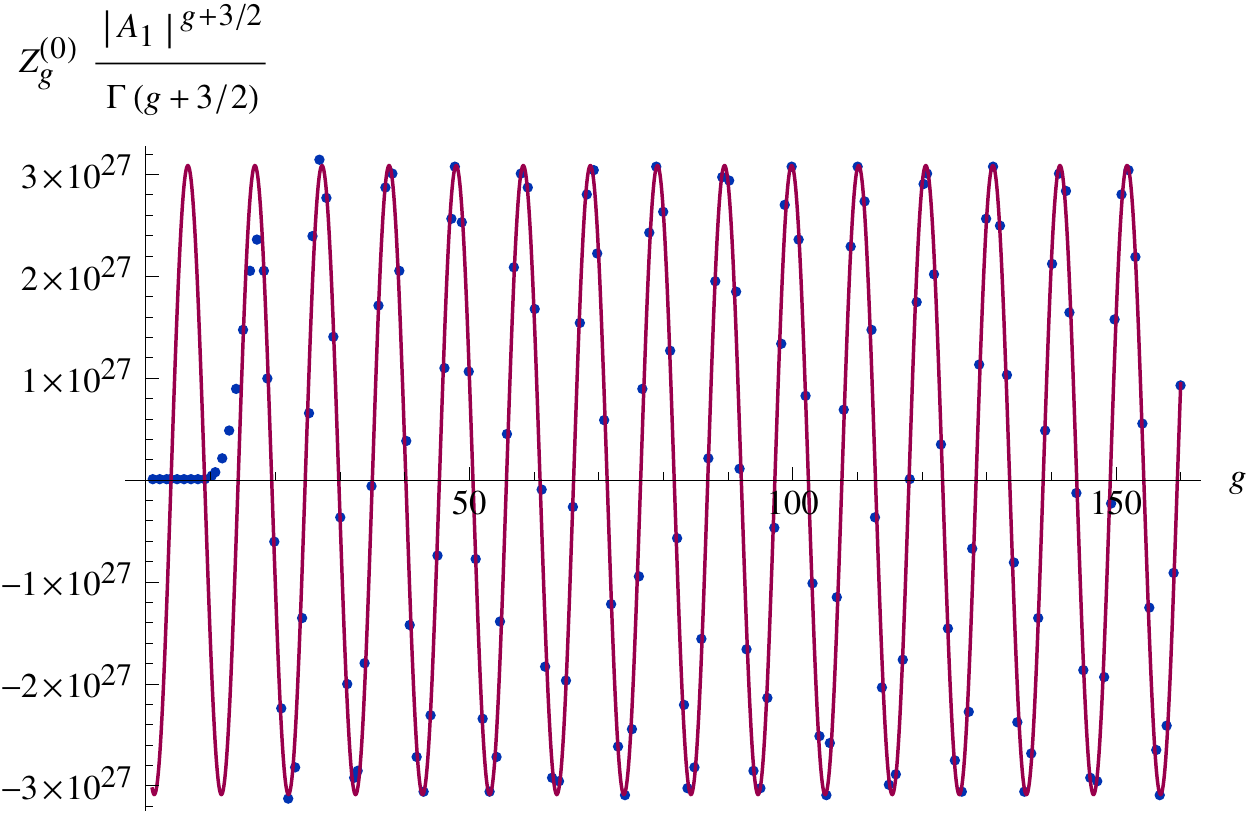} & \includegraphics[scale=0.6]{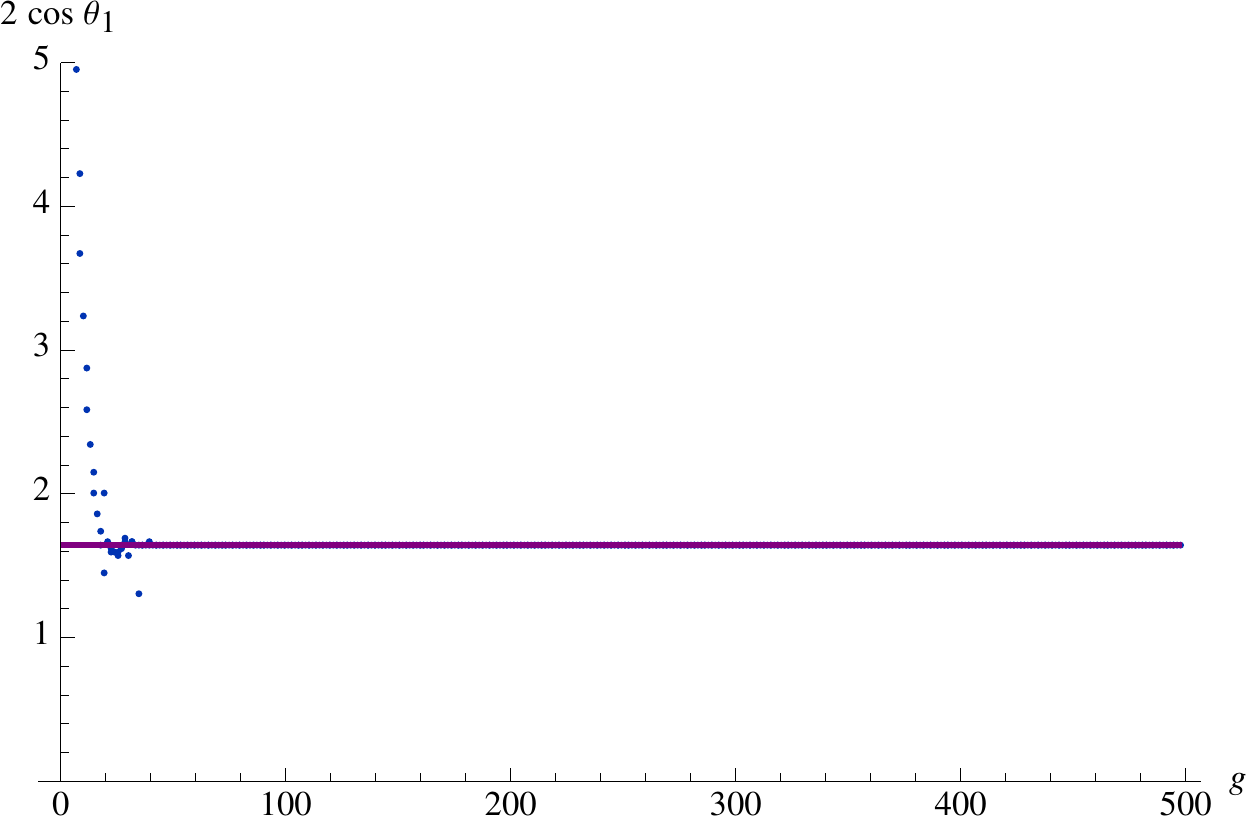}
\end{tabular}
\caption{Numerical analysis of the leading large--order behavior of the coefficients  $Z_g^{(0)}$, for values of the mass parameter $M=3.2$, and $A_1 = \frac{1}{2} A \left( 1+M^{2} \right)$. On the left, both the numerical value of these coefficients (the dots) and their predicted analytical behavior (the solid line) are shown. On the right, we plot the convergence towards $2 \cos \theta_{1,+}$ of a particular combination of the coefficients, the one given in \eqref{eq:2star-pertseries-conv-first-angle}, where $\theta_{1,+}$ is the angle associated with the first instanton action $s_1$. The agreement is excellent.
\label{fig:N2star-pert-1inst}}
\end{figure}
%%%%%%%%%%%%%%%%%%%%%%%%%%%%%%%%%%%%%%%%%%%%%%%%%%%%%%%%%%%%%%%%%

With all the information gathered above, we can now finish by presenting the expression describing the resurgent behavior of the asymptotic series $\mathcal{Z}_{0} (\lambda)$ in \eqref{eq:2star-pert-series}, alongside with  large--order numerical checks. First, from \eqref{eq:2star-Pert-Partfunc-Disc-Stokes} and \eqref{eq:2star-pert-series} in the usual dispersion relation, we find
\begin{eqnarray}
Z_{g}^{(0)} &\simeq& \sum_{n=1}^{+\infty}\, \sum_{\ell=0}^{n-1} \frac{1}{\Gamma (n-\ell)\, \ell!}\, \frac{\Gamma \left(g+n-\ell+\frac{1}{2}\right)}{\left|s_n\right|^{g+n-\ell+\frac{1}{2}}}\, \sum_{\pm} f_{\ell,\pm}^{(n)}\, \rme^{\rmi \theta_{n,\pm} \left( \ell-g-n-\frac{1}{2} \right)}.
\label{eq:2star-pertcoeff-prediction}
\end{eqnarray}
\noindent
In the above expression we use the notation $\left|s_n\right| \equiv \left|s_{n,\pm}\right|$, which is the same for both signs. The leading factorial growth, and next--to--leading exponential growth are very distinguishable when $g\gg1$,
\begin{eqnarray}
Z_{g}^{(0)} &\simeq& \frac{\Gamma \left( g+\frac{3}{2} \right)}{\left(\frac{1}{2} A \left(1+M^{2}\right)\right)^{g+\frac{3}{2}}} \left( f_{0,+}^{(1)}\, \rme^{-\rmi\theta_{1,+} \left( g+\frac{3}{2} \right)} + f_{0,-}^{(1)}\, \rme^{-\rmi\theta_{1,-} \left( g+\frac{3}{2} \right)} \right) + \mathcal{O} \left( \left(s_{2,\pm}\right)^{-g-\frac{5}{2}} \right),
\label{eq:2star-pertseries-coeff-1-inst}
\end{eqnarray}
\noindent
where the coefficients $f_{0,\pm}^{(1)} = \left|f_{0}^{(1)}\right| \rme^{\pm\rmi\theta_{f_{0}}}$ may be found in appendix \ref{sec:App-N2-data}, and the angles $\theta_{1,\pm}$ are given in \eqref{eq:2star-sing-direc-pertBorel}. It is the existence of complex ``instanton actions'' which gives rise to the oscillatory behavior of the perturbative coefficients, now clearly seen in \eqref{eq:2star-pertseries-coeff-1-inst} above. In particular, from a numerical standpoint, for large order $g$ the leading behavior of the coefficients $Z_{g}^{(0)}$ is no longer a series in $1/g$, and one cannot use the method of Richardson transforms. Nevertheless, one can still perform a numerical check of the predicted behavior \eqref{eq:2star-pertseries-coeff-1-inst}. The left plot of figure \ref{fig:N2star-pert-1inst} shows the numerical coefficients $Z_{g}^{(0)}$ alongside their predicted behavior \eqref{eq:2star-pertseries-coeff-1-inst}, for values of $2A = (4\pi)^{2}$ and $M=3.2$. The right plot of figure \ref{fig:N2star-pert-1inst} shows the large--order convergence towards the value of $2\cos\theta_{1,+}$ via a numerical determination of the combination of ratios
\be
\frac{Z_{g+1}^{(0)}}{Z_{g}^{(0)}}\, \frac{\frac{1}{2} A \left(1+M^{2}\right)}{g+\frac{3}{2}} + \frac{Z_{g-1}^{(0)}}{Z_{g}^{(0)}}\, \frac{g+\frac{1}{2}}{\frac{1}{2} A \left(1+M^{2}\right)} \simeq 2 \cos \theta_{1,+}.
\label{eq:2star-pertseries-conv-first-angle}
\ee
\noindent
In both cases the coincidence between numerical and analytical results is excellent.

%%%%%%%%%%%%%%%%%%%%%%%%%%%%%%%%%%%%%%%%%%%%%%%%%%%%%%%%%%%%%%%%%
\begin{figure}
\centering{}
\begin{tabular}{cc}
\includegraphics[scale=0.6]{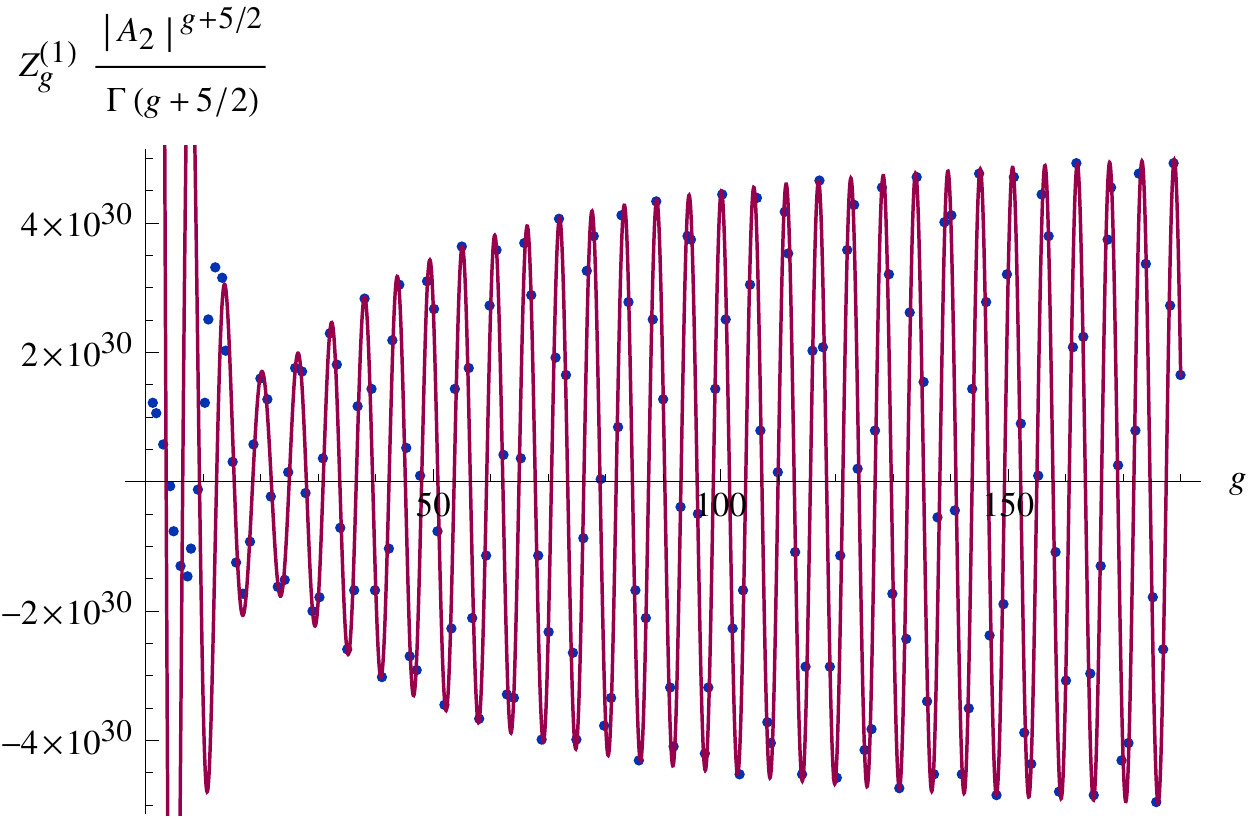} & \includegraphics[scale=0.6]{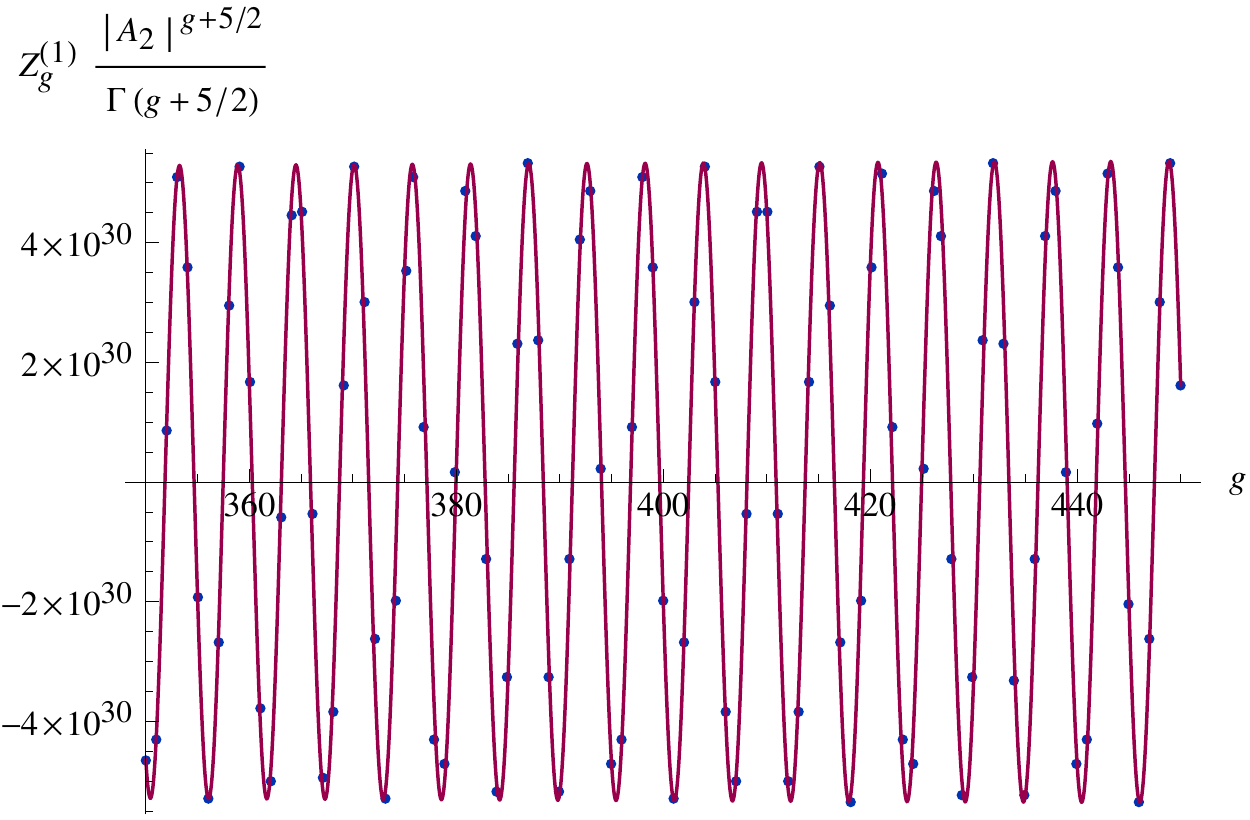}
\end{tabular}
\caption{Numerical analysis of the exponentially subleading large--order behavior of the  coefficients $Z_g^{(0)}$, for a value of the mass parameter $M=3.2$ and $A_2 = \frac{1}{2} A \left(4+M^{2}\right)$, corresponding to the \textit{leading} behavior of the redefined coefficients $Z_g^{\{1\}}$. Left and right plots show both the numerical value of these coefficients (the dots) as well as their analytically predicted large--order behavior (the solid line), for small and large $g$, respectively. The agreement is clearly excellent at large order.
\label{fig:N2star-pert-2inst}}
\end{figure}
%%%%%%%%%%%%%%%%%%%%%%%%%%%%%%%%%%%%%%%%%%%%%%%%%%%%%%%%%%%%%%%%%

One can go one step further in the numerical checks of the predicted resurgent behavior, \eqref{eq:2star-pertcoeff-prediction}. Defining
\be
Z_{g}^{\{1\}} \equiv Z_{g}^{(0)} - \frac{\Gamma \left(g+\frac{3}{2}\right)}{\left(\frac{1}{2} A \left(1+M^{2}\right)\right)^{g+\frac{3}{2}}} \left| f_{0}^{(1)} \right| \left( \rme^{-\rmi\theta_{1,+} \left( g+\frac{3}{2} \right) + \rmi\theta_{f_{0}}} + \rme^{-\rmi\theta_{1,-} \left( g+\frac{3}{2} \right) - \rmi \theta_{f_{0}}} \right),
\ee
\noindent
it is simple to verify that the large--order behaviour of $Z_{g}^{\{1\}}$ is the exponentially suppressed ``two--instanton'' sector, given by
\be
Z_{g}^{\{1\}} \simeq \frac{\Gamma \left(g+\frac{5}{2}\right)}{\left|\frac{A}{2} \left(4+M^{2}\right)\right|^{g+\frac{5}{2}}} \left|f_{0}^{(2)}\right| \sum_{\pm} \rme^{-\rmi\theta_{2,\pm} \left(g+\frac{5}{2}\right) \pm \rmi\theta_{f_{0,2}}} + \frac{\Gamma \left(g+\frac{3}{2}\right)}{\left|\frac{A}{2}\left(4+M^{2}\right)\right|^{g+\frac{3}{2}}} \left|f_{1}^{(2)}\right| \sum_{\pm} \rme^{-\rmi\theta_{2.\pm} \left(g+\frac{3}{2}\right) \pm \rmi\theta_{f_{1,2}}}.
\label{eq:2star-pertseries-coeff-2-inst}
\ee
\noindent
The coefficients $f_{\ell,\pm}^{(2)}$, with $\ell=0,1$, may be found in appendix \ref{sec:App-N2-data}. Figure \ref{fig:N2star-pert-2inst} shows the numerical convergence of the coefficients $Z_{g}^{\{1\}}$ to their analytically predicted behavior, given in \eqref{eq:2star-pertseries-coeff-2-inst}, for a chosen value of the mass parameter of $M=3.2$. As usual, excellent agreement is evident.

Summing things up, the numerical checks we have performed above provide very clean and ample evidence that indeed the resurgent structure of the perturbative series associated with the partition function of $\CN=2^*$ SYM theory, explicitly given by formulae \eqref{eq:2star-full-part-func} and \eqref{eq:2star-pert-series}, is fully and correctly described by the exact resurgence result we derived, \eqref{eq:2star-pertcoeff-prediction}.

%%%%%%%%%%%%%%%%%%%%%%%%%%%%%%%%%%%%%%%%%%%%%%%%%%%%%%%%%%%%%%%%%
%%%%%%%%%%%%%%%%%%%%%%%%%%%%%%%%%%%%%%%%%%%%%%%%%%%%%%%%%%%%%%%%%
\section{Nonperturbative Effects and Borel Singularities}\label{sec:physical-int}
%%%%%%%%%%%%%%%%%%%%%%%%%%%%%%%%%%%%%%%%%%%%%%%%%%%%%%%%%%%%%%%%%
%%%%%%%%%%%%%%%%%%%%%%%%%%%%%%%%%%%%%%%%%%%%%%%%%%%%%%%%%%%%%%%%%

In conventional quantum field theories, the large--order behavior of perturbation theory is typically associated with classical solutions involving large fields. The underlying mechanism was first understood by Lipatov \citep{Lipatov:1976ny}, who studied simple examples where the asymptotic behavior of perturbation theory is controlled by classical solutions (but see also \citep{LeGuillou:1990nq}). In turn, as discussed throughout this paper, these classical solutions imply singularities in the Borel transform. For Yang--Mills gauge theory in four dimensions, the large--order behavior was first argued to be controlled by certain complex instantons \citep{Lipatov:1978en}. The current understanding is that the precise picture must be more intricate: while instanton solutions do lead to singularities in the Borel transform, and thus affect the large--order behavior, other (more dominant) singularities will be present due to renormalons, see, \textit{e.g.}, \citep{Beneke:1998ui}. These latter singularities are associated with terms in the operator product expansion (OPE) and will contribute to the $n!$ behavior of perturbation theory, this time around arising not from a growing number $\sim n!$ of Feynman diagrams, but rather from certain $n$--loop Feynman diagrams individually contributing as $n!$.

In the context of our present work, supersymmetric physical observables are given in terms of analytic formulae, which in particular encapsulate their exact coupling dependence. This allows for a complete analytic description of the singularities in the complex Borel plane, as described in detail in previous sections, providing for a clean set--up to clarify the interplay between large--order behavior and semiclassical field configurations.

In many examples (\textit{e.g.}, \citep{Aniceto:2011nu}) instanton sectors are responsible (at least in part; see \citep{Cherman:2013yfa}) for the asymptotic behavior of the perturbative series; they resurge in the perturbative series. For the examples considered in this work, one might therefore expect that the resurgence properties of the perturbative sector could be traced back to these gauge--theory instanton sectors  (given by the Nekrasov equivariant partition function), and that this would be seen at the level of the Borel transform as having its singularities located at $s_{n}=n\, A$, with $A$ representing the instanton action. In contrast, however, in previous sections we have seen that resurgence does not mix these different gauge--theory instanton sectors.

Consider, for example,  the case of $\CN=2$ superconformal $\text{S}\text{U}(2)$ Yang--Mills theory. We found that all Borel singularities lied on the negative real axis, at $s_{n}=- (4\pi)^{2}\, n^{2}$, for integer $n\ge1$. On the other hand, the gauge theory instanton action is positive and real, $A=8\pi^{2}$, and consequently in each topological sector the $k$--instanton sector is exponentially suppressed as usual, $\exp \left( -\frac{2A}{g_{\text{YM}}^{2}}\, |k| \right)$. One would thus naively expect to find Borel singularities related to these sectors at multiples of twice the gauge instanton action, \textit{i.e.}, located at $s_{n}^{*}=n\, 2A$. However, this is \textit{not} what we have found in our resurgent analysis. The Borel singularities are not located at positive multiples of the instanton action (where they would lie on the positive real axis), but instead at negative multiples of twice this action (on the negative real axis). In particular, we found in \eqref{eq:N2-SYM-Pert-Partfunc-Disc-pi} that the discontinuity of the perturbative partition function across the Stokes line has the form
\be
\label{parz}
\text{Disc}_{\pi}\, \mathcal{Z}_0 = \sum_{n=1}^{+\infty} \rme^{\frac{16\pi^2n^2}{\lambda}}\, \lambda^{-4n-\frac{1}{2}}\, p_n(\lambda), \qquad \lambda \equiv g_{\text{YM}}^2,
\ee
\noindent
where $p_n$ is a polynomial of degree $4n-1$ in $g_{\rm YM}^2$. Furthermore, we see that not all multiples of the instanton action appear, as one might expect from a standard theory with multi--instantons. Instead, only the multiples corresponding to an integer squared, $s_{n} = -n^{2}\, 2A$, appear. One then concludes that the resurgence properties of the perturbative sector (\textit{i.e.}, its Borel singularity structure) are not governed by the gauge theory instanton sectors, but by some other physical effect. What these effects might be is what we shall discuss in the following.

%%%%%%%%%%%%%%%%%%%%%%%%%%%%%%%%%%%%%%%%%%%%%%%%%%%%%%%%%%%%%%%%%
\subsection{Instantons in Gauge Theories}
%%%%%%%%%%%%%%%%%%%%%%%%%%%%%%%%%%%%%%%%%%%%%%%%%%%%%%%%%%%%%%%%%

In the computation of the partition function by supersymmetric localization, instanton solutions arise by admitting singular field configurations where the gauge field strength is non--vanishing only at the North or South poles of the four--sphere. Instanton solutions with $F^+=0$ are localized at the North pole, whereas anti--instanton solutions with $F^-=0$ are localized at the South pole. In turn, this gives rise to the equivariant instanton partition function of \citep{Nekrasov:2002qd, Nekrasov:2003rj},
\be
Z_{\text{inst}} (\rmi a) = \sum_{k=0}^{+\infty} \rme^{2\pi\rmi k\tau}\, Z_{k} (\rmi a) \equiv \sum_{k=0}^{+\infty} q^k\, Z_{k} (\rmi a).
\ee
\noindent
An interesting question concerns the convergence properties of this series. In the abelian $\text{U}(1)$ case, for $\CN=2$ SYM theory, it was already shown in \cite{Pestun:2007rz} that the instanton partition function is a convergent series (in fact yielding an entire function). On what concerns the nonabelian case, we have studied this problem for the $\text{S}\text{U}(2)$ $\CN=2$ SCF theory. The first few multi--instanton contributions $Z_{k} (\rmi a)$ are listed in table \ref{tab:Nekrasov-instanton-partfuncs} in appendix \ref{sec:App-N2-data}, and for fixed values of $a$ higher instantons can be generated numerically (we have generated terms up to $k=20$). Analyzing ratios of consecutive coefficients in the above series, we find that the series is \textit{not} asymptotic; in fact it has non--zero radius of convergence (of order $q\sim 1$). Another example, which can be carried out analytically \citep{Pestun:2007rz}, is that of $\CN=2^*$ $\text{U}(N)$ theory with unphysical mass parameter $M=\rmi$. In this case one finds
\be
Z_{\text{inst}}^{\CN=2^*} (\rmi a, M=\rmi) = \prod_{n=1}^{+\infty} \frac{1}{(1-q^n)^N},
\ee
\noindent
which also has finite non--zero radius of convergence $|q|=1$.

Returning to the resurgence properties of perturbation series in $\CN=2$ supersymmetric Yang--Mills theories, the fact that gauge--theory instantons do not imply $n!$ large--order behavior is more evident in the case of $\CN=2$ pure $\text{S}\text{U}(2)$ SYM theory, \textit{i.e.}, with no matter multiplets. It is instructive to describe this case in more detail. The partition function for $\CN=2$ $\text{S}\text{U}(2)$ pure SYM is given by
\be
Z^{\text{S}\text{U}(2)}_{\text{Pure}} = \frac{128\pi^{5/2}}{g_{\text{YM}}^3} \int_{-\infty}^{+\infty} \rmd a\, a^2\, \rme^{-\frac{16\pi^2}{g^2}\, a^2}\, H^2(2a) \left|Z_{\text{inst}}(a) \right|^2,
\ee
\noindent
with
\be
H^2(2a) = \prod_{n=1}^{+\infty} \left(1+\frac{4a^2}{n^2}\right)^{2n} \qquad \text{and} \qquad Z_{\text{inst}} = \sum_{k=0}^{+\infty} q^k\, Z_k (a).
\ee
\noindent
The instanton partition function is understood with equivariant parameters $\epsilon_1=\epsilon_2=1$ (recall that we are using units where the radius $R$ of the four--sphere is $R=1$). It can be computed explicitly term by term. In particular \citep{Pestun:2007rz},
\bea
Z_1 &=& \frac{1}{2(1+ a^2)}, \\
Z_2 &=& \frac{8a^2+33}{\left(4+ 4a^2\right) \left(9+4a^2\right)^2}.
\eea
\noindent
Note that the potential pole at $a^2=-1$  cancels against one of the zeros of $H^2(2a)$. Similarly for the potential pole at  $a^2=-9/4$. This pattern holds to all orders, and it is ensured by the structure of the equivariant instanton partition function. Indeed, as observed in \citep{Pestun:2007rz}, the function $Z_{\text{inst}} (a;\epsilon_1, \epsilon_2)$ has simple poles at
\be
a_i - a_j = \rmi \left( n_1 \epsilon_1 + n_2 \epsilon_2 \right), \qquad n_1, n_2 = 1,  2, 3, \ldots,
\ee
\noindent
which, for $\text{S}\text{U}(2)$ gauge theories on the sphere, implies poles at
\be
2a = \pm \rmi n, \qquad n \equiv n_1 + n_2.
\ee
\noindent
There are exactly $n$ of such poles for $n_1=1,2,...,n-1$, leading to a pole of order $n$ in $Z_{\text{inst}}(a)$ and a pole of order $2n$ in $\left| Z_{\text{inst}} (a) \right|^2$. Since the zero of the one--loop factor at $2a = \pm \rmi n$ is of order $2n$, for pure $\text{S}\text{U}(2)$ SYM, the  poles from the instanton factor are completely canceled by the zeroes of the one--loop factor.

Consequently, unlike the cases of the $\CN=2$ theories with matter supermultiplets, which we have studied at length earlier, now the Borel transform has no poles. This implies that perturbation theory does not have an asymptotic $\sim n!$ large--order behavior, \textit{i.e.}, that it is not an asymptotic series. In fact we have numerically found that the perturbation series for $Z^{\text{S}\text{U}(2)}$ has a finite non--zero radius of convergence, of order $|g_{\text{YM}}| < g_{\text{YM}}^0$ with\footnote{The same considerations apply for the vev of the $\frac{1}{2}$BPS circular Wilson loop operator, obtained by insertion of $\rme^{2\pi a}$ in the integrand. It is easy to see that the convergence properties of the perturbative series are the same as for the partition function.} $g_{\text{YM}}^0 \sim  2.8$.

Another example exhibiting similar features is massless SQCD, \textit{i.e.}, $\CN=2$ $\text{S}\text{U}(2)$ SYM with one fundamental and one antifundamental matter multiplet. Akin to $\CN=2$ pure SYM, this theory is also asymptotically free. The perturbative expansion of the partition function can be obtained from the expression for the partition function of the superconformal theory by simply replacing $H^8 (a)$ in the denominator by $H^4 (a)$. For the perturbative part (the $k=0$ instanton sector), we find
\be
\left. Z_{\text{SQCD}}^{\text{S}\text{U}(2)} \right|_{k=0} = \frac{2}{\sqrt{\pi\alpha}} \int_0^{+\infty} \rmd t\, \rme^{-t}\, \CB (\alpha t), \qquad \alpha = \frac{g_{\text{YM}}^2}{16\pi^2},
\ee
\noindent
where the Borel transform is now
\be
\CB (s) = \sqrt{s}\, \prod_{n=1}^{+\infty} \frac{\left( 1 + \frac{4s}{n^2} \right)^{2n}}{\left( 1 + \frac{s}{n^2} \right)^{4n}} = \sqrt{s}\, \prod_{n=1}^{+\infty} \left( 1+ \frac{4s}{\left(2n-1\right)^2} \right)^{4n-2}.
\ee
\noindent
Compared to the case of the $\CN=2$ SCF theory, the change of power in the denominator has produced a dramatic effect: all the poles have now canceled against zeroes of the numerator. What remains has no singularities in the complex plane. As a result, perturbation theory again has a finite non--zero radius of convergence.

In conclusion, despite these theories containing instanton solutions there is no associated asymptotic growth $\sim n!$ in the perturbative coefficients. More generally, it is easy to see that in any $\CN=2$ SYM theory with gauge group $\text{S}\text{U}(2)$ or $\text{U}(2)$ and arbitrary matter content---including cases where perturbation theory exhibits $n!$ large--order behavior---resurgence never mixes the perturbative series with instanton sectors of non--zero instanton number. We expect that this result extends to any gauge group, for similar considerations. The reason that gauge theoretic instantons do not imply $n!$ large--order behavior of the perturbative expansion is most likely $\CN=2$ supersymmetry. For pure SYM, $\CN=2$ supersymmetry seems to ensure a massive cancellation of the $n!$ Feynman diagrams that one can draw at $n$--loop order in the gauge theory. 

On the other hand, for $\CN=2$ theories on $\BS^4$ with general matter content, more Feynman diagrams are added. The $n!$ large--order behavior of the perturbative series indicates that there are $\sim \CO(n!)$ surviving Feynman diagrams, notwithstanding the $\CN=2$ supersymmetry. This diagrammatic argument is also explaining why the resurgence triangle of our $\CN=2$ theories is somewhat in--between the bosonic and the supersymmetric resurgence triangles of \citep{Dunne:2012ae}. Different gauge--theoretic nonperturbative sectors are classified according to their instanton number and topological charge $(k,k')$, respectively (recall \eqref{topchargeZ} and \eqref{instchargeZ}). From a resurgence viewpoint, these sectors are neatly organized into the so--called resurgence triangle \citep{Dunne:2012ae}, as
\begin{equation}
\label{bosonictriangle}
\begin{tikzpicture}[%
  node distance=2cm, auto,
  back line/.style={densely dotted},
  cross line/.style={preaction={draw=white,->,line width=6pt}}]
  \node (00) {$(0,0)$};
  \node (11) [left of=00, below of =00, node distance=1cm] {$(1,1)$};
  \node (1-1) [right of=00, below of =00, node distance=1cm] {$(1,-1)$};
  \node (22) [left of=11, below of =11, node distance=1cm] {$(2,2)$};
  \node (2-2) [right of=1-1, below of =1-1, node distance=1cm] {$(2,-2)$};
  \node (20) [right of =22] {$(2,0)$};
  \node (33) [left of=22, below of =22, node distance=1cm] {$(3,3)$};
  \node (3-3) [right of=2-2, below of =2-2, node distance=1cm] {$(3,-3)$};
  \node (31) [right of =33] {$(3,1)$};
  \node (3-1) [right of =31] {$(3,-1)$};
  \node (44) [left of=33, below of =33, node distance=1cm] {$(4,4)$};
  \node (4-4) [right of=3-3, below of =3-3, node distance=1cm] {$(4,-4)$};
  \node (42) [right of =44] {$(4,2)$};
  \node (40) [right of =42] {$(4,0)$};
  \node (4-2) [right of =40] {$(4,-2)$};
  \node (5dots1) [left of=42, below of =42, node distance=1cm] {$\cdots$};
  \node (5dots2) [right of =5dots1] {$\cdots$};
  \node (5dots3) [right of =5dots2] {$\cdots$};
  \node (5dots4) [right of =5dots3] {$\cdots$};
  \draw[back line] (11) -- (1-1);
  \draw[back line] (22) -- (20) -- (2-2);
  \draw[back line] (33) -- (31) -- (3-1) -- (3-3);
  \draw[back line] (44) -- (42) -- (40) -- (4-2) -- (4-4);
  \draw[->,cross line] (00) -- (20);
  \draw[->,cross line] (20) -- (40);
  \draw[->,cross line] (22) -- (42);
  \draw[->,cross line] (2-2) -- (4-2);
  \draw[->,cross line] (33) -- (5dots1);
  \draw[->,cross line] (11) -- (31);
  \draw[->,cross line] (31) -- (5dots2);
  \draw[->,cross line] (1-1) -- (3-1);
  \draw[->,cross line] (3-1) -- (5dots3);
  \draw[->,cross line] (3-3) -- (5dots4);
\end{tikzpicture}
\end{equation}
\noindent
where the resurgence mixing (and any nonperturbative ambiguities cancelation) occurs only at fixed topological charge, \textit{i.e.}, along the columns of the above triangle (the solid arrows). For the flat--space two--dimensional extended supersymmetric theories considered in \citep{Dunne:2012ae}, this triangle becomes almost trivial as many sectors do not exist,
\begin{equation}
\label{susytriangle}
\begin{tikzpicture}[%
  node distance=2cm, auto,
  back line/.style={densely dotted},
  cross line/.style={preaction={draw=white,->,line width=6pt}}]
  \node (00) {$(0,0)$};
  \node (11) [left of=00, below of =00, node distance=1cm] {$(1,1)$};
  \node (1-1) [right of=00, below of =00, node distance=1cm] {$(1,-1)$};
  \node (22) [left of=11, below of =11, node distance=1cm] {$(2,2)$};
  \node (2-2) [right of=1-1, below of =1-1, node distance=1cm] {$(2,-2)$};
  \node (20) [right of =22] {$\times$};
  \node (33) [left of=22, below of =22, node distance=1cm] {$(3,3)$};
  \node (3-3) [right of=2-2, below of =2-2, node distance=1cm] {$(3,-3)$};
  \node (31) [right of =33] {$\times$};
  \node (3-1) [right of =31] {$\times$};
  \node (4dots1) [left of=31, below of =31, node distance=1cm] {$\cdots$};
  \node (4dots2) [right of =4dots1] {$\cdots$};
  \node (4dots3) [right of =4dots2] {$\cdots$};
  \draw[back line] (11) -- (1-1);
  \draw[back line] (22) -- (20) -- (2-2);
  \draw[back line] (33) -- (31) -- (3-1) -- (3-3);
\end{tikzpicture}
\end{equation}
\noindent
But for the $\CN=2$ theories on the compact four--sphere $\BS^4$ we consider in this paper, these sectors are still there. One could thus be tempted to write a resurgence triangle similar to the ``bosonic'' triangle above, in \eqref{bosonictriangle}. However, it is still the case that supersymmetry/localization ensure that resurgence only acts inside each fixed gauge--theoretic instanton \textit{and} topological sector as
\begin{equation}
\begin{tikzpicture}[node distance=2cm, auto]
\node {$(k,k')$} edge [out=10, in=170, loop] (kk');
\end{tikzpicture}
\end{equation}
\noindent
This essentially means that the resurgence triangle is actually simplified even further, albeit in a different fashion to what occurred in \eqref{susytriangle}: the triangular structure is of course still there, but resurgence only acts within each sector. The discontinuities in the partition function further show that there still is an underlying resurgence triangle, but now mixing sectors with weights $\exp \left( \frac{n^2}{\alpha} \right)$. One basic question still remains: if not gauge theory instantons, what are the classical field configurations associated with the $n!$ large--order behavior?

%%%%%%%%%%%%%%%%%%%%%%%%%%%%%%%%%%%%%%%%%%%%%%%%%%%%%%%%%%%%%%%%%
\subsection{Resurgence of ``Instantons''}
%%%%%%%%%%%%%%%%%%%%%%%%%%%%%%%%%%%%%%%%%%%%%%%%%%%%%%%%%%%%%%%%%

In the gauge theories discussed herein, the discontinuities associated with the $n!$ large--order behavior of perturbation theory are of the form of an exponential of a classical action. Thus, a natural question is whether this $n!$ behavior is associated with any semiclassical field configuration.

As stressed earlier, the $n!$ large--order behavior of perturbation theory cannot be produced by renormalons, since they do not show up in theories with vanishing $\beta$--functions. Therefore, the $n!$ behavior of the perturbative series in the $\CN=2$ SCF theory and in the $\CN=2^*$  theory must indeed be due to the fact that there are approximately $n!$ Feynman diagrams contributing at the $n$th loop order, as argued long ago by Dyson \cite{Dyson:1952tj}. Moreover, in specific examples with non--vanishing $\beta$--function, such as the pure $\CN=2$ SYM or SQCD discussed above, perturbation theory does not exhibit $n!$ growth behavior. This suggests that renormalons do not contribute to $\frac{1}{2}$BPS supersymmetric observables in any $\CN=2$ supersymmetric gauge theory. 

To understand the physical origin of the singularities in the complex Borel plane, it is useful to recall the  structure of the one--loop determinant within the localization procedure. Schematically, fluctuations from the fields associated with vector and hyper multiplets, in a representation $W$, appear at the Lagrangian level as
\bea
\CL_{\text{fluct}}^{\text{vec}} &\sim& \Phi_{\text{V}} \left( \nabla^2 + \left( \alpha \cdot a \right)^2 \right) \Phi_{\text{V}}, \\
\CL_{\text{fluct}}^{\text{hyper}} &\sim& \Phi_{\text{H}} \left( \nabla^2 + \left( \left( w \cdot a \right) \pm M \right)^2 \right) \Phi_{\text{H}},
\eea
\noindent
where $\alpha$ runs over all roots of the Cartan subalgebra $\mathfrak{h}$ of the gauge group, $w$ runs over the weights of $W$ and $a\in \mathfrak{h}$ represents the vev of the scalar of the vector multiplet. The differential operator $\nabla^2$ is the Laplacian on $\mathbb{S}^{4}$ (its eigenvalues given by spin eigenvalues of spherical harmonics), and $M$ represents the mass of the matter multiplets. In this case one finds 
\be
Z_{\text{1-loop}} \propto \frac{\prod_{\alpha} H \left( \rmi \alpha \cdot a \right)}{\prod_w H \left( \rmi w \cdot a + \rmi M \right)\, H \left( \rmi w \cdot a - \rmi M \right)},
\ee
\noindent
where the numerator originates from the vector multiplet and the denominator from the hypermultiplet. In particular, for the $\CN=2$ superconformal theory,
\be
Z_{\text{1-loop}} = \prod_{n=1}^{+\infty} \frac{\left( 1+\frac{4a^2}{n^2} \right)^{2n}}{\left( 1+\frac{a^2}{n^2} \right)^{8n}}.
\ee
\noindent
The poles in this expression thus correspond to points in the Coulomb branch of the moduli space where a given hypermultiplet fluctuation becomes a zero mode. Viewing the matrix theory as a Kaluza--Klein reduction to zero dimensions, these are points in the moduli space where a given Kaluza--Klein excitation becomes a zero mode of the determinant. Restoring the dependence on the four--sphere radius, $R$, this occurs when
\bea
\label{unoz}
&& \CN=2 \text{ SCF}: \qquad \quad \ \, a^2+\frac{n^2}{R^2} = 0, \\
\label{dosz}
&& \CN=2^* : \qquad \ \left( a\pm M \right)^2+\frac{n^2}{R^2} = 0.
\eea
\noindent
The first term in each equation represents the mass of the hypermultiplet in a vacuum labelled by $a_1=-a_2=a$. The second term is the contribution from the angular part of the Laplacian. The zero modes thus appear on the imaginary axis of $a$. It should be noticed that the integral defining the partition function is convergent, due to the fact that one deliberately chooses the integration contour over real $a$. As a result,  singularities do not lie on the integration region and the Gaussian factor renders the integral convergent at infinity. The fact that there are no singularities in the integration region is also underlying the Borel summability of the perturbative series.

An important question is whether the singularities described by \eqref{unoz} and \eqref{dosz} are a consequence of the infrared regularization provided by the four--sphere. In particular, one would like to know if the Borel transform also has the same singularities for the theory on flat spacetime. To address this question, we must separate conformal and non--conformal theories. Recall that for the superconformal theories the partition function is independent of the radius of the sphere. In particular, the perturbation series with $n!$ large--order behavior that one computes at finite radius is exactly the same, term by term, to the one obtained by computing Feynman diagrams on flat spacetime with an appropriate infrared regularization (some explicit calculations can be found in, \textit{e.g.}, \citep{Plefka:2001bu, Arutyunov:2001hs, Andree:2010na}). The four--sphere can simply be viewed as a gauge invariant, supersymmetric infrared regularization which does not affect the calculation of observables.

In a non--conformal theory, observables depend upon the radius of $\BS^4$. This means that the large--order behavior of the perturbative series for the theory on the four--sphere can be potentially different from the behavior of the perturbative series for the theory on $\BR^4$. The precise form of the perturbative expansion at large $R$ depends on the precise way in which the decompactification limit is taken. In $\CN=2^*$ $\text{S}\text{U}(N)$ SYM theory, the partition function depends only on the dimensionless combination $MR$. Therefore, taking $R\to\infty$ at fixed $M$ is equivalent to taking $M\to\infty$ at fixed $R$. The latter limit is known \cite{Seiberg:1994aj, Pestun:2007rz} to decouple the matter hypermultiplets provided we take at the same time the coupling $g_{\text{YM}}^2 \to 0$, keeping fixed
\be
\frac{4\pi^2}{g_{\text{YM,ren}}^2} = \frac{4\pi^2}{g_{\text{YM}}^2} - N \log MR.
\ee
\noindent
In this case the theory flows to pure SYM which, as shown above, has a perturbative series with finite non--zero radius of convergence. Thus the Borel singularities disappear in this case. 

On the other hand, if it is the $R\to\infty $ limit which is taken, at fixed coupling $g_{\text{YM}}^2$, one obtains the $\CN=2^*$ theory on flat space \cite{Russo:2013qaa}. This can be seen explicitly by using the asymptotic expansion of the Barnes $G$--function
\be
\log G (1+z) \approx \frac{z^2}{2} \log z 
\ee
\noindent
(in using this expression, the zeroes of the Barnes $G$--function disappear). This gives rise to the different terms in the one--loop effective action, which on flat $\BR^4$ space are proportional to $m^2 R^2 \log m^2 R^2$, for a field of mass $m$ and an infrared cutoff $1/R$. The mass runs over the mass spectrum of the theory
\begin{eqnarray}
\label{mv}
m_{ij}^{\text{v}} &=& \left|a_i-a_j\right|\ , \\ 
\label{mh}
m_{ij}^{\text{h}} &=& \left|a_i-a_j\pm M\right|\ .
\end{eqnarray}
\noindent
Then the integrand of the zero--instanton part of the partition function for the $\text{S}\text{U}(N)$ $\CN=2^*$ theory becomes $\exp \left( -S_{\text{eff}}[a] \right)$, with
\begin{eqnarray}
\label{Seff}
\frac{S_{\text{eff}}[a]}{R^2} &=& \frac{8\pi ^2}{g_{\text{YM}}^2}\, \sum_{i} a_i^2 -\sum_{i\neq j} \left\{ \frac{1}{4} \left(a_i-a_j+M\right)^2 \log \left(a_i-a_j+M\right)^2 R^2 + \right. \\
&&
\left. +\frac{1}{4} \left(a_i-a_j-M\right)^2 \log \left(a_i-a_j-M\right)^2 R^2 - \frac{1}{2} \left(a_i-a_j\right)^2 \log \left(a_i-a_j\right)^2 R^2 \right\}. \nonumber
\end{eqnarray}
\noindent
The first term originates from the coupling of the scalar $\Phi$ to the curvature of the sphere $\BS^4$, and it scales in the same way with the radius as the one--loop terms. Thus, in the decompactification limit the singularities of the integrand disappear. In agreement with this, it was shown in \citep{Russo:2013qaa, Russo:2013kea} that the large $N$ partition function has a weak coupling expansion in powers of $\exp \left( - \frac{8\pi^2}{N g^2_{\text{YM}}} \right)$, with finite radius of convergence.

Let us now discuss the semiclassical field configurations associated with the $n!$ large--order behavior of the perturbative series for the $\CN=2$ SCF theory. Specifically, one would like to identify the semiclassical field configurations that contribute to the discontinuities across the Stokes lines. In the localized partition function, where all classical gauge fields are set to zero and the scalar of the vector multiplet is set to $\Phi = {\text{diag}} \left( a_1,\ldots,a_N \right)$, the field configurations which contribute to the discontinuities are not manifest, but encrypted in the matrix integral. For $\text{S}\text{U}(2)$, after localization, we end up with an effective action for the constant part of the scalar field and the Coleman--Weinberg one--loop potential of the theory. The one--loop potential incorporates the effects of integrating out all physical fluctuations of the theory. We write
\be
Z \sim \int \rmd a\, \exp \left( -V(a) \right), 
\ee
\noindent
with
\be
\label{potz}
V(a) = \frac{a^2}{\alpha} - 2 \log a - \sum_{n=1}^{+\infty} n \log \left( 1 + \frac{4a^2}{n^2} \right)^2 + \sum_{n=1}^{+\infty} n \log \left( 1 + \frac{a^2}{n^2} \right)^8
\ee
\noindent
and $\alpha \equiv g^2_{\text{YM}}/16\pi^2$. The saddle--point equation is $V'(a)=0$. This has an infinite number of solutions, which for $\alpha \ll 1$  get close to the singular points. They are of the form
\be
\label{agas}
a^2_n \approx - n^2 - 4 n \alpha,  \qquad a^2_m \approx - \left(m+\frac{1}{2}\right)^2 + 4 \alpha \left(m+\frac{1}{2}\right) \ .
\ee
\noindent
For $\alpha\ll 1$, the $a_n$ lie slightly off the poles on the imaginary axis. The 
saddle--point evaluation of $Z$ then gives
\be
\label{npa}
Z \sim \rme^{\frac{n^2}{\alpha}}\, \frac{n^{4n} \rme^{4n}}{(4\alpha)^{4n+\frac{1}{2}}} \left( 1 + \CO(\alpha) \right).
\ee
\noindent
Thus, by deforming the contour of integration, upon crossing a pole one picks the discontinuity \eqref{npa}. This is in \textit{precise} agreement with the leading term in \eqref{parz}, showing that the discontinuities can indeed be interpreted in terms of semiclassical values for the vacuum expectation values for the scalar field of the vector multiplet. On the other hand, the $a^2_m$ lie near the zeroes of the Borel transform and give an irrelevant subleading contribution $\alpha^{4m+\frac{1}{2}}\, \rme^{\frac{m^2}{\alpha}}$.

Finally, note that here we have only explored semiclassical solutions of the one--loop effective action for the vev of the scalar field, and shown that these solutions effectively account for the discontinuities. It would be extremely interesting to identify non--trivial (possibly complex) classical field configurations in the non--localized theory, which have classical action proportional to $-n^2/\alpha$, with integer $n$ (and  classify them in terms of a topological number). We will not attempt this in this work, but leave it as an open problem for future research.

The analysis for the case of ABJM is very similar, so we omit  the derivation. The basic idea is as follows. Start with expression \eqref{eq:Part-funct-ABJM-first} for the ABJM partition function. The discontinuities in the Borel transform are related to residues at the double poles of the factor $\tanh^2 u$ in this partition function, upon the imaginary axis. There is an infinite set of saddle--points lying near these poles. They can be found by expanding the factor $1/\sinh (\pi u k)$ in the partition function in terms of exponential functions. In turn, this produces a double sum: a sum over saddles, and the series expansion of $1/\sinh( \pi u k)$ reproducing the second term within parenthesis of $\text{Disc}_{\theta_{\pm}} Z_{\text{ABJM}}\left(\BS^{3}\right) (k)$ in \eqref{discZABJMS3} (this is the leading term at large $\ell, m$ where the saddle--point approximation applies). One can then trace back the origin of the discontinuities to the original formula for the partition function. The poles of $\tanh^2 u$ come from zero modes of the one--loop determinant for the bifundamental matter fluctuations, that occur at specific values of the eigenvalues. In the original, non--localized action, presumably these saddle--points correspond to nonperturbative field configurations describing monopole instantons\footnote{For $\text{U}(2)_k \times \text{U}(2)_{-k}$ ABJM theory on flat spacetime, a family of BPS monopole instanton solutions was found in \cite{Hosomichi:2008ip}. It would be very interesting to find the corresponding solutions on the three--sphere, where the vacuum degeneracy of the moduli space is lifted.}.

%%%%%%%%%%%%%%%%%%%%%%%%%%%%%%%%%%%%%%%%%%%%%%%%%%%%%%%%%%%%%%%%%
%%%%%%%%%%%%%%%%%%%%%%%%%%%%%%%%%%%%%%%%%%%%%%%%%%%%%%%%%%%%%%%%%
\section{Comments and Outlook}\label{sec:comments}
%%%%%%%%%%%%%%%%%%%%%%%%%%%%%%%%%%%%%%%%%%%%%%%%%%%%%%%%%%%%%%%%%
%%%%%%%%%%%%%%%%%%%%%%%%%%%%%%%%%%%%%%%%%%%%%%%%%%%%%%%%%%%%%%%%%

In this work we began the study of the resurgence properties of the partition function (and free energy) in examples of supersymmetric field theories where exact analytic results are known. This allows for precise tests of resurgence properties in the context of gauge theories. Concretely, we investigated the resurgence properties of the $1/N$ expansion of Chern--Simons gauge theory on lens spaces, and weak--coupling expansions in ABJM gauge theory on $\mathbb{S}^3$, $\CN=2$ superconformal Yang--Mills theory on $\mathbb{S}^4$, and $\CN=2^*$ supersymmetric Yang--Mills theory also on $\mathbb{S}^4$. 

All our examples produced asymptotic perturbative expansions, with coefficients which grew factorially fast. The study of the many different asymptotic series at hand always led to associated Borel transforms which turned out to be meromorphic functions. This is an important point as it was the key element which allowed us to perform exact calculations from an analytical standpoint. Usually, the Borel surface may be a very intricate Riemann surface with many branch cuts. But in the meromorphic case the Borel surface is essentially the complex plane, turning the resurgent algebra ``abelian'', in the sense that all resurgence formulae may be computed exactly. In particular, although obviously starting off with asymptotic perturbative series, the remaining (would--be) ``resurgent'' series around the nonperturbative exponential sectors were always found to have a finite number of terms. This is naturally associated with the meromorphicity of the Borel transform, and in particular it implies that the ``instantons'' have trivial large--order behavior. Nontrivial resurgence would require finding an asymptotic series in front of the log branch--cut of the Borel transform (with this asymptotic series representing another sector, which would be ``mixing'' or ``resurging'' with the one of the Borel transform under consideration). This occurs in examples with a ``nonabelian'' branched structure for the Borel surface, see, \textit{e.g.}, \citep{Aniceto:2011nu, Schiappa:2013opa, Couso-Santamaria:2014iia}. 

In our four--dimensional examples, $\mathcal{N}=2$ superconformal and $\mathcal{N}=2^{*}$ supersymmetric Yang--Mills theories, there is never any mixing of the perturbative series with gauge--theory instantons described by the different sectors in the Nekrasov equivariant partition function. In these cases, the perturbative expansion of the partition function corresponds to the zero--instanton sector of the Nekrasov partition function in the integrand. Of course the total partition function includes higher (exponentially suppressed) instanton sectors, and loop expansions around each of these sectors will also lead to asymptotic series. At first, one would think that all these asymptotic series would relate to each other via resurgence. But what we have found is that resurgence acts within each sector alone. Each of these nonperturbative sectors has its own resurgence properties, described by their own meromorphic Borel transform. Thus, at fixed rank and small gauge coupling, there seems to be no resurgent mixing between different sectors of Nekrasov instantons (in principle, there could be some mixing if the Borel transform had poles on the positive real axes at specific points, but this is not the case). One of the lessons of the examples examined in this paper is that the perturbative series does not always permit the reconstruction of all nonperturbative effects, \textit{i.e.}, in the present case of $\CN=2$ theories it does not allow for a reconstruction of sectors computed around higher Nekrasov instantons.

Let us summarize the location and nature of the Borel singularities we found:
\begin{itemize}
\item Chern--Simons localized on lens spaces: we addressed the $1/N$ expansion for the free energy, and found a countable infinity of Stokes lines (condensing closer to the real axis, and reflective symmetric with respect to both real and imaginary axes). The Borel singularities along these lines consisted of first and second order poles.
\item $\text{U}(2)\times \text{U}(2)$ ABJM on $\mathbb{S}^3$: we addressed the perturbative (large level) expansion of the partition function. We found two Stokes lines, positive and negative imaginary axis, with the Borel singularities consisting of first and second order poles.
\item $\mathcal{N}=2$ superconformal Yang--Mills on $\mathbb{S}^4$: at fixed rank $N=2$, we addressed the partition function at small gauge coupling, $g_{\text{YM}}$. We found a single Stokes line, along the negative real axis. The Borel singularities consisted of poles of different order, with an increasing distance between consecutive poles.
\item $\mathcal{N}=2^{*}$ supersymmetric Yang--Mills on $\mathbb{S}^4$: again, at fixed rank $N=2$, we addressed the partition function at small gauge coupling, $g_{\text{YM}}$. This time around we found a countable infinity of Stokes lines, whose direction depended upon the mass parameter of this theory. Each Stokes line only had one singularity, a pole whose order was found to be higher the closer the Stokes line was to the negative real axis.
\end{itemize}

Along the way we produced many (analytical) resurgence formulae, for Stokes discontinuities and (exact) large--order relations. These formulae may be checked numerically against the perturbative expansions themselves, and we found very precise and ample agreement between the analytical resurgent analysis and the numerical large--order results, up to ``three--instantons'' level. We have also discussed the semiclassical, physical origin of the many Borel singularities we found. Their origin lies in zero modes within the one--loop determinant, that appear at certain complex values of the vacuum expectation values of the scalar field. At these points in the moduli space there is a cancellation between the squared mass of the hypermultiplet fluctuation in the vev background and the contribution $n^2/R^2$ coming from the Laplacian on $\mathbb{S}^4$. For the $\text{U}(2) \times \text{U}(2)$ ABJM model on $\mathbb{S}^3$, Borel singularities again occur due to the appearance of zero modes, this time associated with fluctuations of bifundamental matter.

One natural extension of the results in this work is to consider other localizable observables. For example, it would be of great interest to extend the resurgent analysis to Wilson loops in gauge theories, both in three dimensions, Chern--Simons and ABJM, or in four dimensions in $\CN=2$ gauge theories; for instance Wilson loops with less supersymmetry in $\CN=4$ supersymmetric Yang--Mills theory on $\mathbb{S}^4$ \citep{Giombi:2012ep}, or within ABJM theory following the results in \citep{Drukker:2009hy}.

Within the four--dimensional realm there are many other theories which are worth to explore, in particular $\CN=2$ gauge theories with different matter contents and different gauge structures, including quivers. It would be very interesting to investigate how changing the amount of matter (and eventually of supersymmetry) would interplay with the asymptotic and resurgence properties of specific observables.
Examples were given in section 4, were it was shown that  both $\CN=2$ pure SYM and $\CN=2$ SQCD with one fundamental and one antifundamental hyper have a 
perturbation series for the partition function with finite non--zero radius of convergence.

In general, finding exact expressions for gauge theoretic observables (in all of their parameters, coupling constants and rank of the gauge group) is a very ambitious goal. However, due to lack of integrability in generic gauge theories, many observables are only accessible via their (different) asymptotic expansions. Resurgence and transseries thus open a door into understanding nonperturbative phenomena in broad classes of theories: while closed--form expressions may not be available, and the asymptotic series at hand may not even be Borel summable, resurgent transseries allow for proper resummations (\textit{e.g.}, the median resummation recently discussed in \citep{Aniceto:2013fka}) and for adequate nonperturbative definitions of observables starting out with perturbation theory. Obtaining resurgent transseries descriptions of many gauge theoretic observables may be a key step in the study of diverse dualities and of many other fascinating nonperturbative phenomena inside gauge and field theories.

%%%%%%%%%%%%%%%%%%%%%%%%%%%%%%%%%%%%%%%%%%%%%%%%%%%%%%%%%%%%%%%%%
\acknowledgments
We would like to thank G\"ok\c ce Ba{\c s}ar, Aleksey Cherman, Ricardo Couso--Santamar\'\i a, Daniele Dorigoni, Gerald Dunne, Marcos Mari\~no, David Sauzin, Mithat \"Unsal and Andr\'e Voros for useful discussions and/or comments. In particular, we thank Gerald Dunne for comments on the draft. IA and RS would further like to thank CERN TH--Division for hospitality, where a large part of this work was conducted. IA was partially supported by the FCT--Portugal fellowship SFRH/BPD/69696/2010 and by the NCN grant 2012/06/A/ST2/00396. The research of RS was partially supported by the FCT--Portugal grants PTDC/MAT/119689/2010 and EXCL/MAT-GEO/0222/2012.
%%%%%%%%%%%%%%%%%%%%%%%%%%%%%%%%%%%%%%%%%%%%%%%%%%%%%%%%%%%%%%%%%

\newpage

%%%%%%%%%%%%%%%%%%%%%%%%%%%%%%%%%%%%%%%%%%%%%%%%%%%%%%%%%%%%%%%%%
%%%%%%%%%%%%%%%%%%%%%%%%%%%%%%%%%%%%%%%%%%%%%%%%%%%%%%%%%%%%%%%%%
\appendix
%%%%%%%%%%%%%%%%%%%%%%%%%%%%%%%%%%%%%%%%%%%%%%%%%%%%%%%%%%%%%%%%%
%%%%%%%%%%%%%%%%%%%%%%%%%%%%%%%%%%%%%%%%%%%%%%%%%%%%%%%%%%%%%%%%%

%%%%%%%%%%%%%%%%%%%%%%%%%%%%%%%%%%%%%%%%%%%%%%%%%%%%%%%%%%%%%%%%%
%%%%%%%%%%%%%%%%%%%%%%%%%%%%%%%%%%%%%%%%%%%%%%%%%%%%%%%%%%%%%%%%%
\section{Analytical Data for Large--Order Asymptotics}\label{sec:App-N2-data}
%%%%%%%%%%%%%%%%%%%%%%%%%%%%%%%%%%%%%%%%%%%%%%%%%%%%%%%%%%%%%%%%%
%%%%%%%%%%%%%%%%%%%%%%%%%%%%%%%%%%%%%%%%%%%%%%%%%%%%%%%%%%%%%%%%%

In this appendix we present the analytical data which was required for the expansions of the Borel transforms we computed in the main body of this paper; for the case of $\CN=2$ superconformal Yang--Mills theory in section \ref{sec:N2}, and for the case of $\CN=2^{*}$ supersymmetric Yang--Mills theory in section \ref{sec:N2star}. For the former, we first present the rational functions arising from the Nekrasov instanton partition functions in Table \ref{tab:Nekrasov-instanton-partfuncs}, which are required throughout the analysis. Then, we present in Table \ref{tab:N2-SYM-Coeff-Taylor-pert-series} the Laurent coefficients required for describing the singular behavior of the Borel transform of the perturbative series, which appears in \eqref{eq:N2-SYM-borel-taylor-exp-each-pole}. The subsequent Laurent coefficients for the Borel transforms 
of one and two instanton sectors, appearing in \eqref{eq:N2-SYM-borel-k-inst-Taylor-each-pole}, are presented in Table \ref{tab:N2-SYM-Coeff-Taylor-inst-series}. In Table \ref{tab:2star-borel-coeffs-0-1-inst} we then turn to the latter case of $\CN=2^{*}$ supersymmetric Yang--Mills theory, presenting the respective Laurent coefficients for describing the singular behavior of the Borel transform of the perturbative series, appearing in \eqref{eq:2star-borel-taylor-exp-each-pole}.

\bigskip

%%%%%%%%%%%%%%%%%%%%%%%%%%%%%%%%%%%%%%%%%%%%%%%%%%%%%%%%%%%%%%%%%
\begin{table}[h]
%%%%%%%%%%%%%%%%%%%%%%%%%%%%%%%%%%%%%%%%%%%%%%%%%%%%%%%%%%%%%%%%%
\centering
%%%%%%%%%%%%%%%%%%%%%%%%%%%%%%%%%%%%%%%%%%%%%%%%%%%%%%%%%%%%%%%%%
\begin{tabular}{|c|c|}
%%%%%%%%%%%%%%%%%%%%%%%%%%%%%%%%%%%%%%%%%%%%%%%%%%%%%%%%%%%%%%%%%
\hline 
$k$ & $Z_{k}\left(\rmi a\right)$\tabularnewline
\hline 
\hline 
0 & $1$\tabularnewline
\hline 
1 & $\frac{1}{2}\left(a^{2}-3\right)$\tabularnewline
\hline 
2 & $\frac{8a^{8}+a^{6}-91a^{4}-60a^{2}+132}{4\left(4a^{2}+9\right)^{2}}$\tabularnewline
\hline 
3 & $\frac{\left(a^{2}-1\right)\left(8a^{8}+11a^{6}-90a^{4}-207a^{2}-90\right)}{24\left(4a^{2}+9\right)^{2}}$\tabularnewline
\hline 
4 & $\frac{256a^{16}+3776a^{14}+13836a^{12}-30881a^{10}-277915a^{8}-491643a^{6}-33525a^{4}+569160a^{2}+330480}{384\left(4a^{2}+9\right)^{2}\left(4a^{2}+25\right)^{2}}$\tabularnewline
\hline 
5 & $\frac{\left(a^{2}+1\right)\left(256a^{16}+4416a^{14}+22460a^{12}-9485a^{10}-420951a^{8}-1240351a^{6}-876825a^{4}+1285800a^{2}+1652400\right)}{3840\left(4a^{2}+9\right)^{2}\left(4a^{2}+25\right)^{2}}$\tabularnewline
\hline 
6 & $\begin{array}{cc}
 & \frac{1}{23040\left(4a^{2}+9\right)^{2}\left(4a^{2}+25\right)^{4}\left(4a^{2}+49\right)^{2}}\left(32768a^{28}+1978368a^{26}+51087872a^{24}+\right.\\
 & +732417088a^{22}+6242760464a^{20}+29889128428a^{18}+44014800773a^{16}\\
 & -371126428388a^{14}-2708497959742a^{12}-8452958293056a^{10}-13389026503795a^{8}\\
 & \left.-6268361042400a^{6}+12446978984100a^{4}+20072072714400a^{2}+8724701160000\right)
\end{array}$
\tabularnewline
\hline 
7 & $\begin{array}{cc}
 & \frac{3+a^{2}}{322560\left(4a^{2}+9\right)^{2}\left(4a^{2}+25\right)^{4}\left(4a^{2}+49\right)^{2}}\left(32768a^{28}+2101248a^{26}+58196480a^{24}+\right.\\
 & +905593024a^{22}+8524396016a^{20}+46808453524a^{18}+102391212911a^{16}\\
 & -457776318917a^{14}-4531662127960a^{12}-16676922708654a^{10}-31270664940265a^{8}\\
 & \left.-21220269432225a^{6}+24481569794850a^{4}+52656021280800a^{2}+26358049620000\right)
\end{array}$
\tabularnewline
\hline 
8 & $\begin{array}{cc}
 & \frac{1}{20643840\left(4a^{2}+9\right)^{2}\left(4a^{2}+25\right)^{4}\left(4a^{2}+49\right)^{2}\left(4a^{2}+81\right)^{2}}\left(2097152a^{36}+243269632a^{34}+12587302912a^{32}\right.\\
 & +384342622208a^{30}+7707957613056a^{28}+106562568185856a^{26}+1029641436380928a^{24}\\
 & +6778934872979664a^{22}+26832406593705870a^{20}+20713659134020547a^{18}\\
 & -478489859018087120a^{16}-3408622894238047142a^{14}-12496805889021145798a^{12}\\
 & -27452324916541292565a^{10}-32182941054844974000a^{8}-3206638500090973200a^{6}\\
 & \left.+45064330148340720000a^{4}+56461877336011392000a^{2}+23000296271500800000\right)
\end{array}$
\tabularnewline
\hline 
%%%%%%%%%%%%%%%%%%%%%%%%%%%%%%%%%%%%%%%%%%%%%%%%%%%%%%%%%%%%%%%%%
\end{tabular}
%%%%%%%%%%%%%%%%%%%%%%%%%%%%%%%%%%%%%%%%%%%%%%%%%%%%%%%%%%%%%%%%%
\caption{Nekrasov instanton partition functions \citep{Nekrasov:2002qd, Alday:2009aq}, for $k=0,1,2,3,4,5,6,7,8$ instantons.
\label{tab:Nekrasov-instanton-partfuncs}}
%%%%%%%%%%%%%%%%%%%%%%%%%%%%%%%%%%%%%%%%%%%%%%%%%%%%%%%%%%%%%%%%%
\end{table}
%%%%%%%%%%%%%%%%%%%%%%%%%%%%%%%%%%%%%%%%%%%%%%%%%%%%%%%%%%%%%%%%%

%%%%%%%%%%%%%%%%%%%%%%%%%%%%%%%%%%%%%%%%%%%%%%%%%%%%%%%%%%%%%%%%%
\begin{landscape}
%%%%%%%%%%%%%%%%%%%%%%%%%%%%%%%%%%%%%%%%%%%%%%%%%%%%%%%%%%%%%%%%%
\begin{table}[t]
%%%%%%%%%%%%%%%%%%%%%%%%%%%%%%%%%%%%%%%%%%%%%%%%%%%%%%%%%%%%%%%%%
\begin{tabular}{cc|c|c|c|}
%%%%%%%%%%%%%%%%%%%%%%%%%%%%%%%%%%%%%%%%%%%%%%%%%%%%%%%%%%%%%%%%%
$f_{\ell}^{(n)}$ &  & \multicolumn{1}{c}{} & \multicolumn{1}{c}{$\qquad\qquad n$ ~ } & \tabularnewline
\cline{3-5} 
&  & {\footnotesize 1} & {\footnotesize 2} & {\footnotesize 3}\tabularnewline
\hline 
& {\footnotesize 0} & {\footnotesize $134217728\rmi\pi^{17/2}$} & {\footnotesize $2^{68}3^{4}\rmi\pi^{33/2}$} & {\footnotesize $2^{91}3^{25}5^{4}\rmi\pi^{49/2}$}\tabularnewline
\cline{2-5} 
& {\footnotesize 1} & {\footnotesize $46137344\rmi\pi^{13/2}$} & {\footnotesize $-2^{61}3^{3}127\rmi\pi^{29/2}$} & {\footnotesize $-2^{86}3^{23}5^{3}479\rmi\pi^{45/2}$}\tabularnewline
\cline{2-5} 
& {\footnotesize 2} & {\footnotesize $12058624\rmi\pi^{9/2}$} & {\footnotesize $2^{54}3^{3}7^{1}677\rmi\pi^{25/2}$} & {\footnotesize $2^{83}3^{21}5^{2}7^{1}7559\rmi\pi^{41/2}$}\tabularnewline
\cline{2-5} 
& {\footnotesize 3} & {\footnotesize $196608\rmi\pi^{5/2}\left(2\zeta(3)-11\right)$} & {\footnotesize $2^{47}3^{2}\rmi\pi^{21/2}\left(13824\zeta(3)-458585\right)$} & {\footnotesize $2^{78}3^{23}5\rmi\pi^{37/2}\left(3000\zeta(3)-264191\right)$}\tabularnewline
\cline{2-5} 
& {\footnotesize 4} & {\footnotesize -} & {\footnotesize $-2^{43}3\rmi\pi^{17/2}\left(862272\zeta(3)-4669063\right)$} & {\footnotesize $-2^{70}3^{19}\rmi\pi^{33/2}\left(410616000\zeta(3)-6974481067\right)$}\tabularnewline
\cline{2-5} 
& {\footnotesize 5} & {\footnotesize -} & {\footnotesize $2^{38}15\rmi\pi^{13/2}\left(3933792\zeta(3)+62208\zeta(5)-5133935\right)$} & {\footnotesize $2^{65}3^{17}5\rmi\pi^{29/2}\left(17979580800\zeta(3)+116640000\zeta(5)-91647354373\right)$}\tabularnewline
\cline{2-5} 
& {\footnotesize 6} & {\footnotesize -} & {\footnotesize $\hspace{-20pt}\begin{array}{cc}
& 2^{35}3^{2}5\rmi\pi^{9/2}\left(96\zeta(3)\left(864\zeta(3)-53533\right)-\right.\\
& \qquad\left.-316224\zeta(5)+1793831\right)
\end{array}$} & {\footnotesize $\begin{array}{cc}
& 2^{61}3^{15}5\rmi\pi^{25/2}\left(12960\zeta(3)\left(648000\zeta(3)-111026707\right)-\right.\\
& \qquad\qquad\qquad\left.-32939136000\zeta(5)+2657114966989\right)
\end{array}$}\tabularnewline
\cline{2-5} 
& {\footnotesize 7} & {\footnotesize -} & {\footnotesize $\begin{array}{cc}
& -2^{31}3^{2}35\rmi\pi^{5/2}\left(423360\zeta(3)^{2}-4198228\zeta(3)-\right.\\
& \qquad\qquad\left.-701568\zeta(5)-11664\zeta(7)+368031\right)
\end{array}$} & {\footnotesize $\begin{array}{cc}
& -2^{56}3^{13}35\rmi\pi^{21/2}\left(1296\zeta(3)\left(611064000\zeta(3)-20098385161\right)-\right.\\
& \qquad\qquad-\left.25\left(57049183872\zeta(5)+377913600\zeta(7)-761757221755\right)\right)
\end{array}$}\tabularnewline
\cline{2-5} 
$\ell$ & {\footnotesize 8} & {\footnotesize -} & {\footnotesize -} & {\footnotesize $\hspace{-20pt}\begin{array}{cc}
& 2^{50}3^{11}35\rmi\pi^{17/2}\left(10368\zeta(3)\left(53024576400\zeta(3)+699840000\zeta(5)\right)-\right.\\
& \qquad\qquad-5392800794086272\zeta(3)-600805335536640\zeta(5)-\\
& \qquad\qquad\left.-14084839872000\zeta(7)+1608834024697183\right)
\end{array}$}\tabularnewline
\cline{2-5} 
& {\footnotesize 9} & {\footnotesize -} & {\footnotesize -} & {\footnotesize $\hspace{-20pt}\begin{array}{cc}
& 2^{45}3^{11}35\rmi\pi^{13/2}\left(1492992\zeta(3)\left(45\zeta(3)\left(864000\zeta(3)-215951329\right)\right)-\right.\\
& \qquad\qquad-1492992\zeta(3)\left(453843000\zeta(5)-34245118364\right)+10706745108367872\zeta(5)\\
& \qquad\qquad\left.-6109832556501749+602668643673600\zeta(7)+4283020800000\zeta(9)\right)
\end{array}$}\tabularnewline
\cline{2-5} 
& {\footnotesize 10} & {\footnotesize -} & {\footnotesize -} & {\footnotesize $\hspace{-20pt}\begin{array}{cc}
& -2^{43}3^{9}5^{2}7\rmi\pi^{9/2}\left(3911863299995520\zeta(7)+98723629440000\zeta(9)+\right.\\
& \qquad+11664\left(116391168000\zeta(3)^{3}-5557270743648\zeta(3)^{2}-2099520000\zeta(5)^{2}\right)+\\
& \qquad+11664\left(\zeta(3)(-623016748800\zeta(5)-4199040000\zeta(7)+7662317088271)\right)+\\
& \qquad\qquad\left.+33979972678509312\zeta(5)-3967112417729327\right)
\end{array}$}\tabularnewline
\cline{2-5} 
& {\footnotesize 11} & {\footnotesize -} & {\footnotesize -} & {\footnotesize $\hspace{-20pt}\begin{array}{cc}
& 2^{37}3^{7}5^{2}77\rmi\pi^{5/2}\left(629055664909144704\zeta(5)+137112357991560192\zeta(7)-\right.\\
& \qquad-12708681666562711+4299816960\left(546750\zeta(5)-384051611\right)\zeta(3)^{2}+\\
& \qquad+1944\left(60039590860800\zeta(3)^{3}-2329627392000\zeta(5)^{2}\right)+\\
& \qquad-1944\zeta(3)\left(194693650375680\zeta(5)+4657155264000\zeta(7)-469564295444861\right)\\
& \qquad\qquad\left.+8345808670464000\zeta(9)+63262736640000\zeta(11)\right)
\end{array}$}\tabularnewline
\hline 
%%%%%%%%%%%%%%%%%%%%%%%%%%%%%%%%%%%%%%%%%%%%%%%%%%%%%%%%%%%%%%%%%
\end{tabular}
%%%%%%%%%%%%%%%%%%%%%%%%%%%%%%%%%%%%%%%%%%%%%%%%%%%%%%%%%%%%%%%%%
\caption{Coefficients $f_{\ell}^{(n)}$ of the Laurent expansion of the Borel
transform for the perturbative series, given in \eqref{eq:N2-SYM-borel-taylor-exp-each-pole}, near each of the pole singularities $s_{n} = -n^{2}2A = - \left(4\pi n\right)^{2}$, with $n=1,2,3$.
\label{tab:N2-SYM-Coeff-Taylor-pert-series}}
%%%%%%%%%%%%%%%%%%%%%%%%%%%%%%%%%%%%%%%%%%%%%%%%%%%%%%%%%%%%%%%%%
\end{table}
%%%%%%%%%%%%%%%%%%%%%%%%%%%%%%%%%%%%%%%%%%%%%%%%%%%%%%%%%%%%%%%%%
\end{landscape}
%%%%%%%%%%%%%%%%%%%%%%%%%%%%%%%%%%%%%%%%%%%%%%%%%%%%%%%%%%%%%%%%%

%%%%%%%%%%%%%%%%%%%%%%%%%%%%%%%%%%%%%%%%%%%%%%%%%%%%%%%%%%%%%%%%%
\begin{landscape}
%%%%%%%%%%%%%%%%%%%%%%%%%%%%%%%%%%%%%%%%%%%%%%%%%%%%%%%%%%%%%%%%%
\begin{table}
%%%%%%%%%%%%%%%%%%%%%%%%%%%%%%%%%%%%%%%%%%%%%%%%%%%%%%%%%%%%%%%%%
\centering
%%%%%%%%%%%%%%%%%%%%%%%%%%%%%%%%%%%%%%%%%%%%%%%%%%%%%%%%%%%%%%%%%
\begin{tabular}{cc|c|c|c|c|}
%%%%%%%%%%%%%%%%%%%%%%%%%%%%%%%%%%%%%%%%%%%%%%%%%%%%%%%%%%%%%%%%%
$f_{\ell}^{(n)[k]}$ &  & \multicolumn{1}{c}{} & \hspace{-90pt}$k=1$ & \multicolumn{1}{c}{} & \hspace{-90pt}$k=2$\tabularnewline
\cline{3-6} 
&  & \textit{\footnotesize $n=1$} & \textit{\footnotesize $n=2$} & \textit{\footnotesize $n=1$} & \textit{\footnotesize $n=2$}\tabularnewline
\hline 
& {\footnotesize 0} & \textit{\footnotesize $2^{29}\rmi\pi^{17/2}$} & \textit{\footnotesize $2^{66}3^{4}7^{2}\rmi\pi^{33/2}$} & \textit{\footnotesize $\frac{2^{27}3^{6}}{5^{4}}\rmi\pi^{17/2}$} & \textit{\footnotesize $\frac{2^{68}3^{8}5^{4}}{7^{4}}\rmi\pi^{33/2}$}\tabularnewline
\cline{2-6} 
& {\footnotesize 1} & \textit{\footnotesize $-2^{26}3\rmi\pi^{13/2}$} & \textit{\footnotesize $-2^{59}3^{3}6559\rmi\pi^{29/2}$} & \textit{\footnotesize $-\frac{2^{24}3^{4}157}{5^{5}}\rmi\pi^{13/2}$} & \textit{\footnotesize $-\frac{2^{61}3^{7}5^{2}25057}{7^{5}}\rmi\pi^{29/2}$}\tabularnewline
\cline{2-6} 
& {\footnotesize 2} & \textit{\footnotesize $2^{18}229\rmi\pi^{9/2}$} & \textit{\footnotesize $2^{52}3^{3}261043\rmi\pi^{25/2}$$ $} & \textit{\footnotesize $\frac{2^{16}3^{2}132233}{5^{5}}\rmi\pi^{9/2}$} & \textit{\footnotesize $\frac{2^{54}3^{6}567196937}{7^{6}}\rmi\pi^{25/2}$}\tabularnewline
\cline{2-6} 
& {\footnotesize 3} & \textit{\footnotesize $2^{13}3\rmi\pi^{5/2}\left(64\zeta(3)-547\right)$} & \textit{\footnotesize $2^{45}3^{2}\rmi\pi^{21/2}\left(677376\zeta(3)-27393881\right)$} & \textit{\footnotesize $\begin{array}{c}
\frac{2^{11}3^{2}}{5^{6}}\rmi\pi^{5/2}\left(388800\zeta(3)-\right.\\
\left.-6308041\right)
\end{array}$} & \textit{\footnotesize $\frac{2^{47}3^{6}5^{2}}{7^{7}}\rmi\pi^{21/2}\left(118540800\zeta(3)-6120082807\right)$}\tabularnewline
\cline{2-6} 
$\ell$ & {\footnotesize 4} & \textit{\footnotesize -} & \textit{\footnotesize $-2^{41}21\rmi\pi^{17/2}\left(6367680\zeta(3)-44274409\right)$} & \textit{\footnotesize -} & \textit{\footnotesize $-\frac{2^{43}3^{5}}{7^{7}}\rmi\pi^{17/2}\left(208828670400\zeta(3)-1950646403753\right)$}\tabularnewline
\cline{2-6} 
& {\footnotesize 5} & \textit{\footnotesize -} & \textit{\footnotesize $\begin{array}{c}
2^{36}3^{3}5\rmi\pi^{13/2}\left(24136800\zeta(3)+338688\zeta(5)-\right.\\
\left.-43700291\right)
\end{array}$} & \textit{\footnotesize -} & \textit{\footnotesize $\begin{array}{c}
\frac{2^{38}3^{4}5}{7^{8}}\rmi\pi^{13/2}\left(23194962457632\zeta(3)+\right.\\
\left.+280052640000\zeta(5)-60462377285165\right)
\end{array}$}\tabularnewline
\cline{2-6} 
& {\footnotesize 6} & \textit{\footnotesize -} & \textit{\footnotesize $\begin{array}{c}
2^{33}3^{2}5\rmi\pi^{9/2}\left(4064256\zeta(3)^{2}-\right.\\
-308616864\zeta(3)-\\
\left.-16365888\zeta(5)+169110935\right)
\end{array}$} & \textit{\footnotesize -} & \textit{\footnotesize $\begin{array}{c}
\frac{2^{36}3^{4}5}{7^{9}}\rmi\pi^{9/2}\left(3920736960000\zeta(3)^{2}-\right.\\
-382275476863056\zeta(3)\\
\left.-16930582351200\zeta(5)+332351429558641\right)
\end{array}$}\tabularnewline
\cline{2-6} 
& {\footnotesize 7} & \textit{\footnotesize -} & \textit{\footnotesize $\begin{array}{c}
-2^{31}3^{2}35\rmi\pi^{9/2}\left(5476464\zeta(3)^{2}-\right.\\
-70398829\zeta(3)-9716544\zeta(5)-\\
\left.-142884\zeta(7)+12070272\right)
\end{array}$} & \textit{\footnotesize -} & \textit{\footnotesize $\begin{array}{c}
-\frac{2^{31}3^{3}5}{7^{10}}\rmi\pi^{5/2}\left(36395461881533161+\right.\\
+259308\zeta(3)\left(25686838800\zeta(3)-447803368031\right)\\
\left.-12858187028787072\zeta(5)-162097968690000\zeta(7)\right)
\end{array}$}\tabularnewline
\hline 
%%%%%%%%%%%%%%%%%%%%%%%%%%%%%%%%%%%%%%%%%%%%%%%%%%%%%%%%%%%%%%%%%
\end{tabular}
%%%%%%%%%%%%%%%%%%%%%%%%%%%%%%%%%%%%%%%%%%%%%%%%%%%%%%%%%%%%%%%%%
\caption{Coefficients $f_{\ell}^{(n)[k]}$ of the Laurent expansion of the Borel transform for the perturbative expansions around each instanton sector $k=1,2$, given in \eqref{eq:N2-SYM-borel-k-inst-Taylor-each-pole}, and near each of the pole singularities  $s_{n} = -n^{2}2A = -\left(4\pi n\right)^{2}$, with $n=1,2$.
\label{tab:N2-SYM-Coeff-Taylor-inst-series}}
%%%%%%%%%%%%%%%%%%%%%%%%%%%%%%%%%%%%%%%%%%%%%%%%%%%%%%%%%%%%%%%%%
\end{table}
%%%%%%%%%%%%%%%%%%%%%%%%%%%%%%%%%%%%%%%%%%%%%%%%%%%%%%%%%%%%%%%%%
\end{landscape}
%%%%%%%%%%%%%%%%%%%%%%%%%%%%%%%%%%%%%%%%%%%%%%%%%%%%%%%%%%%%%%%%%

%%%%%%%%%%%%%%%%%%%%%%%%%%%%%%%%%%%%%%%%%%%%%%%%%%%%%%%%%%%%%%%%%
\begin{table}
%%%%%%%%%%%%%%%%%%%%%%%%%%%%%%%%%%%%%%%%%%%%%%%%%%%%%%%%%%%%%%%%%
\hspace{-35pt}
%%%%%%%%%%%%%%%%%%%%%%%%%%%%%%%%%%%%%%%%%%%%%%%%%%%%%%%%%%%%%%%%%
\begin{tabular}{cc|c|c|}
%%%%%%%%%%%%%%%%%%%%%%%%%%%%%%%%%%%%%%%%%%%%%%%%%%%%%%%%%%%%%%%%%
$f_{\ell,\pm}^{(n)}$ &  & \multicolumn{1}{c}{} & \hspace{-255pt}$n$\tabularnewline
\cline{3-4} 
&  & \textit{\footnotesize $1$} & \textit{\footnotesize $2$}\tabularnewline
\hline 
$\ell=0$ &  & $\pm32\rmi\,\rme^{2M^{2}(1+\gamma)}\left(M\pm\rmi\right)^{2}\pi^{\frac{5}{2}}\frac{G\left(2\mp\rmi M\right)^{2}G\left(\pm\rmi M\right)^{2}}{G\left(2\mp2\rmi M\right)G\left(\pm2\rmi M\right)}$ & $256\,\rme^{2M^{2}(1+\gamma)}\left(M\pm2\rmi\right)^{3}\pi^{\frac{9}{2}}\frac{G\left(3\mp\rmi M\right)^{2}G\left(-1\pm\rmi M\right)^{2}}{G\left(3\mp2\rmi M\right)G\left(-1\pm2\rmi M\right)}$\tabularnewline
\cline{2-4} 
$\ell=1$ &  & \textit{\footnotesize --} & $\begin{array}{c}
-64\,\rme^{2M^{2}(1+\gamma)}\left(M\pm2\rmi\right)\pi^{\frac{5}{2}}\frac{G\left(3\mp\rmi M\right)^{2}G\left(-1\pm\rmi M\right)^{2}}{G\left(3\mp2\rmi M\right)G\left(-1\pm2\rmi M\right)}\times\\
\times\left(5\mp3M\rmi-2\left(2\mp M\rmi\right)\gamma\right.\\
+\left(M\pm2\rmi\right)^{2}\left(\psi\left(3\mp\rmi M\right)+\psi\left(-1\pm\rmi M\right)\right)\\
\left.+\left(M\pm2\rmi\right)\left(M\pm\rmi\right)\left(\psi\left(3\mp2\rmi M\right)+\psi\left(-1\pm2\rmi M\right)\right)\right)
\end{array}$\tabularnewline
\hline 
%%%%%%%%%%%%%%%%%%%%%%%%%%%%%%%%%%%%%%%%%%%%%%%%%%%%%%%%%%%%%%%%%
\end{tabular}
%%%%%%%%%%%%%%%%%%%%%%%%%%%%%%%%%%%%%%%%%%%%%%%%%%%%%%%%%%%%%%%%%
\caption{Coefficients $f_{\ell,\pm}^{(n)}$ of the Laurent expansion of the Borel transform for the perturbative series given in \eqref{eq:2star-borel-taylor-exp-each-pole}, near each of the pole singularities $s_{n,\pm}(M)$ given in \eqref{eq:2star-sing-pertBorel}, with $n=1,2$.
\label{tab:2star-borel-coeffs-0-1-inst}}
%%%%%%%%%%%%%%%%%%%%%%%%%%%%%%%%%%%%%%%%%%%%%%%%%%%%%%%%%%%%%%%%%
\end{table}
%%%%%%%%%%%%%%%%%%%%%%%%%%%%%%%%%%%%%%%%%%%%%%%%%%%%%%%%%%%%%%%%%

\clearpage{}

%%%%%%%%%%%%%%%%%%%%%%%%%%%%%%%%%%%%%%%%%%%%%%%%%%%%%%%%%%%%%%%%%
\bibliographystyle{my-JHEP-3}
%%%%%%%%%%%%%%%%%%%%%%%%%%%%%%%%%%%%%%%%%%%%%%%%%%%%%%%%%%%%%%%%%
\bibliography{resurgentobservables_bib}
%%%%%%%%%%%%%%%%%%%%%%%%%%%%%%%%%%%%%%%%%%%%%%%%%%%%%%%%%%%%%%%%%

\end{document}